\documentclass[12pt]{article}

\usepackage[a4paper, margin=1in]{geometry}
\usepackage{graphicx}
\usepackage{tikz}
\usetikzlibrary{positioning,arrows.meta}
\tikzset{
    box/.style={
        draw,
        rounded corners,
        minimum width=3.5cm,
        minimum height=0.8cm,
        align=center,
        font=\sffamily
    },
    arrow/.style={
        -{Latex[length=3mm]},
        thick
    }
}
\usetikzlibrary{positioning}
\usepackage{amsmath, amssymb}
\usepackage{hyperref}
\usepackage{caption}
\usepackage{subcaption}
\usepackage{cleveref}
\usepackage{float}
\usepackage{wrapfig}
\usepackage{geometry}
\usepackage{booktabs}
\usepackage{siunitx}
\usepackage[
    backend=biber,
    sorting=none,
    maxnames=10
]{biblatex}
\newcommand{\mycaptionfont}{\fontsize{6}{7}\selectfont\sffamily}
\DeclareCaptionFont{customsize}{\mycaptionfont}
\title{\textbf{Spatio-Temporal Signatures of Intermittency in Helically Rotating Turbulence through Topological Data Analysis}}

\author{
Snigdhashree Mallick,  
Yashwanth Ramamurthi,  
Shiva Kumar Malapaka*, \\[1ex]
and Amit Chattopadhyay  \\
\href{mailto:malapaka@iiitb.ac.in}{* corresponding author : malapaka@iiitb.ac.in}
}

\date{
International Institute of Information Technology, Bangalore, India \\[1ex]}

\begin{document}
\maketitle

\begin{abstract}
A central challenge in hydrodynamic turbulence is identifying precisely when, and at which length scales, strong turbulent fluctuations (STFs) emerge and develop into intermittent events, which are often obscured by conventional statistical diagnostics. We address this problem by applying a Topological Data Analysis (TDA) framework to reveal the spatiotemporal signatures of intermittency in low-resolution ($128^3$) helically rotating turbulent flows. Vorticity magnitude and length-scale (eddy size) fields are used as scalar observables for TDA: vorticity characterizes rotational dynamics that generate multiscale flow structures, while length-scale fields encode the scales at which intermittent activity arises. Their evolving topology is quantified using persistence diagrams and Wasserstein-distance metrics. Compared with traditional statistical approaches, this framework is more sensitive to localized and short-lived flow variations, enabling clearer detection of intermittent behavior. Pronounced variations in Wasserstein-distance heatmaps provide direct signatures of STFs across space and time. Together, these results demonstrate that TDA offers an effective complementary tool for detecting STFs that lead to intermittency within turbulent regime.


\end{abstract}

\section{Introduction}

Intermittency remains one of the foremost challenges in turbulence research. It manifests as irregular, burst-like fluctuations in velocity, dissipation, and other flow quantities that strongly deviate from Gaussian and self-similar statistics. These rare, extreme events dominate high-order moments and play a disproportionate role in shaping the turbulent energy cascade. It is the non-linear nature of these intermittent events and their interactions with the flow, which create and destroy various structures and  drive the turbulent flow. 

Traditionally, intermittency has been quantified using statistical tools such as structure functions, probability density functions, kurtosis, and correlation functions, which characterize fluctuations through moments of physical observables in real space~\cite{tsinober,frisch1995turbulence}. These approaches primarily describe the degree of intermittency—for example, whether a flow is highly intermittent, weakly intermittent, or nearly Gaussian—while the occurrence and intensity of intermittent events depend sensitively on control parameters such as Reynolds number, rotation, and helicity. Several phenomenological frameworks have been proposed to interpret intermittency in terms of these parameters, including the log-Poisson model~\cite{She1994} and the Politano--Pouquet model~\cite{PolitanoPouquet1998}. More broadly, classical models such as the Kolmogorov--Obukhov (1962) log-normal model~\cite{kolmogorov1962,Obukhov1962}, the $\beta$-model~\cite{Frisch1978}, the multifractal formalism~\cite{Parisi1985,frisch1995turbulence}, and the She--Leveque model~\cite{She1994} have advanced the understanding of turbulence by capturing scale-dependent statistics and deviations from Gaussianity. Despite these developments, existing approaches remain fundamentally statistical, relying on global scaling laws and ensemble-averaged quantities. Consequently, a comprehensive understanding of intermittency in turbulent flows—and a first-principles description of turbulence—remains elusive~\cite{sreenivasan2025turbulence}.

Since, existing statistical models of turbulence do not explicitly incorporate the topology of coherent structures; we propose a topology-based framework, built upon existing TDA tools, to provide a more detailed characterization of intermittency. This approach captures the evolution of topological features in both space and time, moving beyond global statistical descriptors to directly track the emergence, interaction, and disappearance of coherent structures associated with strong fluctuations leading to intermittent events. Our approach primarily addresses the questions of \emph{when} and \emph{at which length scales} STFs emerge and evolve, enabling a clearer detection of the spatiotemporal signatures associated with intermittency in three-dimensional turbulent flows. By addressing these questions, the proposed framework offers a pathway toward understanding how intermittent events arise from the underlying turbulent dynamics.

\textbf{Motivation to use TDA:} TDA employs tools such as persistence diagrams, Wasserstein-distance metrics, and contour trees to extract global structural information from complex, high-dimensional data~\cite{carlsson2009topology, dey2022computational}. These methods provide a data-driven framework for identifying and quantifying physically interpretable topological features—such as clusters, loops, and voids—that may be overlooked by conventional statistical approaches. Unlike traditional methods that characterize intermittency through global moments or scaling laws, TDA captures the geometry, connectivity, and temporal evolution of coherent structures across multiple scales. Furthermore, the topological invariance of these features under continuous deformations makes TDA robust to noise and well suited for analyzing nonlinear, multiscale systems such as three-dimensional turbulent flows~\cite{makarenko2018, zafar2024topological, carreras2008topological}. Importantly, TDA is applied to the entire simulated dataset, from the beginning to the end of the simulation, enabling each time frame to be analyzed without omission; an approach rarely adopted in conventional turbulence analyses. Although computationally intensive, this comprehensive approach provides deeper insight into the evolving topology of turbulent structures and establishes a unified framework for uncovering structural signatures of intermittency.

\textbf{Previous Work:} In fluid dynamics, TDA has been applied extensively to two-dimensional turbulent flows~\cite{vivodtzev2022locally, nauleau2022topological, bridelbertomeu2023topological, banesh2020}, whereas only a limited number of studies extend these techniques to three-dimensional turbulence~\cite{lian2017, suzuki2021flow}, primarily demonstrating feasibility. Lian et al.~\cite{lian2017} analyzed 2D time-resolved particle image velocimetry data in a ``box of turbulence'' and showed that rate-of-strain tensor extraction requires median filtering at a \(3\times3\) vector spacing. Makarenko et al.~\cite{makarenko2018} applied Betti numbers and persistence diagrams to time-varying multiphase MHD ISM simulations, capturing magnetic-field-induced structural changes beyond autocorrelation methods. Nauleau et al.~\cite{nauleau2022topological} used persistent homology on an ensemble of 180 Kelvin--Helmholtz simulations and showed that Wasserstein distance outperforms the \(L_2\) norm in distinguishing solver configurations and detecting vortex formation. Bridel-Bertomeu et al.~\cite{bridelbertomeu2023topological} combined merge trees with topological persistence to track eddy evolution in cylindrical pipe flow, while Vivodtzev et al.~\cite{vivodtzev2022locally} employed persistence-based Morse--Smale segmentation to identify vortices in early-stage instabilities. Banesh et al.~\cite{banesh2020} used contour trees and persistence filtering in 2D kinetic plasma simulations to detect multiscale magnetic reconnection dynamics, including nested magnetic islands and X-points, representing one of the earliest applications of TDA to MHD reconnection, albeit restricted to two dimensions.

Turbulent flow simulations are mostly three-dimensional, and a complete characterization of intermittency requires information from all spatial directions. Since the motivation of this work is to understand the spatio-temporal origins of STFs using TDA, we perform our analysis on the full three-dimensional dataset, in contrast to many previous studies that rely on two-dimensional representations. 

The main contributions of this study are summarized as follows:

\begin{itemize}

\item Identification of STFs leading to intermittent events via tracking the continuous evolution of topological features in time, by applying TDA techniques to seven low-resolution ($128^3$) helically rotating hydrodynamic turbulent flow datasets generated by systematically varying helicity and rotation.
\item Computation of persistence diagrams from physically meaningful scalar fields such as vorticity and local length scale (eddy size) at each time frame, followed by quantitative comparison of the persistence diagrams using Wasserstein-distance metric, and the computation of contour trees in order to show the emergence and evolution of STFs leading to intermittent events in temporal frame.
\item Computation and literature-based classification of local length scales, together with identification of the most strongly affected scales and the physical quantities showing the largest deviations from their mean behaviour, thereby supporting the onset of STFs leading to intermittency in spatial frame.
\end{itemize}

The proposed workflow clearly localizes STFs, which can be attributed to intermittency in the flow, both in space (in terms of which turbulent length scales are most strongly affected) and in time (i.e., at exactly what points STFs emerge within the simulation data). To the best of our knowledge, this represents the first demonstration of intermittent event localization using a data-driven topological framework. We find that the employed TDA techniques effectively classify and quantify turbulent structures and their associated scales as coherent topological features evolving across space and time, thereby providing meaningful physical insights into the mechanisms underlying intermittency.

The results not only reaffirm the established role of energy dissipation ~\cite{She1994} in characterizing intermittency, but also demonstrate that quantities such as kinetic helicity are influenced by intense turbulent fluctuations. Thus, TDA provides a robust, multiscale description of how STFs emerge, interact, and vanish.

The remainder of this paper is organized as follows. Section~\ref{sec:theory}
describes the numerical setup used to generate the turbulent flow datasets, introduces the required theoretical background and provides the data description. Section~\ref{sec:results} presents a detailed analysis of one representative flow configuration, while the results for the remaining six cases are included in the Appendix. Section~\ref{sec:discussion} offers a comparative discussion across all configurations and Section~\ref{sec:conclusion} concludes with final remarks. Section~\ref{sec:methods} outlines the specific TDA methodology implemented and the TDA techniques utilized in this study followed by the supplementary material for this article.

\section{Data Description}
\label{sec:theory}

The helically rotating turbulent flows are governed by the incompressible Navier–Stokes equation in a rotating frame,
\begin{equation}
\frac{\partial \mathbf{u}}{\partial t}
+ \mathbf{u}\cdot\nabla\mathbf{u}
+ 2\boldsymbol{\Omega}\times\mathbf{u}
= -\frac{\nabla p}{\rho_f}
+ \nu\nabla^2\mathbf{u}
+ \mathbf{f}_e + \mathbf{f}_h,
\end{equation}
where $\mathbf{u}$ is the velocity field, $\mathbf{\nu}$ the kinematic viscosity, $\mathbf{f}_e$ the large-scale energy forcing, and $\mathbf{f}_h$ the helicity forcing. Vorticity is defined as $\boldsymbol{\omega}=\nabla\times\mathbf{u}$.

The Taylor-scale Reynolds number is
$Re_\lambda=\frac{u'\lambda}{\nu}$ with
$\lambda=\sqrt{\frac{15\nu u'^2}{\varepsilon}}$,
where $u'$ is the root–mean–square velocity fluctuation and $\varepsilon$ the mean energy dissipation rate. For reference, the integral-scale Reynolds number is $Re_L=\frac{u'L}{\nu}$, where $L$ denotes the energy-containing length scale~\cite{pope2000turbulent}.

Rotational effects are characterized by the Rossby number
$Ro=\frac{u}{fL}$ with $f=2\Omega\sin\varphi$ and the Coriolis number
$Co=\frac{\Omega_0 L}{\sqrt{\langle u^2\rangle}}$
~\cite{brissaud1973helicity,davidson2013}. Kinetic helicity,
$\int_V \mathbf{u}\cdot\boldsymbol{\omega}\,dV$,
measures velocity–vorticity alignment and the topological complexity of vortex structures
~\cite{moffatt1,frisch1975inverse}.

All datasets are generated using a parallel pseudo-spectral solver on a $128^3$ periodic domain with standard $2/3$ de-aliasing and Adams–Bashforth time stepping. The viscosity is fixed at $\nu=0.005$ and the external energy forcing $\mathbf{f}_e$ is injected in large scales and is identical in all runs. The Coriolis number and helicity forcing ($-1,0,1$) are varied to produce seven distinct turbulent configurations summarized in \cref{tab:helicity_rotation}.

\begin{table}[h!]
\centering
\renewcommand{\arraystretch}{1.1}
\setlength{\tabcolsep}{4pt} 
\small 
\begin{tabular}{|c|c|c|p{4cm}|c|}
\hline
\textbf{S.No} & \textbf{Helicity} & \textbf{Rotation (Coriolis No.)} & \textbf{Configuration} & \textbf{Nickname} \\
\hline
1 & None (0)     & None (0)      & No rotation, No helicity           & norotnohel \\
\hline
2 & None (0)     & Maximum (1)   & Maximum rotation, No helicity      & maxrotnohel \\
\hline
3 & Maximum (1)  & Maximum (1)   & Maximum rotation, Maximum helicity & maxrotmaxhel \\
\hline
4 & Maximum (1)  & None (0)      & No rotation, Maximum helicity      & norotmaxhel \\
\hline
5 & Minimum (-1) & Maximum (1)   & Maximum rotation, Minimum helicity & maxrotminhel \\
\hline
6 & Minimum (-1) & None (0)      & No rotation, Minimum helicity      & norotminhel \\
\hline
7 & Minimum (-1) & Very High (3) & Very high rotation, Minimum helicity & highrotminhel \\
\hline
\end{tabular}
\caption{Simulated configurations with varied helicity and rotation parameters, along with their assigned nicknames.}
\label{tab:helicity_rotation}
\end{table}

\newpage
We begin with a three-dimensional turbulent velocity field defined on a spatial grid, from which we construct two complementary scalar observables: the vorticity magnitude and a local characteristic length scale associated with energy transfer. The vorticity magnitude captures regions of intense rotational activity, often linked to coherent structures and extreme events, while the length-scale field encodes the spatial hierarchy over which these structures evolve. Together, these fields provide a physically grounded representation of multiscale turbulent dynamics.

To extract topological information from the scalar fields, we employ tools from TDA (definitions of TDA can be found in section~\ref{subsec:tda}). As the scalar fields are varied across different thresholds, topological features such as connected components (0D), loops or tunnels (1D), and enclosed voids (2D) emerge (birth) and later disappear (death), reflecting the multiscale organization of the flow. At each threshold, the scalar field defines level sets (isosurfaces in three dimensions) whose connectivity encodes the underlying structure of the flow. This evolution can be examined through both sublevel and superlevel filtrations, which respectively track structures formed below and above a given threshold, thereby capturing complementary aspects of the underlying field. This information is summarized using persistence diagrams, which record the birth and death of these features and provide a compact representation of their evolution across scales (i.e., scalar field thresholds). Features that persist over a wide range between birth and death correspond to more robust and dynamically relevant structures in the flow. Using the vorticity magnitude and local length scale as input fields, we compute persistence diagrams at each time step.

Having computed persistence diagrams for each timestep, we quantify temporal changes in topology by comparing them using the Wasserstein distance, which measures the optimal transport cost required to match topological features between two diagrams. Since, each dataset spans a minimum of 40 snapshots, separated by approximately $10^4$ timesteps; this produces $40 \times 40$ pairwise distance matrices, which are visualized as heatmaps to capture spatio-temporal structural differences in the evolving turbulent flow. Large Wasserstein distances therefore indicate significant structural differences in the underlying flow fields and are treated as regions of interest for further analysis.

In parallel, we analyze the global connectivity of the scalar field using contour trees, which track how connected regions form, merge, and split as the scalar threshold is varied. This provides an interpretable description of large-scale structural transitions and complements the persistence-based analysis. Further, to spatially localize these changes, the length-scale field is partitioned into regimes corresponding to characteristic turbulent scales, allowing us to identify the spatial ranges in which intermittent events emerge.

\begin{figure}[H]
\centering
\begin{tikzpicture}[
    scale=0.8, transform shape,
    node distance=1.0cm and 1.8cm,
    every node/.style={
        draw, rounded corners,
        font=\scriptsize,
        align=center,
        minimum width=2.8cm,
        minimum height=0.7cm
    }
]

\node (vel) {3D Velocity Field};

\node[below left=0.4cm and 1.8cm of vel] (vort) {Vorticity $\nabla \times u$};
\node[below=0.8cm of vort] (vortmod) {Vorticity modulus};

\node[below=0.5cm of vel] (kh) {Kinetic helicity $H_k$, \\ Kinetic energy $E_k$};

\node[below right=0.4cm and 1.8cm of vel] (ed) {Energy dissipation rate $\epsilon$};
\node[below=0.5cm of ed] (len) {Length scale $L$};

\node[below=0.4cm and 0.1cm of len] (bin) {Binning};
\node[below right=0.4cm and 0.2cm of len] (ct) {Contour trees};

\node[below=3.2cm of vel] (pd) {Persistence diagrams};
\node[below=0.8cm of pd] (wd) {Wasserstein distances};

\draw[->] (vel) -- (vort);
\draw[->] (vort) -- (vortmod);
\draw[->] (vortmod) -- (pd.north west);

\draw[->] (vel) -- (kh);
\draw[->] (vel) -- (ed);
\draw[->] (ed) -- (len);

\draw[->] (len.west) -- (pd.north east);
\draw[->] (len) -- (bin);
\draw[->] (len) -- (ct);

\draw[->] (pd) -- (wd);

\end{tikzpicture}

\caption{Schematic overview of our approach}
\label{fig:tda_pipeline_clean}
\end{figure}
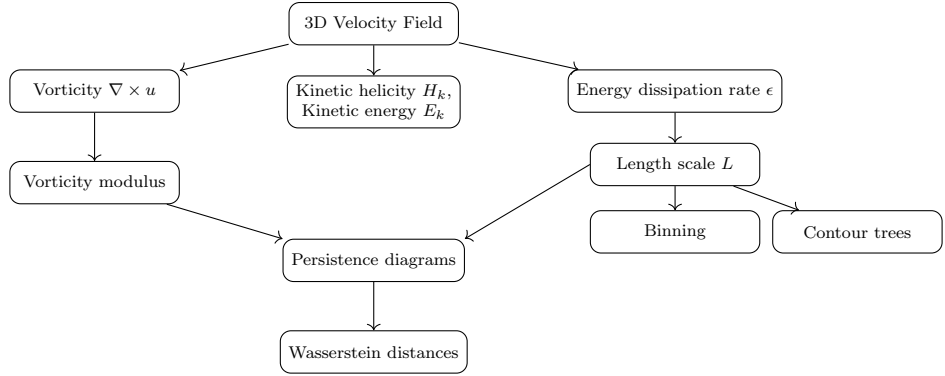

Finally, we examine the temporal evolution of topological feature counts alongside key physical observables, including energy dissipation rate, kinetic energy, and kinetic helicity. By correlating changes in topological features with variations in these physical quantities, we assess the extent to which topological signatures capture the onset and evolution of intermittent events in the flow. A schematic overview of the approach is shown in Fig ~\ref{fig:tda_pipeline_clean}.

\section{Results}
\label{sec:results}
We now present our analysis for a single representative case, namely \textit{highrotminhel} (entry no.~7 in \cref{tab:helicity_rotation}), to maintain clarity and focus on the techniques and results discussed herein. For the remaining cases, presented in the same order as described in the subsequent sections, are included in the appendix. 

\subsection{Identification of STFs leading to intermittent events using Persistence Diagrams: Highrotminhel case}
\label{subsec:highrotminhel}

\begin{figure}[H]
\centering

\begin{subfigure}[b]{0.48\linewidth}
    \centering
    \includegraphics[width=\linewidth]{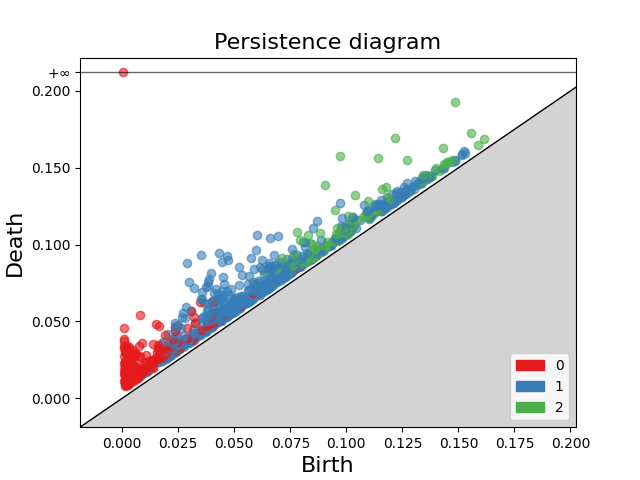}
    \caption{PD for vorticity at $t = 52$}
    \label{fig:pd-highrotminhel-a}
\end{subfigure}
\hfill
\begin{subfigure}[b]{0.48\linewidth}
    \centering
    \includegraphics[width=\linewidth]{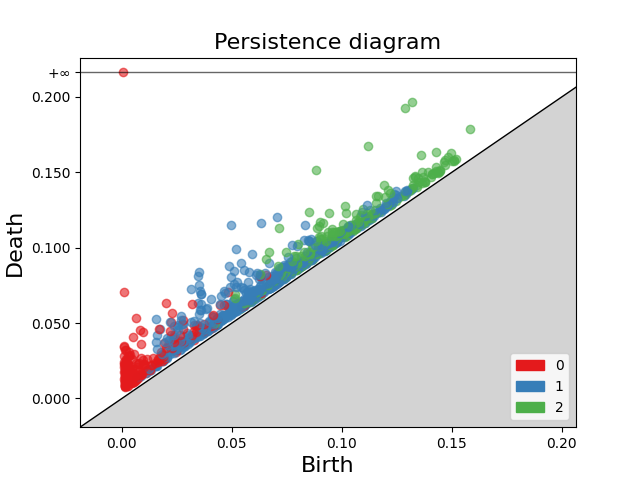}
    \caption{PD for vorticity at $t = 58$}
    \label{fig:pd-highrotminhel-b}
\end{subfigure}

\vspace{1em}

\begin{subfigure}[b]{0.45\linewidth}
    \centering
    \includegraphics[width=\linewidth]{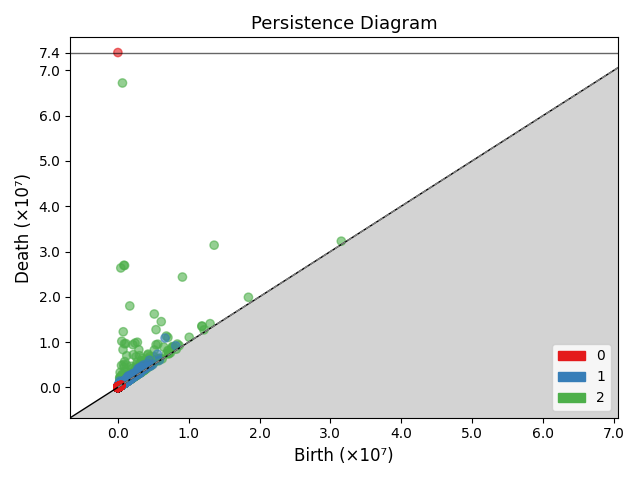}
    \caption{PD for length scale at $t = 52$}
    \label{fig:pd-highrotminhel-c}
\end{subfigure}
\hfill
\begin{subfigure}[b]{0.46\linewidth}
    \centering
    \includegraphics[width=\linewidth]{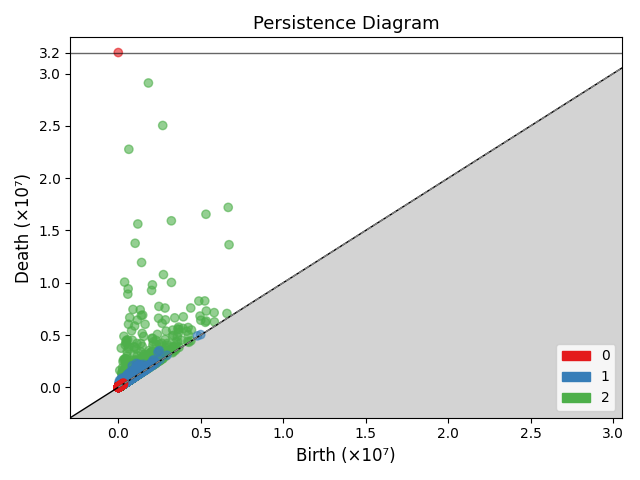}
    \caption{PD for length scale at $t = 58$}
    \label{fig:pd-highrotminhel-d}
\end{subfigure}

\caption{Persistence diagrams for selected timesteps of the \textit{highrotminhel} case.}
\label{fig:pd-highrotminhel}
\end{figure}

The persistence diagrams for the \textit{highrotminhel} configuration are examined for two scalar fields: vorticity (\cref{fig:pd-highrotminhel-a,fig:pd-highrotminhel-b}) and length scale (\cref{fig:pd-highrotminhel-c,fig:pd-highrotminhel-d}). In the vorticity diagrams, a large number of one-dimensional (1D) features appear compared to the zero-dimensional (0D) and two-dimensional (2D) features, while in the length-scale diagrams, most features observed are two-dimensional. This distinction reflects the contrasting geometric nature of the two scalar fields under TDA.

The vorticity field, which exhibits relatively small variations across time steps, is dominated by filamentary and tunnel-like structures, such as the vortex tubes and columnar vortices typically observed in rotating turbulence~\cite{biferale2016coherent, mininni2010part2}. These coherent filaments give rise to numerous 1D features. However, most of the 0D, 1D, and 2D features in the vorticity persistence diagrams lie close to the diagonal line, indicating that they are short-lived and likely represent topological noise rather than physically significant structures.

In contrast, the length-scale field varies over a much broader range and forms volumetric, bubble-like regions of coherence. Such three-dimensional coherent structures are consistent with the formation and breakup of vortex sheets and helical regions in rotating flows~\cite{moffatt2014helicity, mininni2009helicity}. This produces a predominance of persistent 2D voids far from the diagonal, revealing the underlying topology of the field and indicating that the scalar field spans multiple orders of magnitude, unlike the vorticity field. This pattern is characteristic across all other flow configurations examined. The diagrams also reveal the progressive emergence of large-scale structures as the simulation advances, a phenomenon well documented in high-resolution rotating DNS studies~\cite{pouquet2010interplay}.

Beyond this qualitative observation, however, the persistence diagrams alone do not provide sufficient insight into the flow dynamics. Therefore, we proceed to compute the Wasserstein distances between the persistence diagrams for the entire dataset to gain a more quantitative understanding of the underlying flow behavior.

\begin{figure}[H]
    \centering
    \begin{subfigure}[b]{0.48\linewidth}
        \centering
        \includegraphics[width=\linewidth]{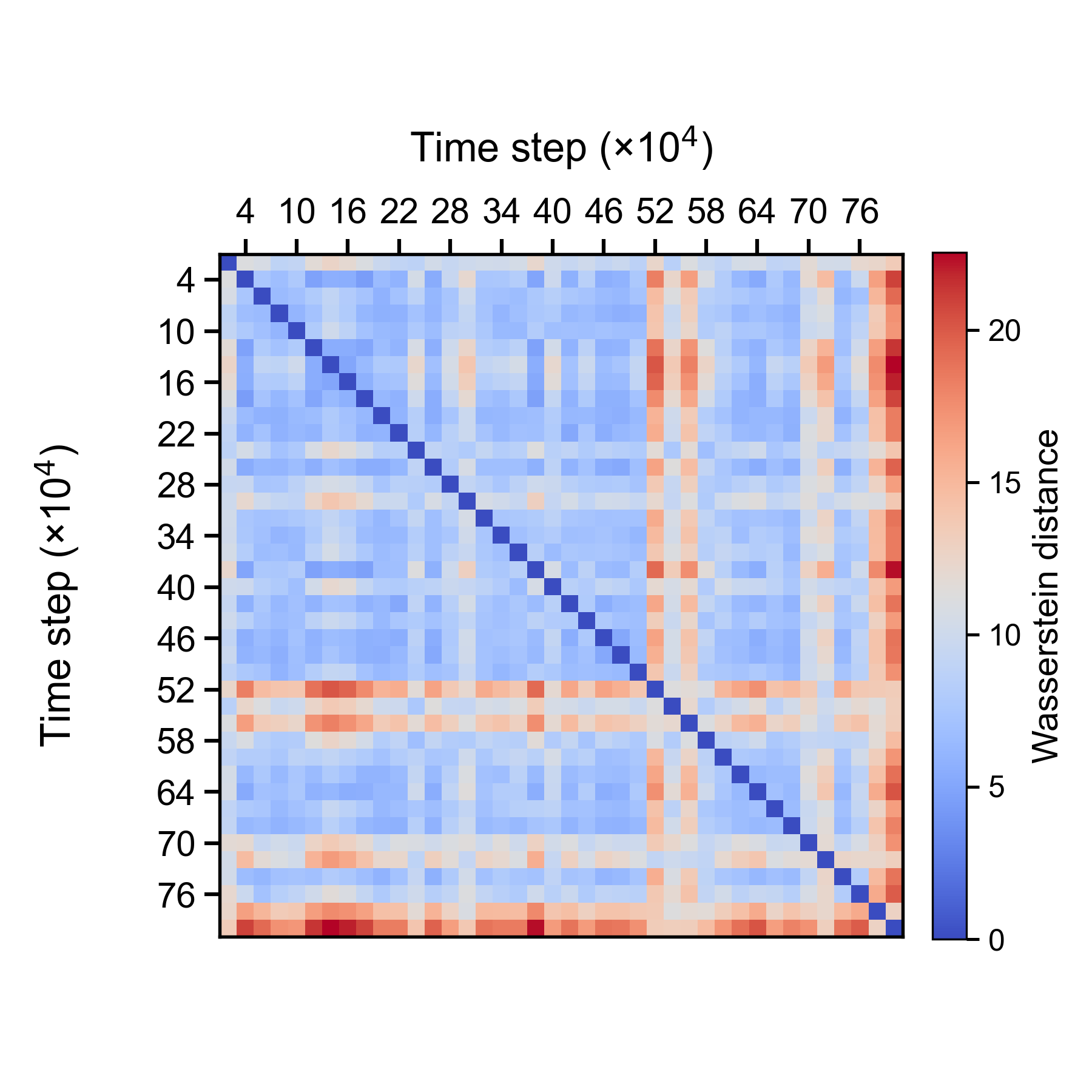}
        \caption{Vorticity as the scalar function}
        \label{fig:wass-vorticity-highrotminhel}
    \end{subfigure}
    \hfill
    \begin{subfigure}[b]{0.48\linewidth}
        \centering
        \includegraphics[width=\linewidth]{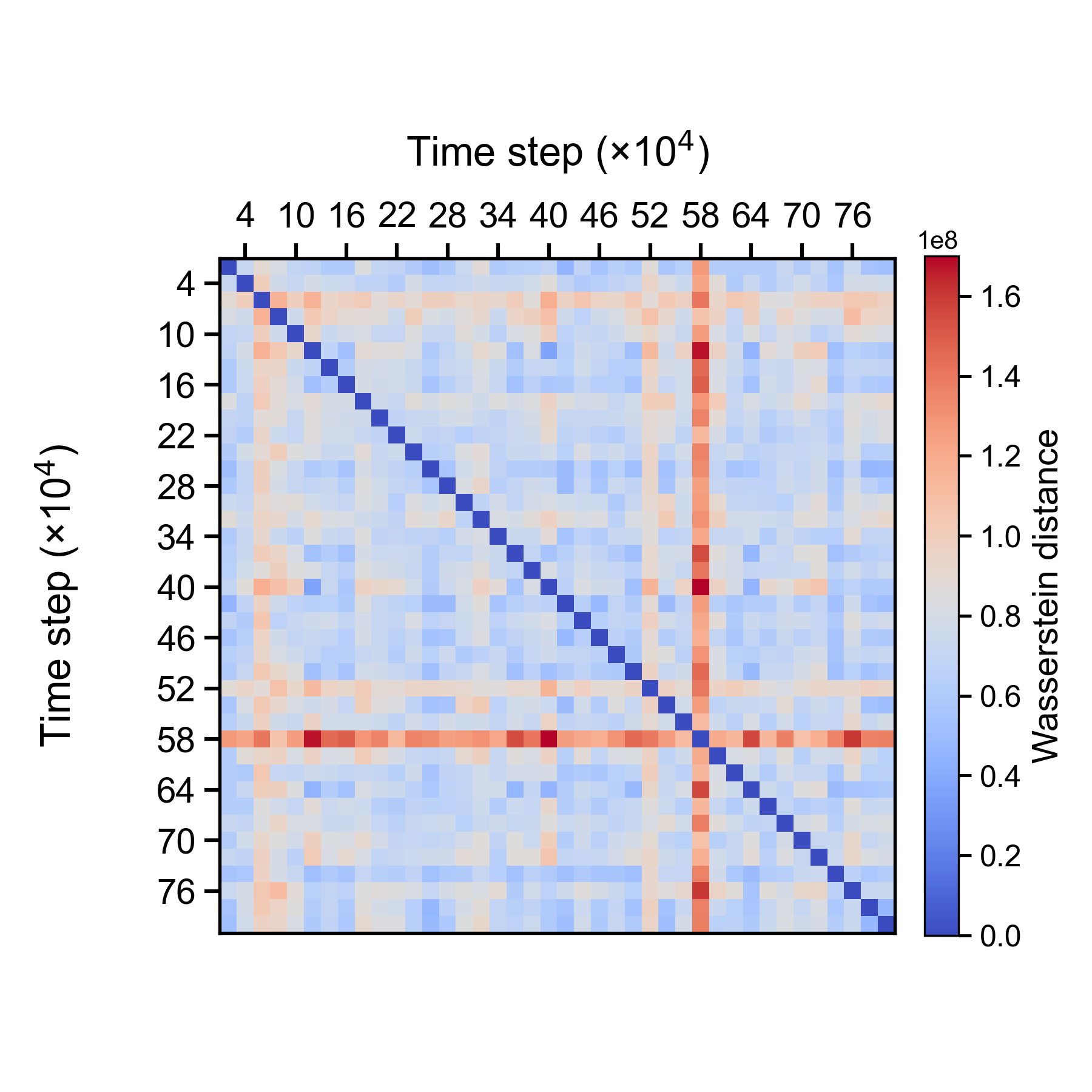}
        \caption{Length scale as the scalar function}
        \label{fig:wass-lengthscale-highrotminhel}
    \end{subfigure}
    
    \caption{Comparison of Wasserstein distance heatmaps for the \textit{highrotminhel} case. The left panel corresponds to \textbf{vorticity} and the right to \textbf{length scale} as the scalar function. Both exhibit similar deviations at specific timesteps, indicating STFs during those intervals.We have divided the time frames into certain tick labels to get an overall clearer picture.}
    \label{fig:wass-comparison-highrotminhel}
\end{figure}

In \cref{fig:wass-comparison-highrotminhel}, we present these Wasserstein distance matrices computed between 40 persistence diagrams, producing a $40\times40$ pairwise distance heatmap, allowing spatio-temporal structural differences in the evolving turbulent flow to be visualized. The heatmaps are shown for the two scalar fields considered in this study: vorticity magnitude (left panels) and the length-scale field (right panels). In both cases, a distinct band of timesteps between $t=52$ and $t=58$ exhibits significantly larger Wasserstein distances relative to the rest of the matrix, appearing as prominent red stripes in \cref{fig:wass-vorticity-highrotminhel,fig:wass-lengthscale-highrotminhel}. This pattern marks temporal intervals in which the underlying field undergoes significant structural or dynamical reorganization, likely driven by STFs (a hallmark of intermittency).  From this point onward, we interpret such intervals as indicators of \textbf{strong intermittent events}.

Having identified the time intervals at which STFs, and hence intermittency, might have occurred, we next determine the corresponding length scales at which these events arise in order to identify precisely the spatial scales which are most affected during the intermittent activity.

\subsection{Spatial detection of intermittent events using length scale classification: Highrotminhel Case}
\label{subsec:ls-classify-highrotminhel}
\begin{figure}[H]
    \centering
    \includegraphics[width=0.45\linewidth]{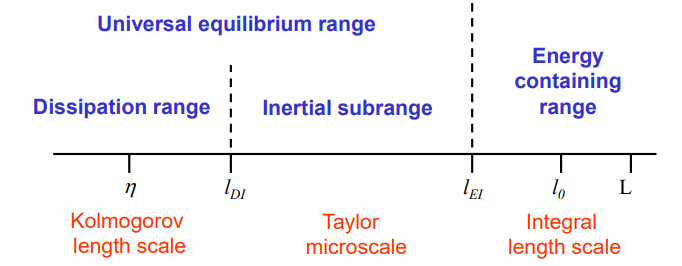}
    \caption{Length scale hierarchy used for classification, reflecting Richardson’s cascade model~\cite{pope2000turbulent}.}
    \label{fig:length-scales}
\end{figure}

The whole range of local length scales (\( L \)) are subsequently categorized into five regimes based on the definitions of different turbulent scales present in the literature~\cite{pope2000turbulent}. All these scales usually are related to each other through the Kolmogorov's scale $\eta = \left(\frac{\nu^3}{\epsilon}\right)^{1/4}$ which is the
 smallest dissipative scale. The dissipative-intermediate scale $l_{DI} = 60\eta$, the energy-intermediate scale $l_{EI} = l_0/6$, and the integral scale $l_0 = \eta \cdot Re_l^{3/4}$ , are the three other scales used for categorization (as shown in ~\cref{fig:length-scales}). A total of five regimes are considered: (1) $l \leq \eta$, (2) $\eta < L \leq l_{DI}$, (3) $l_{DI} < L \leq l_{EI}$, (4) $l_{EI} < L \leq l_0$, and (5) $L > l_0$.\\ 

This classification allows a clearer distinction of flow structures across different length-scale ranges and enables a systematic examination of how flow behavior varies between them. After assigning all $L$ values to the five predefined regimes, we compute the absolute change in the number density of the binned $L$ values across successive timesteps. This analysis reveals how different length-scale regimes evolve relative to one another and identifies the scales that exhibit the most significant variations, potentially associated with STFs leading to intermittent events.

\begin{figure}[H]
    \centering
    \includegraphics[width=0.6\linewidth]{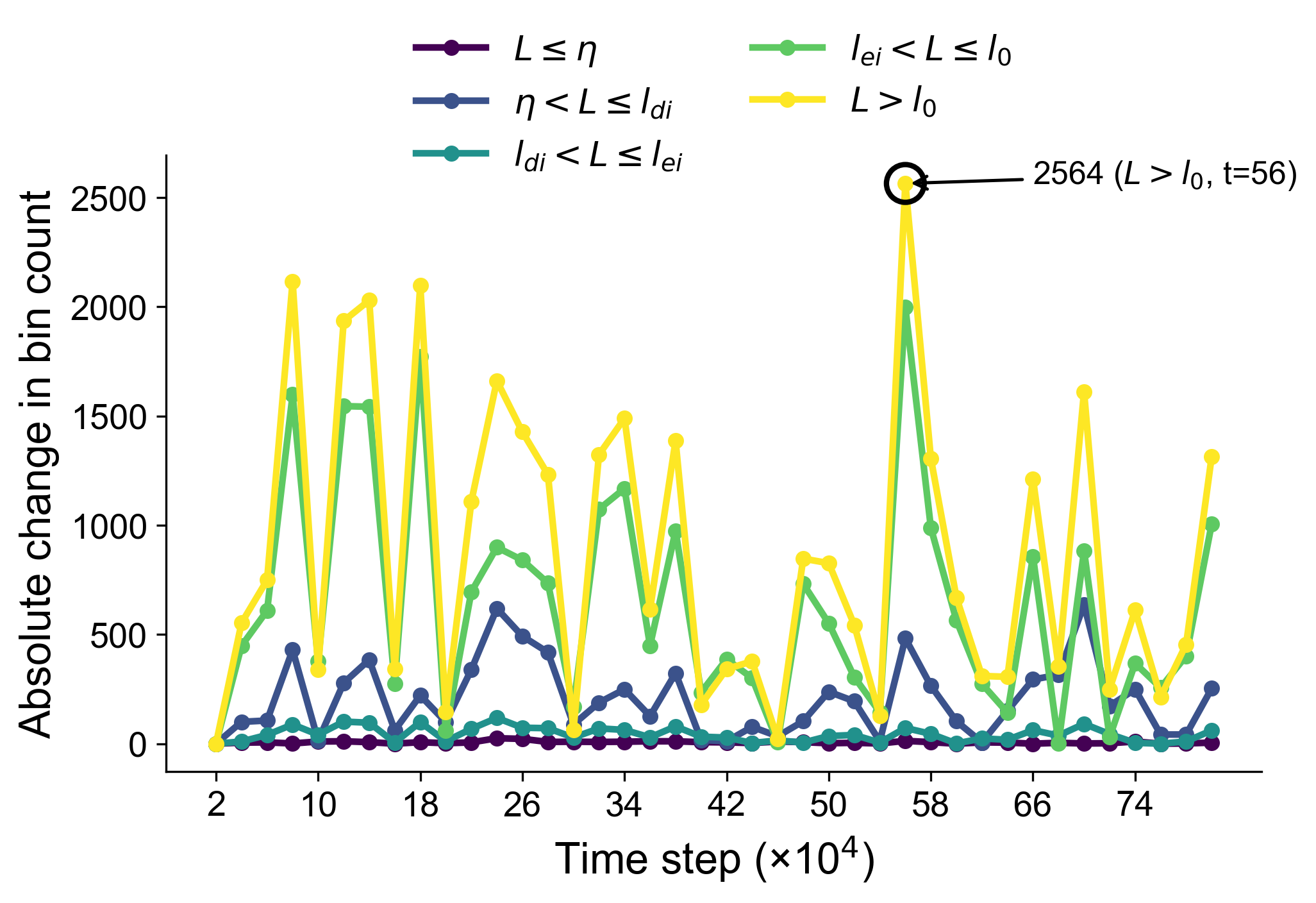}
    \caption{Temporal evolution of the absolute change in the number of binned $L$ values for the \textit{highrotminhel} configuration. The peak indicates dynamical transitions in the flow.}
    \label{fig:change-binned-l-highrotminhel}
\end{figure}

From \cref{fig:change-binned-l-highrotminhel}, we observe that the most significant variations occur at length scales larger than \( l_0 \). The next most prominent changes are found in the range \( l_{EI} < L \leq l_0 \). These variations coincide with the timesteps at which intermittent events were identified in both the vorticity and length-scale heatmaps discussed earlier. The remaining length scales exhibit minimal variation throughout the simulation period. 

From these observations, we infer that the intermittent events in our configuration predominantly affect the flow structures corresponding to the energy-intermediate and energy-containing scales, with the latter being the most strongly influenced. This provides a potential answer to the question of \textbf{which} length scales are most impacted during intermittent activity.

Having successfully identified the spatio-temporal signatures of STFs leading to intermittent events, we now examine how the topological features of different dimensions evolve across timesteps, thereby providing further support for our analysis.

\subsection{Evolution of topological features during intermittent events}
\label{subsec:topo-vs-t-highrotminhel}

We examine how the number of topological features of different dimensions (0D, 1D, and 2D) evolves during the STFs leading to intermittent events. This analysis allows us to identify which class of topological features contributes most significantly to the observed turbulent fluctuations. Our results indicate that the 1D features exhibit the largest variations, suggesting that loop-like structures, corresponding to vortex tubes in the flow, play a dominant role in the emergence of STFs.

\begin{figure}[H]
    \begin{subfigure}[b]{0.48\linewidth}
        \centering
        \includegraphics[width=\linewidth]{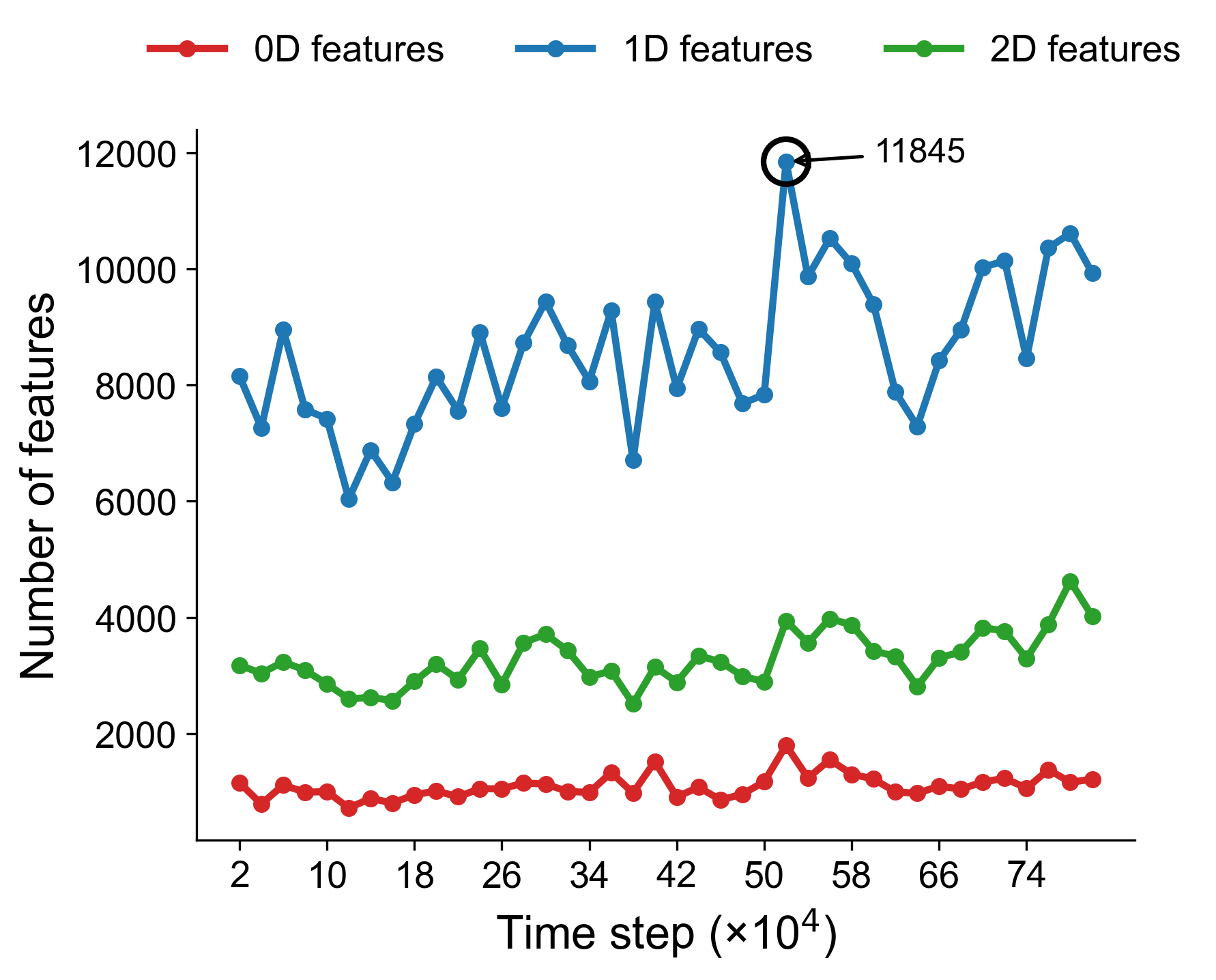}
        \caption{Sublevel set (Vorticity)}
        \label{fig:sublevel-vorticity}
    \end{subfigure}
    \hfill
    \begin{subfigure}[b]{0.48\linewidth}
        \centering
        \includegraphics[width=\linewidth]{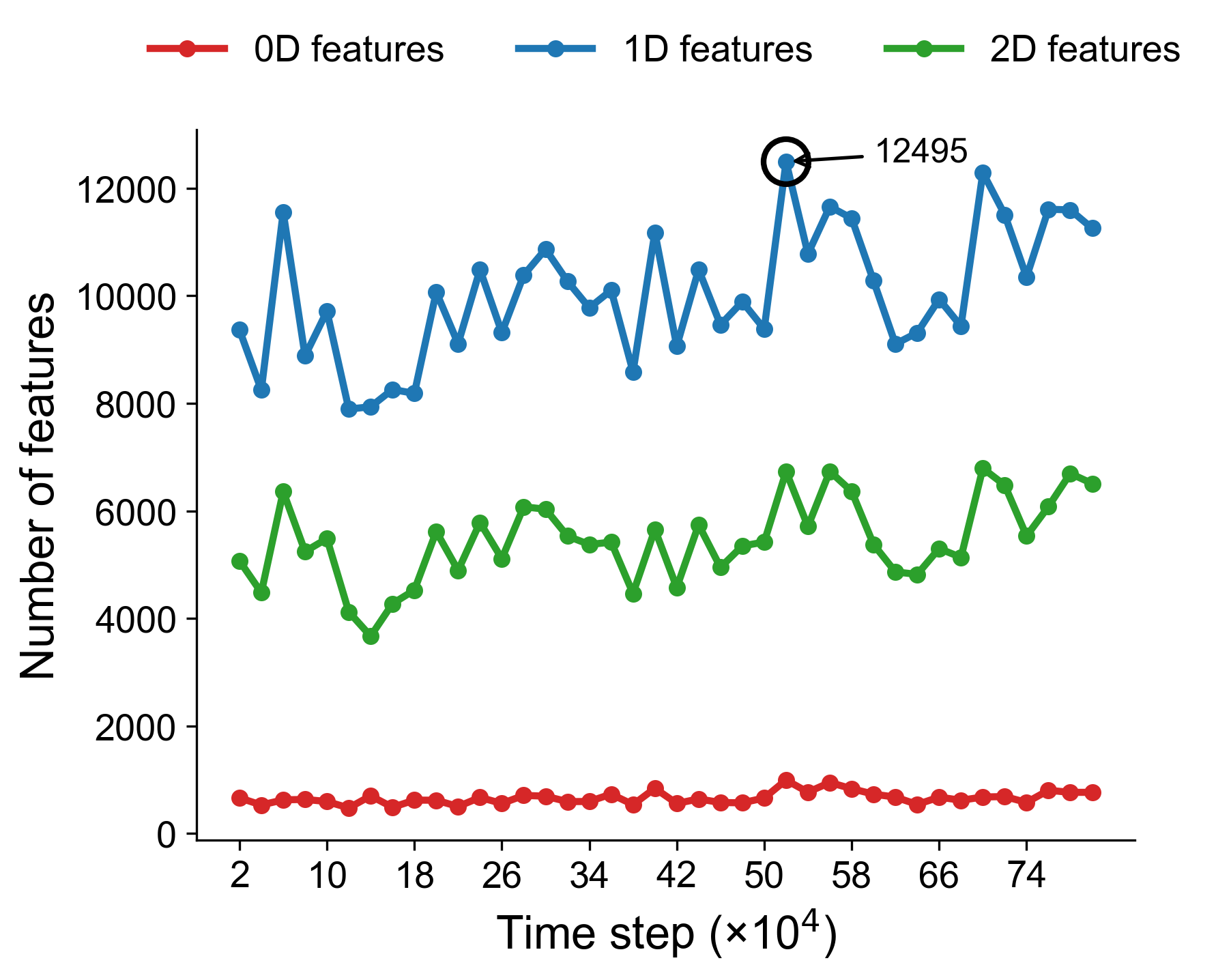}
        \caption{Superlevel set (Vorticity)}
        \label{fig:superlevel-vorticity}
    \end{subfigure}
    \centering
    \begin{subfigure}[b]{0.48\linewidth}
        \centering
        \includegraphics[width=\linewidth]{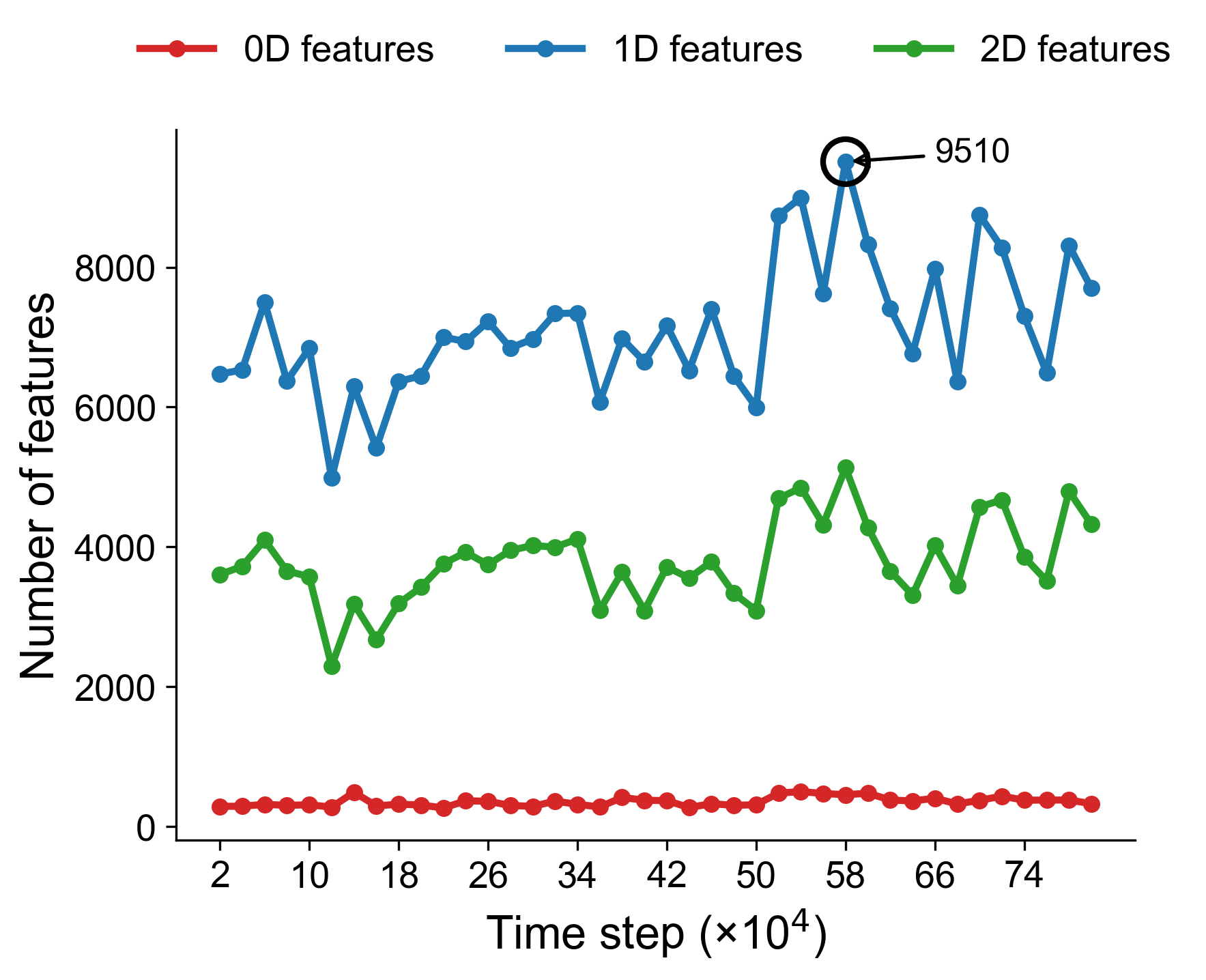}
        \caption{Sublevel set (Length scale)}
        \label{fig:sublevel-lengthscale}
    \end{subfigure}
    \hfill
    \begin{subfigure}[b]{0.48\linewidth}
        \centering
        \includegraphics[width=\linewidth]{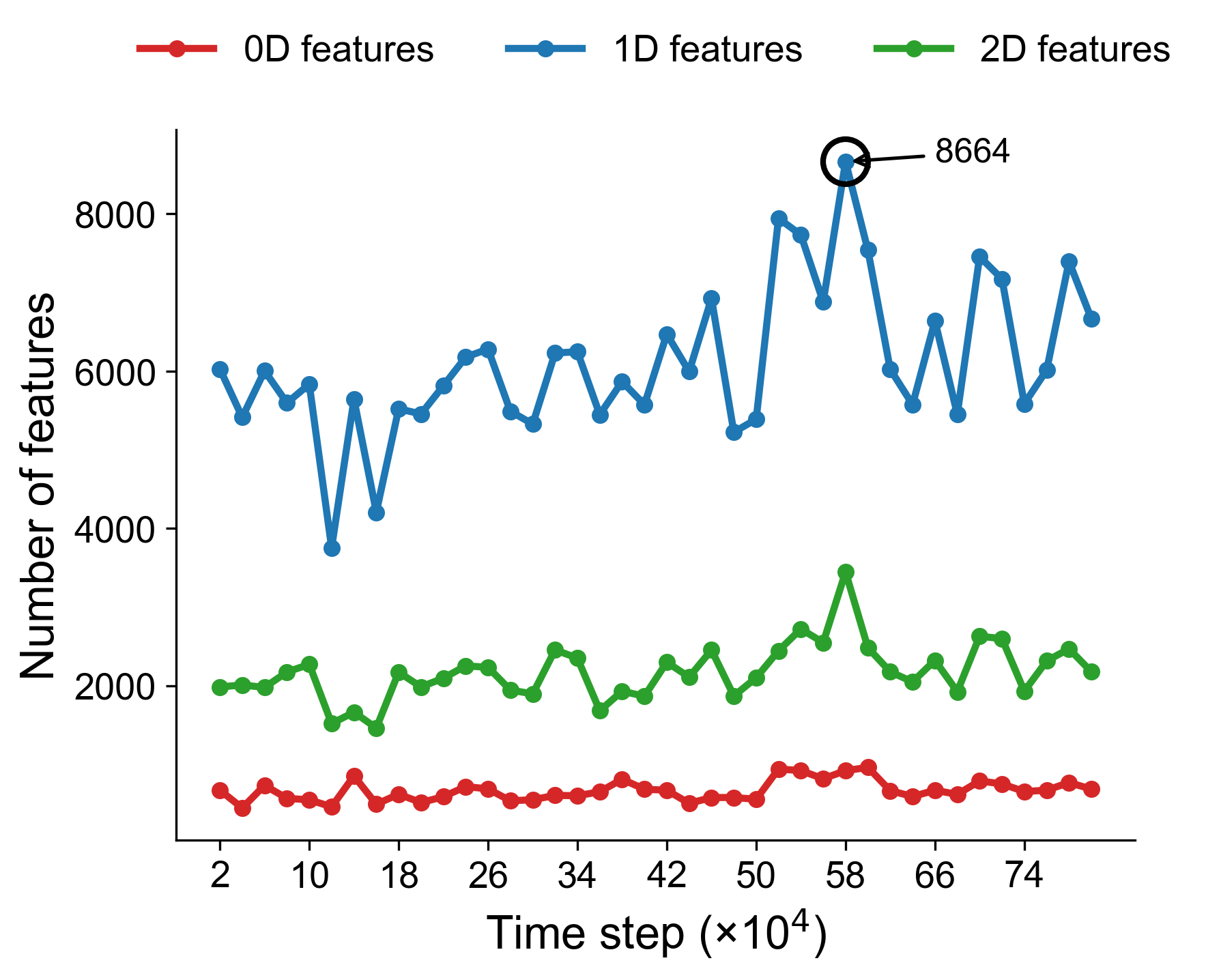}
        \caption{Superlevel set (Length scale)}
        \label{fig:superlevel-lengthscale}
    \end{subfigure}

    \caption{Temporal evolution of the number of persistence features for the \textit{highrotminhel} case. The top row corresponds to the vorticity scalar field (sublevel and superlevel filtrations), while the bottom row corresponds to the length scale scalar field.}

    \label{fig:feature-counts-highrotminhel}
\end{figure}

From \cref{fig:feature-counts-highrotminhel}, the 1D features exhibit the most pronounced variations compared to the 0D and 2D features. Notably, both the vorticity- and length-scale–based analyses reveal that these 1D features vary significantly at the \textbf{same timesteps}, where there were significant Wasserstein distance variations; indicating STFs observed across different scalar fields. Furthermore, the predominance of 1D features is consistent with the expected physics of three-dimensional turbulence, underscoring the vortex-dominated topology of the velocity field.\\

\subsection{Evolution of physical quantities during the intermittent events}
\label{subsec:phy-quantities-highrotminhel}
To further substantiate this interpretation, we analyze the temporal evolution of key physical quantities associated with intermittency. Intermittency in turbulent flows is characterized by irregular and burst-like variations in quantities such as the energy dissipation rate. Motivated by this, we examine the behavior of the energy dissipation rate along with other relevant physical observables, namely kinetic energy and kinetic helicity, which provide complementary indicators of turbulent activity. By studying how these quantities evolve during the identified STFs, we assess whether their temporal variations corroborate the occurrence of intermittent events in the flow.

We analyze the temporal evolution of these three quantities with particular attention to the timesteps corresponding to the onset of intermittency. To assess whether conventional visualization techniques can capture such behavior, we perform volume rendering of these fields at consecutive time steps. For each scalar quantity, we compute the difference between two successive time steps and visualize the resulting structures. This approach highlights regions exhibiting abrupt spatial variations that may correspond to intermittent activity. Representative examples for the energy dissipation rate, kinetic energy, and kinetic helicity are shown in \cref{fig:energy-dissipation-highrotminhel,fig:kinetic-energy-highrotminhel,fig:kinetic-helicity-highrotminhel}.

\subsection*{1.Energy Dissipation Rate}

\begin{figure}[H]
    \centering
    \makebox[\textwidth][c]{%
    \begin{minipage}[t]{0.57\textwidth}
        \centering
        \begin{subfigure}[t]{0.47\linewidth}
            \centering
            \includegraphics[width=\linewidth]{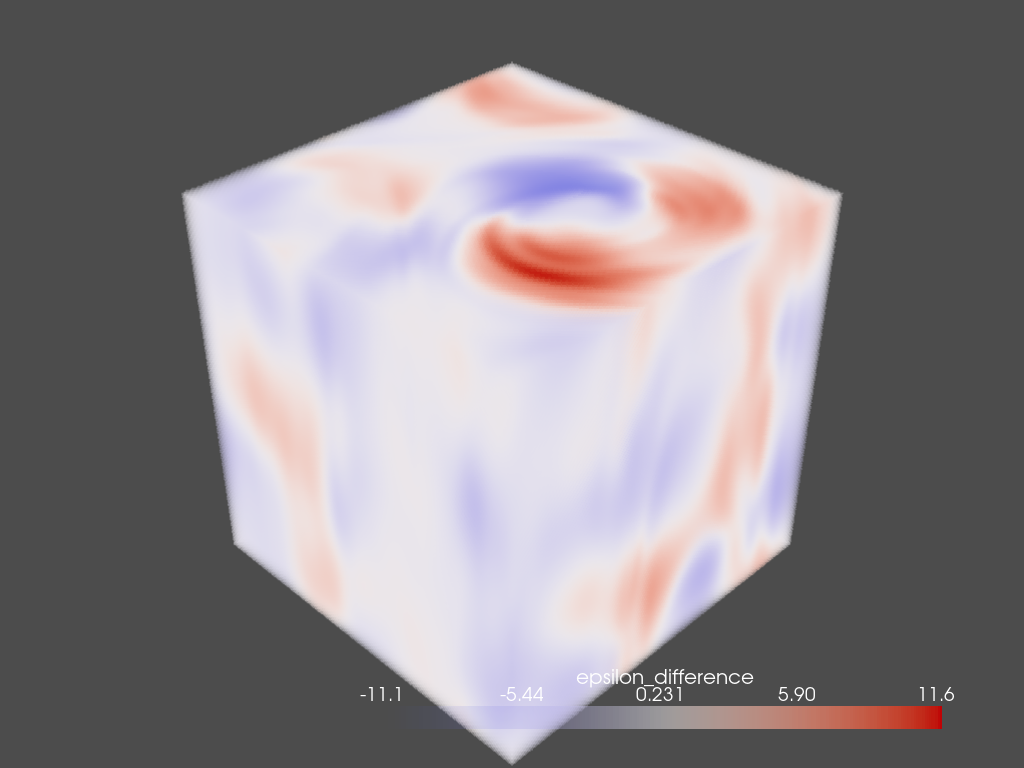}
            \caption{$|52-54|$}
            \label{fig:epsilon-520000-540000}
        \end{subfigure}\hfill
        \begin{subfigure}[t]{0.47\linewidth}
            \centering
            \includegraphics[width=\linewidth]{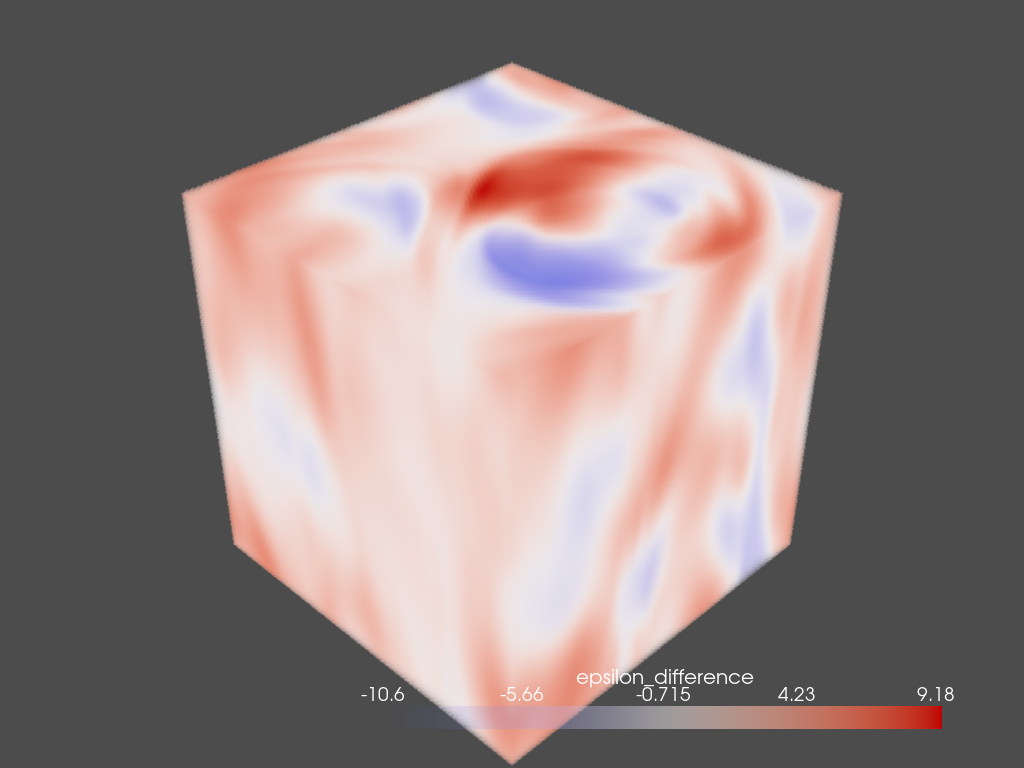}
            \caption{$|54-56|$}
            \label{fig:epsilon-540000-560000}
        \end{subfigure}

        \begin{subfigure}[t]{0.47\linewidth}
            \centering
            \includegraphics[width=\linewidth]{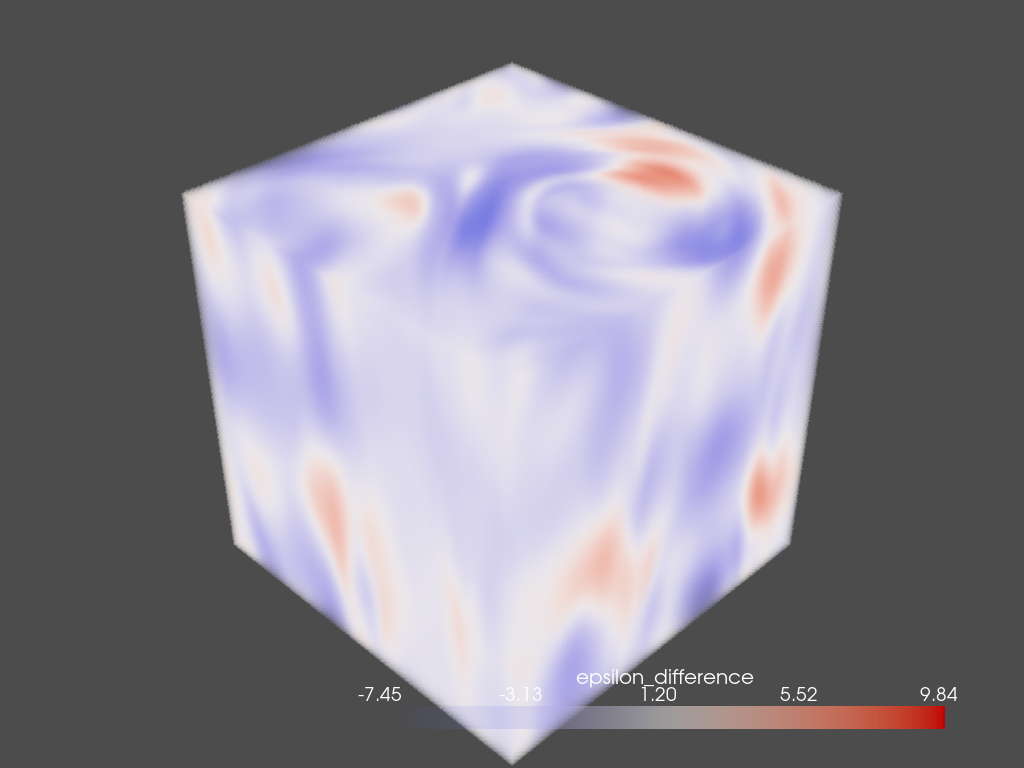}
            \caption{$|56-58|$}
            \label{fig:epsilon-560000-580000}
        \end{subfigure}\hfill
        \begin{subfigure}[t]{0.47\linewidth}
            \centering
            \includegraphics[width=\linewidth]{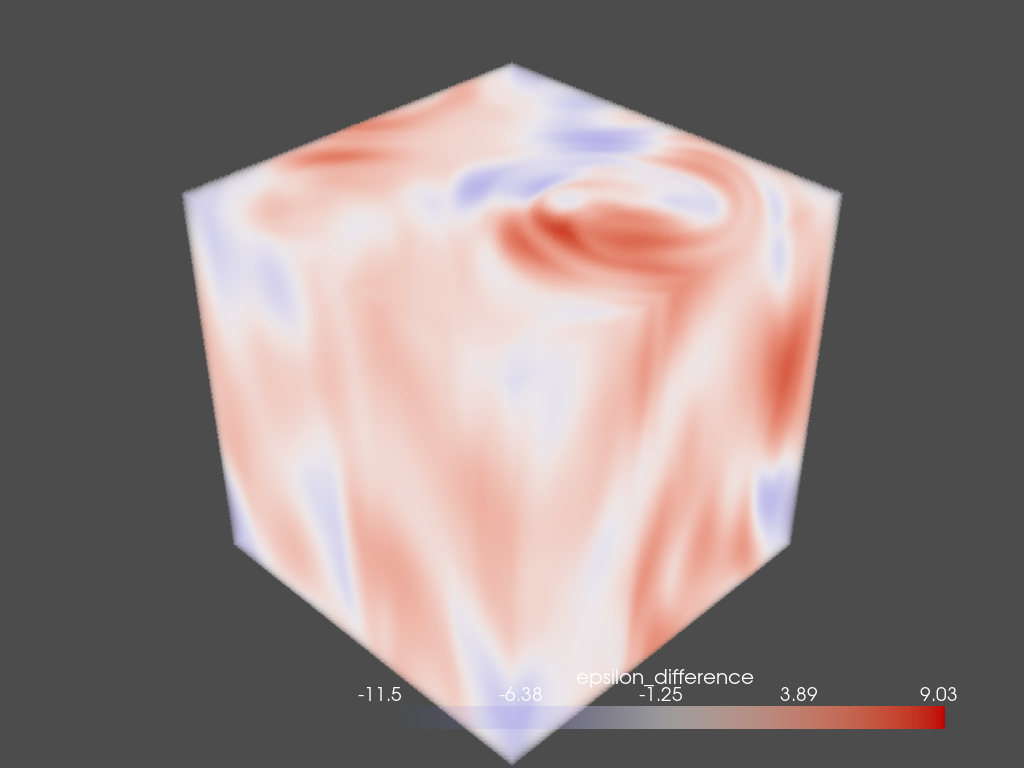}
            \caption{$|58-60|$}
            \label{fig:epsilon-580000-600000}
        \end{subfigure}
    \end{minipage}\hfill
    \begin{subfigure}[t]{0.35\textwidth}
        \centering
        \vspace{0pt}
        \includegraphics[width=\linewidth]{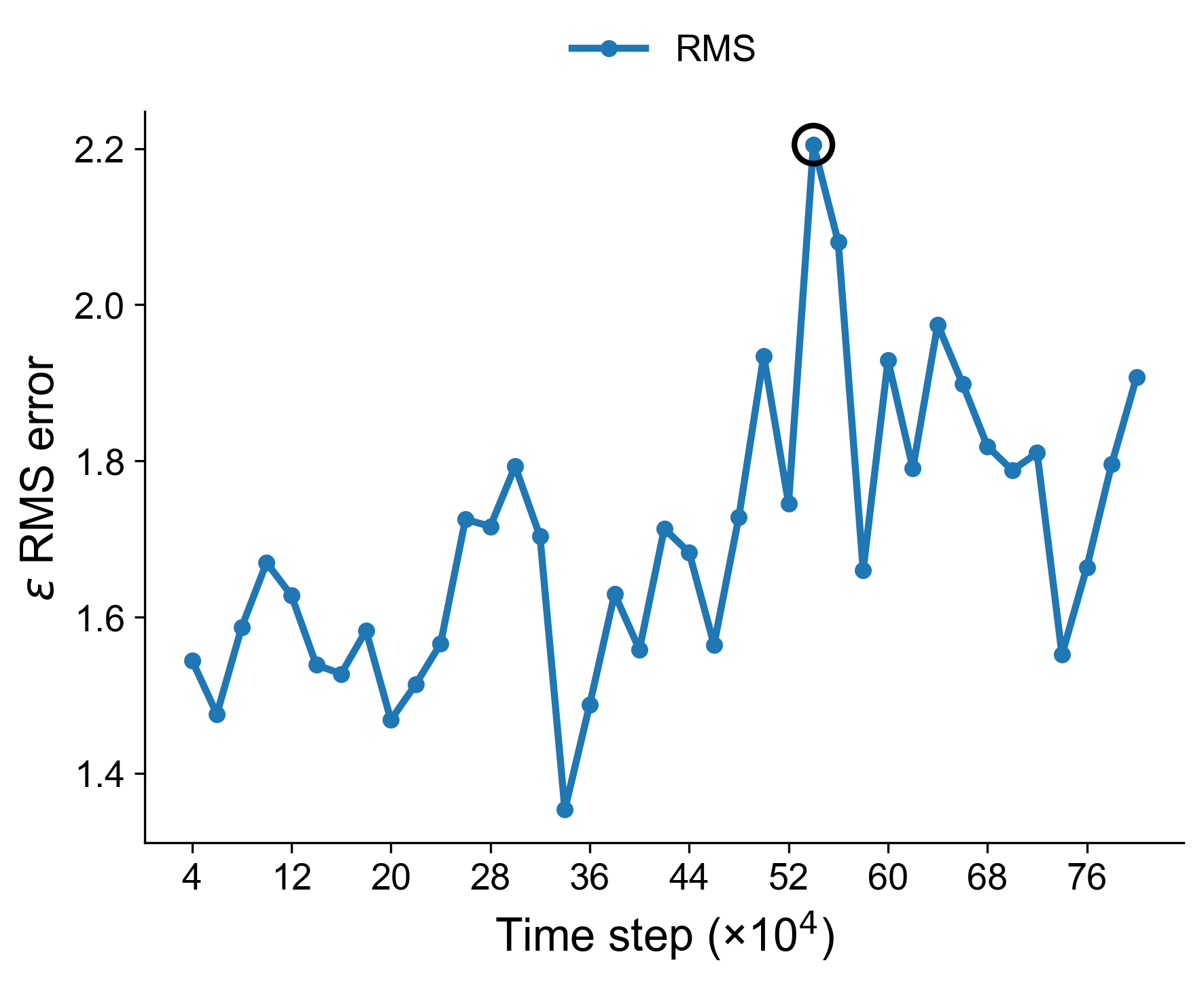}
        \caption{RMS variation of $\epsilon$ throughout the simulation}
        \label{fig:epsilon-rms-highrotminhel}
    \end{subfigure}
    }

   \caption{Differences in the energy dissipation rate between consecutive timesteps. The left panel (2$\times$2) visualizes the absolute differences in energy dissipation rate between successive timestep intervals: $|52-54|$, $|54-56|$, $|56-58|$, and $|58-60|$, where each term represents the change in the energy dissipation rate between the corresponding pair of timesteps. The right panel shows the root-mean-square (RMS) variation of the energy dissipation rate over the entire simulation.}
    \label{fig:energy-dissipation-highrotminhel}
\end{figure}

Consistent with our topological findings, the energy dissipation rate shows marked changes (as can be seen in \cref{fig:epsilon-520000-540000,fig:epsilon-540000-560000,fig:epsilon-560000-580000,fig:epsilon-580000-600000})
 at the same timesteps \textbf{(52--58)} where the Wasserstein-distance heatmaps exhibit sharp variations. This indicates that these intervals correspond to dynamically significant turbulent fluctuations leading to intermittent events. To quantify this behaviour, we compute the root-mean-square (RMS) variation of $\varepsilon$ across consecutive timesteps. The resulting time series (\cref{fig:epsilon-rms-highrotminhel})
 exhibits clear peaks that align with the topologically detected transitions. This demonstrates that abrupt geometric reorganizations in the flow coincide with large physical changes in the magnitude of energy dissipation rate.
These observations are well-aligned with established descriptions of small-scale intermittency, which emphasizes that dissipation is not space-filling but concentrated in sparse, intense bursts arising from a highly intermittent, scale-dependent cascade \cite{MeneveauSreenivasan1991,frisch1995turbulence}. In the log-Poisson refinement of this picture, these bursts are associated with one-dimensional vortex filaments of fractal dimension close to unity, forming the core assumption of the She--Leveque model \cite{She1994}. Although such models are formulated in terms of spatial statistics rather than explicit temporal evolution, their central implication is that dissipation can undergo sharp reorganizations in both space and time when these localized structures form or collapse. The distinct jumps detected in our energy dissipation rate signal, along with the corresponding variations in the Wasserstein-distance heatmaps, represent a time-resolved manifestation of the classical intermittency picture. This observation is consistent with both experimental and numerical evidence \cite{sreenivasan1997phenomenology}.

\subsection*{2.Kinetic Energy}

\begin{figure}[H]
    \centering
    \makebox[\textwidth][c]{
    \begin{minipage}[t]{0.57\textwidth}
        \centering
        \begin{subfigure}[t]{0.47\linewidth}
            \centering
            \includegraphics[width=\linewidth]{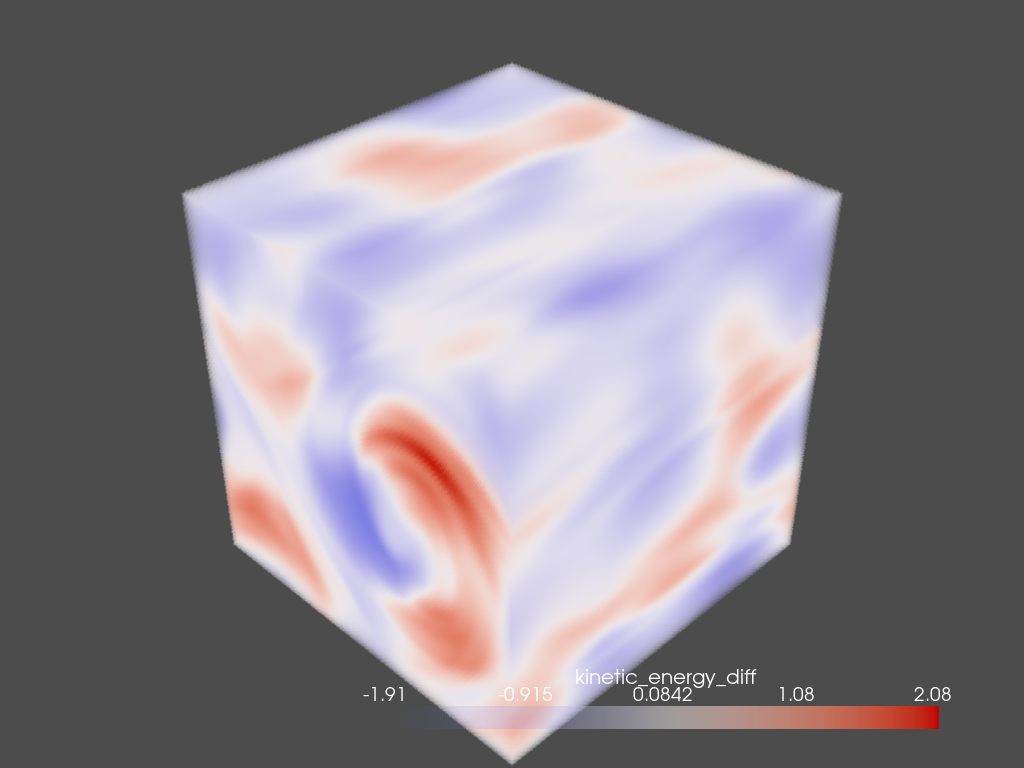}
            \caption{$|52-54|$}
            \label{fig:ke-520000-540000}
        \end{subfigure}\hfill
        \begin{subfigure}[t]{0.47\linewidth}
            \centering
            \includegraphics[width=\linewidth]{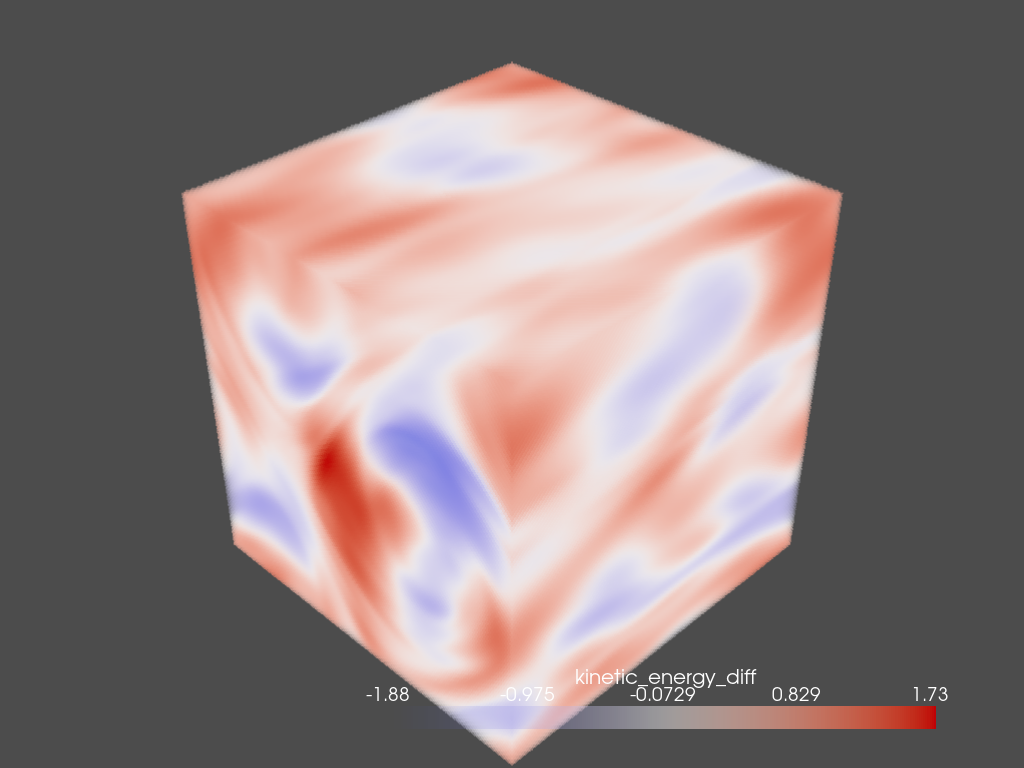}
            \caption{$|54-56|$}
            \label{fig:ke-540000-560000}
        \end{subfigure}

        \begin{subfigure}[t]{0.47\linewidth}
            \centering
            \includegraphics[width=\linewidth]{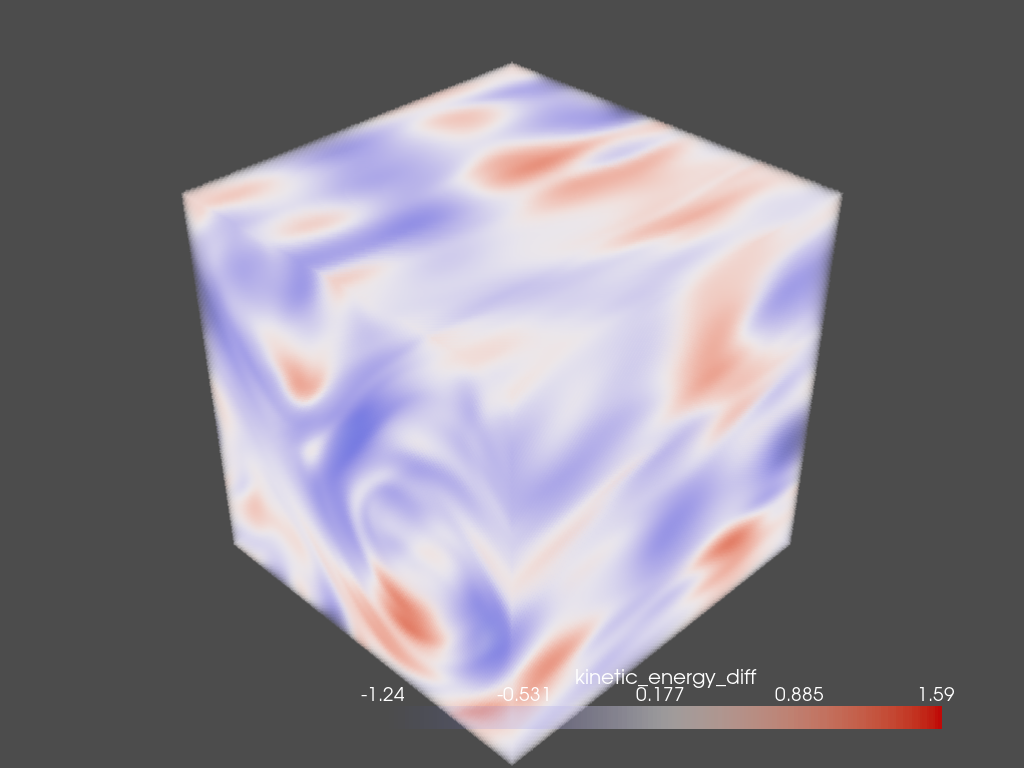}
            \caption{$|56-58|$}
            \label{fig:ke-560000-580000}
        \end{subfigure}\hfill
        \begin{subfigure}[t]{0.47\linewidth}
            \centering
            \includegraphics[width=\linewidth]{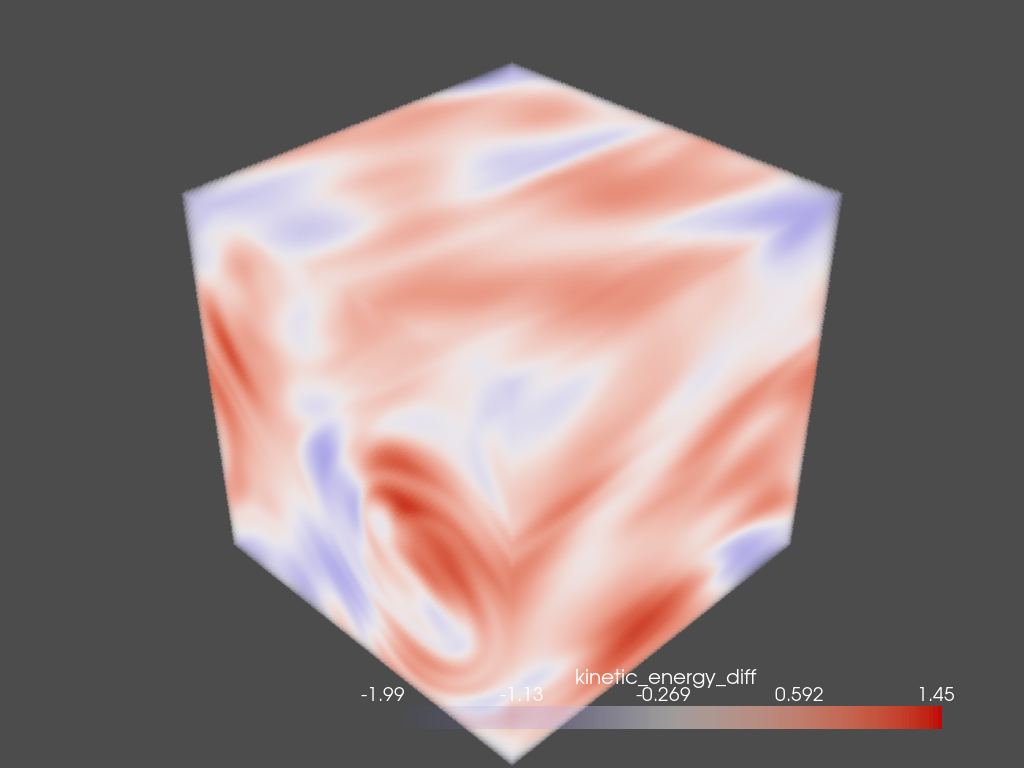}
            \caption{$|58-60|$}
            \label{fig:ke-580000-600000}
        \end{subfigure}
    \end{minipage}\hfill
    \begin{subfigure}[t]{0.35\textwidth}
        \centering
        \vspace{0pt}
        \includegraphics[width=\linewidth]{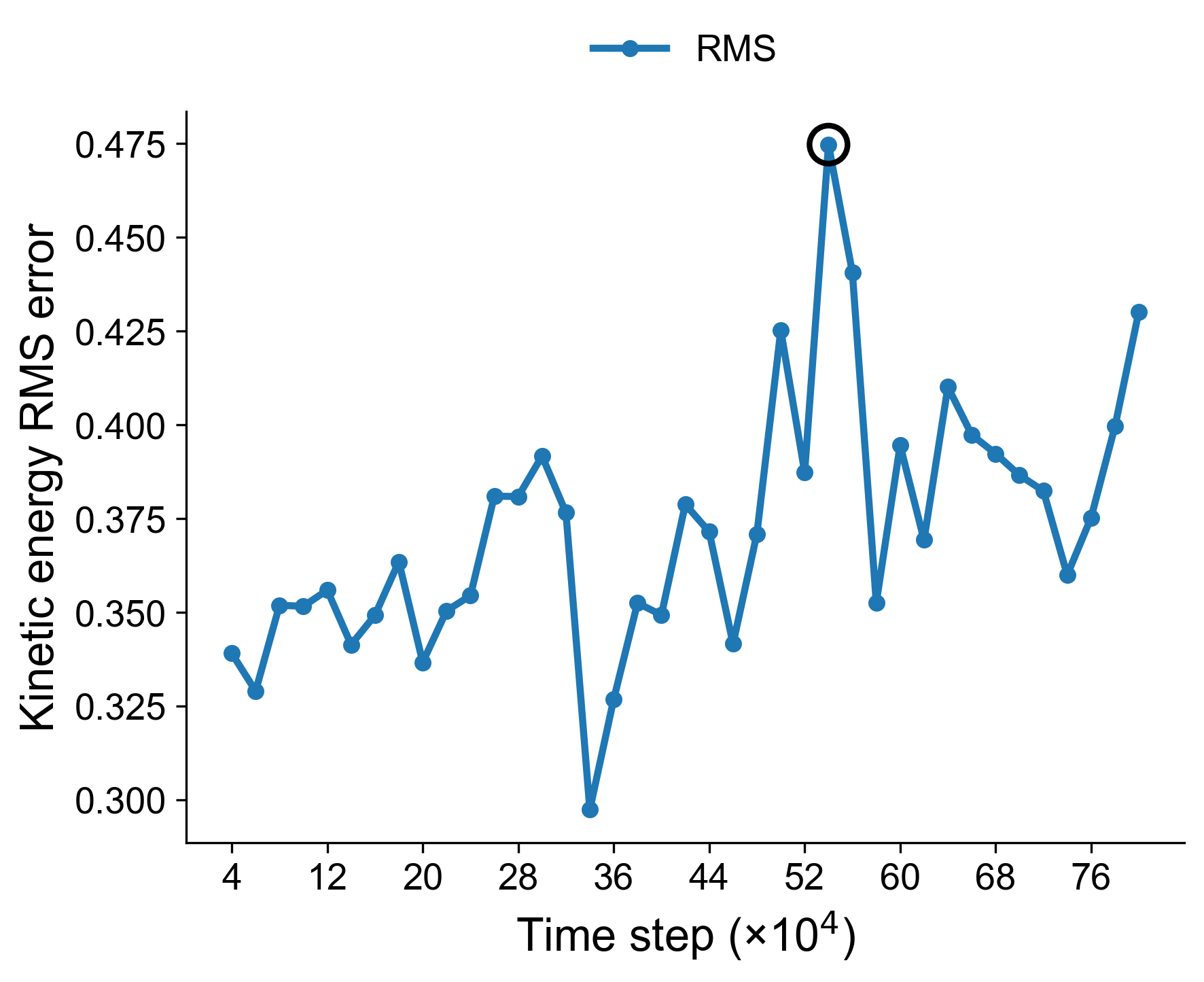}
        \caption{RMS variation of kinetic energy throughout the simulation}
        \label{fig:ke-rms-highrotminhel}
    \end{subfigure}
    }

    \caption{Differences in kinetic energy between consecutive timesteps. The left panel (2$\times$2) visualizes the absolute differences in kinetic energy between successive timestep intervals: $|52-54|$, $|54-56|$, $|56-58|$, and $|58-60|$, where each term represents the change in kinetic energy between the corresponding pair of timesteps. The right panel shows the root-mean-square (RMS) variation of kinetic energy over the entire simulation.}

    \label{fig:kinetic-energy-highrotminhel}
\end{figure}

Similar to the energy dissipation rate, the kinetic energy displays pronounced deviations (as can be seen in  \cref{fig:ke-520000-540000,fig:ke-540000-560000,fig:ke-560000-580000,fig:ke-580000-600000}) at the same timesteps \textbf{(52--58)}, providing additional evidence that these intervals correspond to STFs. Furthermore, the root-mean-square (RMS) variation of the kinetic energy (\cref{fig:ke-rms-highrotminhel}) also exhibit their largest fluctuations at these times, reinforcing the interpretation that the flow undergoes dynamically significant reorganization.

It is important to note that classical intermittency models do not predict discrete jumps in the total kinetic energy of the flow. Instead, they describe the kinetic-energy cascade as an intermittent process in which energy is transferred unevenly across scales through rare, intense events. This behaviour leads to anomalous scaling in velocity increments and heavy-tailed statistics \cite{Parisi1985,frisch1995turbulence,She1994}. Thus, the sharp reorganization detected by TDA in the kinetic-energy field should be interpreted as the topological imprint of intermittent energy transfer, rather than as a fluctuation of the global kinetic-energy content.

\subsection*{3.Kinetic Helicity}

\begin{figure}[H]
    \centering
    \makebox[\textwidth][c]{

    \begin{minipage}[t]{0.57\textwidth}
        \centering
        \begin{subfigure}[t]{0.47\linewidth}
            \centering
            \includegraphics[width=\linewidth]{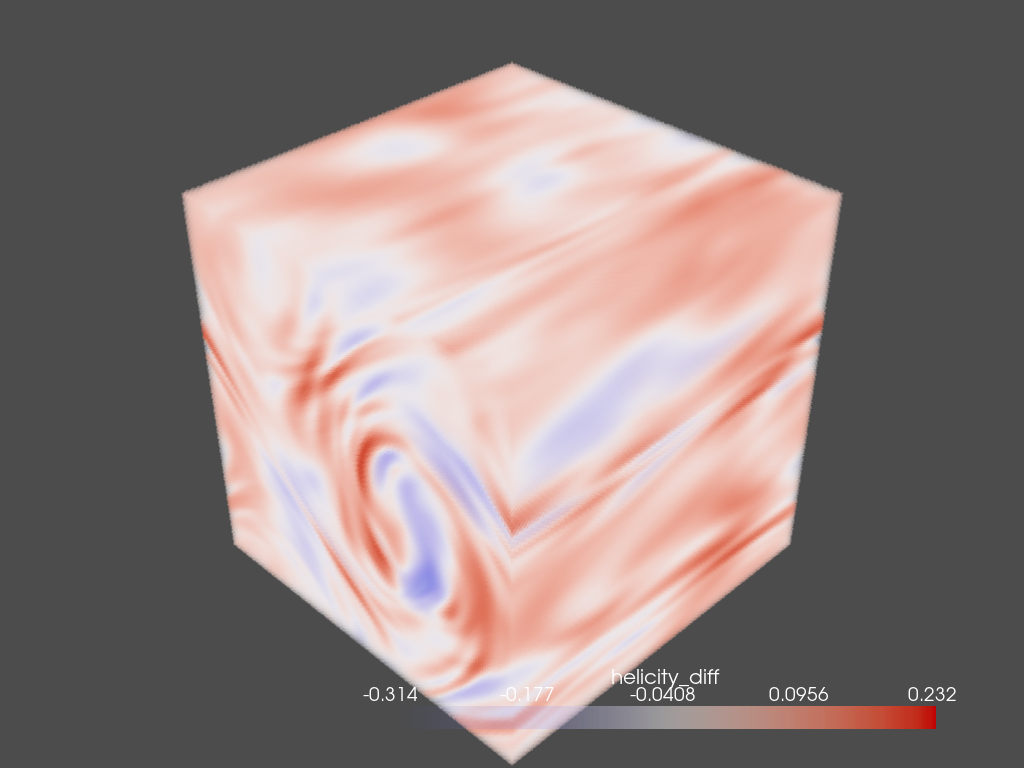}
            \caption{$|52-54|$}
            \label{fig:helicity-520000-540000}
        \end{subfigure}\hfill
        \begin{subfigure}[t]{0.47\linewidth}
            \centering
            \includegraphics[width=\linewidth]{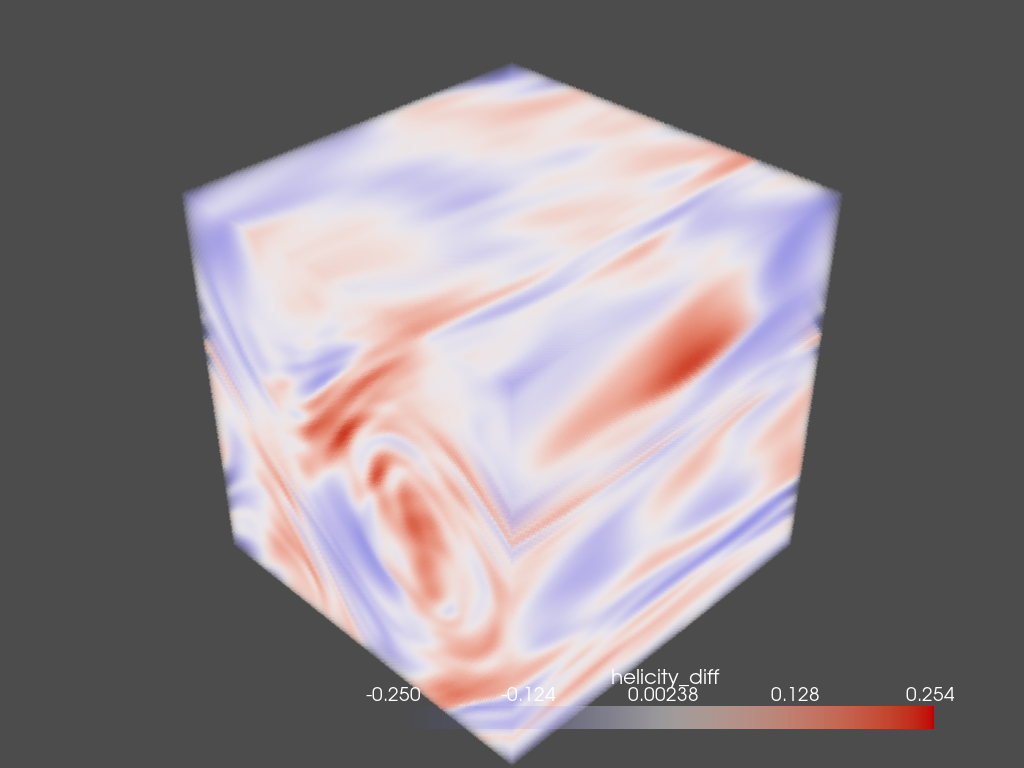}
            \caption{$|54-56|$}
            \label{fig:helicity-540000-560000}
        \end{subfigure}

        \begin{subfigure}[t]{0.47\linewidth}
            \centering
            \includegraphics[width=\linewidth]{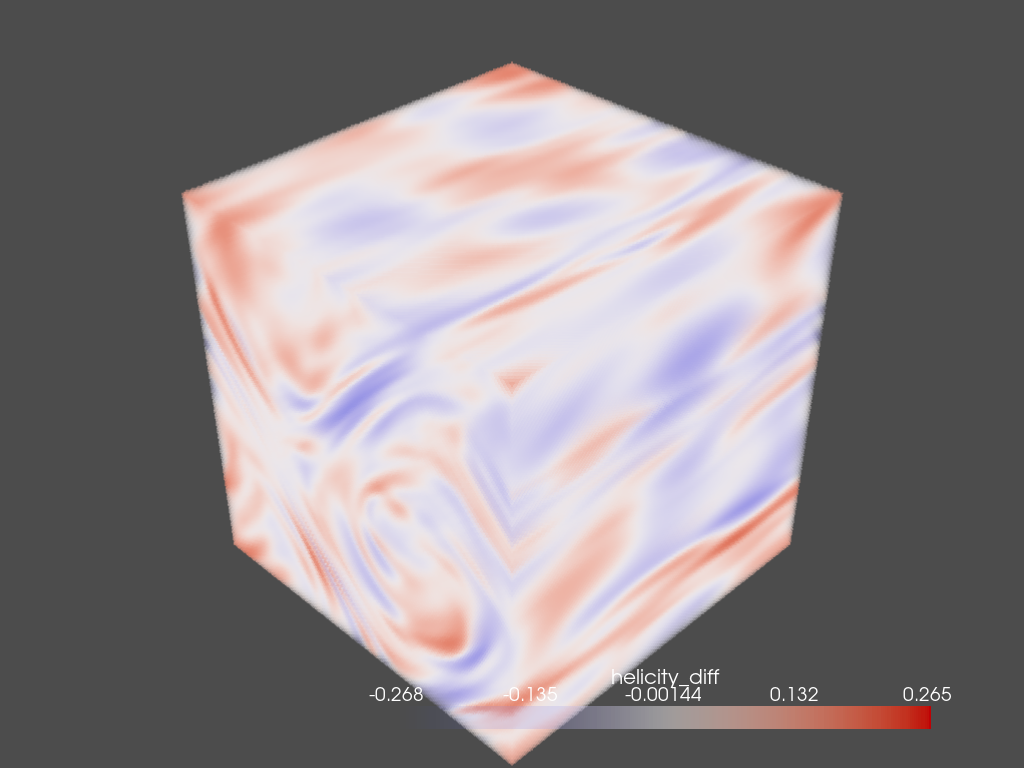}
            \caption{$|56-58|$}
            \label{fig:helicity-560000-580000}
        \end{subfigure}\hfill
        \begin{subfigure}[t]{0.47\linewidth}
            \centering
            \includegraphics[width=\linewidth]{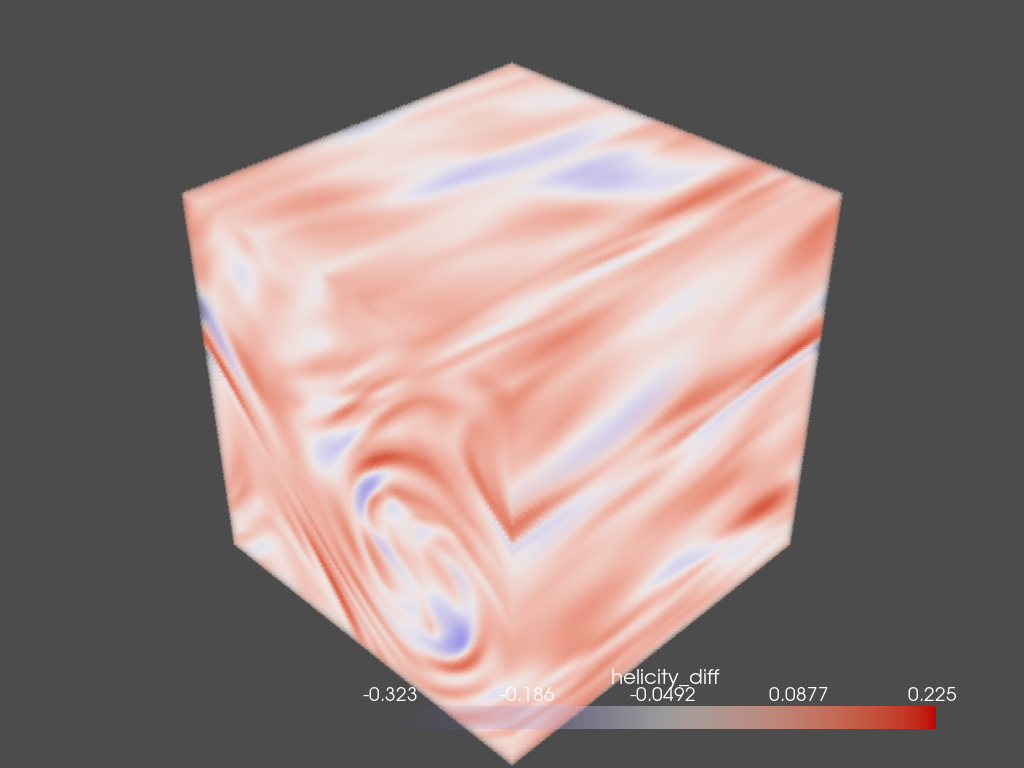}
            \caption{$|58-60|$}
            \label{fig:helicity-580000-600000}
        \end{subfigure}
    \end{minipage}\hfill

    \begin{subfigure}[t]{0.35\textwidth}
        \centering
        \vspace{0pt}
        \includegraphics[width=\linewidth]{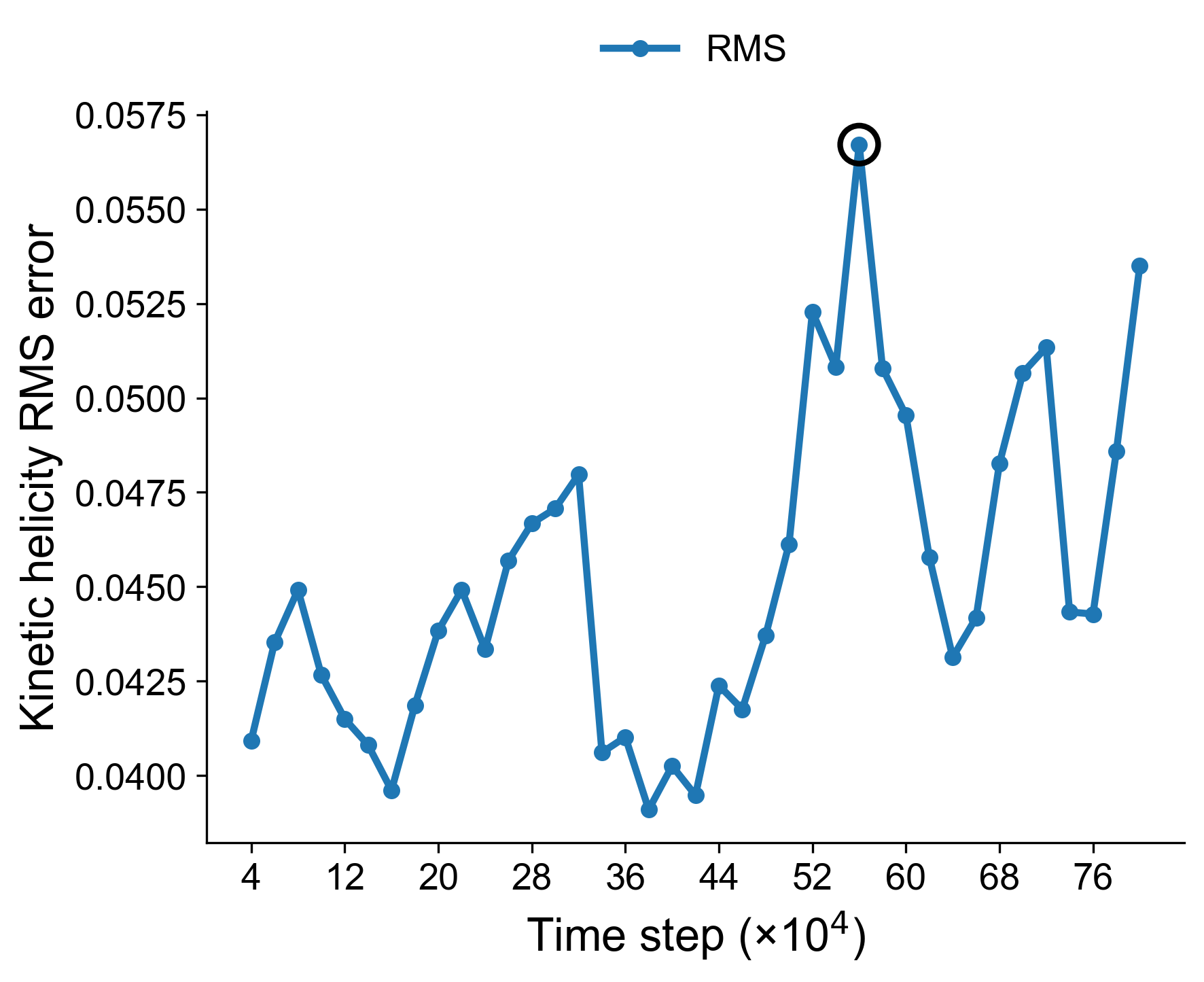}
        \caption{RMS variation of kinetic helicity throughout the simulation}
        \label{fig:helicity-rms-highrotminhel}
    \end{subfigure}
    }

    \caption{Differences in kinetic helicity between consecutive timesteps. The left panel (2$\times$2) visualizes the absolute differences in kinetic helicity between successive timestep intervals: $|52-54|$, $|54-56|$, $|56-58|$, and $|58-60|$, where each term represents the change in kinetic helicity between the corresponding pair of timesteps. The right panel shows the root-mean-square (RMS) variation of kinetic helicity over the entire simulation.}
    \label{fig:kinetic-helicity-highrotminhel}
\end{figure}
Similar to the energy dissipation rate and kinetic energy, the kinetic helicity exhibits strong fluctuations at the same timesteps \textbf{(52--58)}, as evident from the figures (\cref{fig:helicity-520000-540000,fig:helicity-540000-560000,fig:helicity-560000-580000,fig:helicity-580000-600000}). This further confirms that these intervals correspond to periods of intense turbulent fluctuations. Additionally, the root-mean-square (RMS) analysis of kinetic helicity across the full time series (\cref{fig:helicity-rms-highrotminhel}) highlights these peaks clearly, underscoring the dynamical significance of the underlying flow reorganization.

These quantitative observations are mirrored in the topological signatures of the kinetic helicity field, indicating sharp reorganizations of the helical cascade. Physically, kinetic helicity couples velocity and vorticity, and both theory and direct numerical simulations demonstrate that its flux can be as intermittent—and at times even more localized—than the energy cascade, with strong alignment events generating compact helical structures \cite{brissaud1973helicity,ChenEtAl2003PRL}. Refined-similarity arguments and DNS results further reveal that such helical bursts can occur more sporadically than energy-transfer events \cite{ChenEtAl2003PRL}, while studies of rotating or strongly helical flows report enhanced intermittency and the emergence of distinct coherent structures \cite{mininni2010part2}. Within this context, the abrupt creation or annihilation of components and loops in high $f_h$ regions provides a geometric marker of intermittent helical events, and the corresponding Wasserstein-distance variations offer an objective metric of these regime shifts, thereby directly linking the TDA results to established turbulence theory.

We now turn to an additional topological measure to further assess the consistency of our interpretation. Specifically, we compute contour trees, using the length-scale values as the scalar function.

\subsection{Identification of STFs through Contour Trees}
\label{contour-trees-highrotminhel}

Contour trees \cite{carr2003computing} are constructed using the local length scale as the scalar field. In particular, we examine whether the emergence or breaking of branches in the contour trees corroborates the occurrence of STFs. 
We identify critical points of $L$, including minima, maxima, and saddle points, by tracking how connected regions appear, merge, or split as the scalar threshold is varied. These structural transitions are encoded as nodes and edges in the contour tree, yielding a compact summary of the field’s global organization.(More about contour trees can be found in sec~\cref{subsec:tda}).

\begin{figure}[h!]
  \centering
  \begin{subfigure}[t]{0.23\linewidth}
    \centering
    \includegraphics[width=\linewidth]{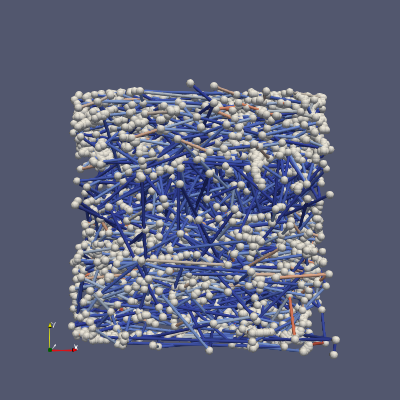}
    \caption{$t=58$}
    \label{fig:ct-58}
  \end{subfigure}\hfill
  \begin{subfigure}[t]{0.23\linewidth}
    \centering
    \includegraphics[width=\linewidth]{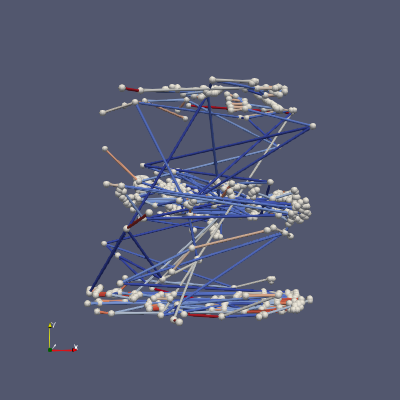}
    \caption{$t=60$}
    \label{fig:ct-60}
  \end{subfigure}\hfill
  \begin{subfigure}[t]{0.23\linewidth}
    \centering
    \includegraphics[width=\linewidth]{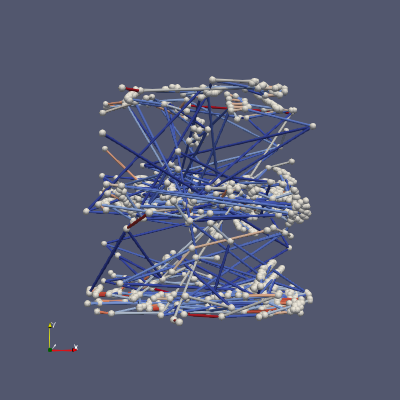}
    \caption{$t=62$}
    \label{fig:ct-62}
  \end{subfigure}\hfill
  \begin{subfigure}[t]{0.23\linewidth}
    \centering
    \includegraphics[width=\linewidth]{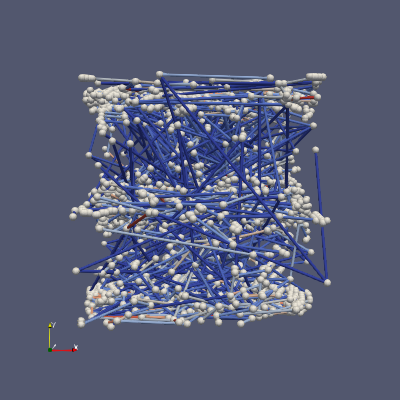}
    \caption{$t=80$}
    \label{fig:ct-80}
  \end{subfigure}

 \caption{Contour trees computed from the length scale scalar field of the \textit{highrotminhel} simulation at selected timesteps. White spheres denote critical points (minima, maxima, and saddles), while colored tubular arcs represent the connectivity of level-set components. Arc color reflects the scalar value: blue indicates low length scale values and brown-red indicates higher values.}

  \label{fig:contour-trees-highrotminhel}
\end{figure}

We observe that the trees reveal clear structural reorganizations at the same timesteps (\cref{fig:contour-trees-highrotminhel}) highlighted independently by the Wasserstein-distance analysis . In the rendered visualization, nodes correspond to critical points and colored tubular edges represent the connectivity of level-set components, providing a concise summary of how flow structures evolve. 

These contour trees provide a direct topological view of how the flow reorganizes during intermittent fluctuations. As the system evolves from $t=58$ to $t=60$, the trees exhibit a pronounced structural break, with several branches disconnecting or rearranging (\cref{fig:ct-58}). At $t=62$, the structure begins to merge and rebuild, indicating a recovery of connectivity (\cref{fig:ct-60}). By later times (e.g., $t=80$), the contour tree regains a more stable and coherent shape, similar to its pre-intermittent configuration (\cref{fig:ct-80}). This sequence offers a visual clue to the intrinsic transition of the flow: from a well-organized state, through intermittent disruption, and back toward statistical steady behavior.

\section{Discussion}
\label{sec:discussion}
Understanding the role of different topological features in relation to STFs, persistence diagrams provide a natural decomposition of coherent structures in three-dimensional turbulence into 0D, 1D, and 2D features. \textbf{0D features} correspond to isolated connected components of intense vorticity or dissipation, representing localized bursts of enstrophy embedded within smoother regions of the flow~\cite{MeneveauSreenivasan1991}. \textbf{1D features} appear as filamentary or loop-like structures and are physically associated with vortex tubes, widely recognized as the dominant dissipative structures in fully developed turbulence. This interpretation underpins the log-Poisson She--Leveque intermittency model, which assumes dissipation is concentrated in filamentary sets of fractal dimension close to one~\cite{She1994,frisch1995turbulence}. \textbf{2D features} manifest as sheet-like structures, corresponding to thin regions of large velocity gradients that often envelop or separate vortex filaments, and are associated with strong strain and localized dissipation. Together, these features—isolated bursts (0D), vortex tubes (1D), and vortex sheets (2D)—provide a unified geometric and topological framework for understanding the organization of intermittent dissipation in hydrodynamic turbulence.

Coming to Wasserstein distance heatmaps, we have already demonstrated the effectiveness of the distance metric in ~\cref{subsec:highrotminhel} to showcase strong intermittent events in ~\emph{highrotminhel} configuration. Collectively, across all flow configurations, the Wasserstein-distance heatmaps and the evolution of classified length-scale measures, together with the root-mean-square fluctuations of physical quantities such as energy dissipation, kinetic energy, and kinetic helicity, consistently identify significant deviations at critical temporal intervals. These results show that the framework is sensitive not only to STFs leading to intermittent events but also to comparatively weaker turbulent fluctuations, enabling a quantitative characterization of intermittency across different flow configurations. This interpretation further supports the kurtosis trends shown in \cref{fig:kurtosis}. Furthermore, the rapid rise in 1D persistent features and the pronounced stripes in the Wasserstein distance heatmaps correspond naturally to the emergence of vortex tubes reported in~\cite{mininni2010part2}, reinforcing the interpretation that the observed topological bursts originate from coherent vortex structures.

Another notable trend emerges in configurations with stronger rotation: enhanced Coriolis forcing appears to amplify dynamical imbalance, coinciding with intensified turbulent fluctuations that lead to strong intermittent events. This behavior is evident in the Wasserstein-distance heatmaps and the corresponding fluctuations in the physical quantities for the \emph{maxhelmaxrot}, \emph{maxrotminhel}, and \emph{nohelmaxrot} cases (\cref{sec:maxhelmaxrot,sec:maxrotminhel,sec:nohelmaxrot})
 as well as in the \emph{highrotminhel} case discussed above in \cref{subsec:highrotminhel}. This observation is consistent with the findings of Maity \textit{et al.}~\cite{maity2024}, who demonstrate that strong rotation enhances intermittency. By tracking Lagrangian particle trajectories in rotating turbulence, they show that decreasing the Rossby number promotes the formation of intense, coherent cyclonic vortices and destabilizes anti-cyclonic structures, a behavior that is also reflected in our configurations with stronger rotations. This structural reorganization amplifies the asymmetry between gradual energy gains (``flights”) and sudden dissipative bursts (``crashes”), producing pronounced skewness in the probability distributions of energy increments and vorticity fluctuations. Their irreversibility measure, $I_r=\langle p^3\rangle/\langle p^2\rangle^{3/2}$, increases with stronger rotation, indicating more violent dissipative events—a clear signature of intermittency that is likewise observed in the Wasserstein-distance measures of our strongly rotating configurations, namely the \emph{max-rotation} and \emph{high-rotation} cases (\cref{fig:maxhelmaxrot1,fig:maxrotminhel1,fig:wass-comparison-highrotminhel}).

To account for the effects of helicity, previous direct numerical simulations have shown that helicity significantly modifies small-scale turbulent dynamics and intermittency by altering the structure, alignment, and scaling properties of intense fluctuations relative to non-helical flows~\cite{mininni2010part2,SahooBiferale2015}. In particular, helicity reorganizes small-scale structures and influences the temporal development of intermittent events, while rotation introduces additional wave-mediated timescales that can suppress, delay, or redistribute these effects~\cite{pouquet2010interplay}. These findings are consistent with our observation that, in the absence of rotational constraints (as in the \emph{nohelnorot}, \emph{norotmaxhel}, and \emph{norotminhel} cases; (~\cref{fig:nohelnorot1}– ~\cref{fig:norotminhel2}), helicity can synchronize local length-scale reorganization with fluctuations in physical quantities, giving rise to an early and coherent onset of intermittent behavior. Moreover, the combined influence of rotation and helicity has been shown to produce distinct dynamical regimes, with negative helicity enhancing the coupling between structural organization and global dynamical response under specific conditions~\cite{TeitelbaumMininni2009}, consistent with the behavior observed in our \emph{norotminhel} case (\cref{sec:norotminhel}).

Despite these variations in the temporal behavior of the absolute changes in the number of length scales and the root-mean-square fluctuations of physical quantities, the contour-tree analysis remains strongly correlated with the patterns observed in the Wasserstein-distance measures. In the flows with substantial rotation, the contour trees undergo visible structural reorganizations clustered around the same temporal window in which the Wasserstein distance heatmaps show the most pronounced change, reinforcing the interpretation that localized topological instabilities correspond to physically meaningful intermittent events.\\

Additional confirmation arises from the fourth-order Wasserstein distance heatmaps (\cref{sec:fourth-order-wd}), which are broadly consistent across scalar fields for most configurations, diverging only in a few cases with complex forcing. The length-scale field shows a similar trend: both 0D and 2D Wasserstein distance heatmaps (\cref{sec:zero-and-two-wd}) identify a narrow temporal band where abrupt changes occur, while the remaining configurations show dominant transitions only during an early stage of the evolution. Finally, the reduced-dimensional (2D) analysis supports the same conclusion: in planar slices, Wasserstein distance signatures remain sharply localized in time, with sublevel and superlevel feature counts diverging in the same window identified by the 3D analysis. Notably, the 0D features stay nearly flat, indicating that the most dramatic changes originate from higher-dimensional structures rather than simple connected components.\\

The proposed framework also enables targeted, localized analysis of specific regions and time intervals that exhibit significant deviations from the average behavior of both the absolute change in the number density of length scales and the associated physical quantities. By focusing on selected time windows of interest (for example, the second peak highlighted using a red circle in \cref{fig:norotminhel1} for the \emph{norotminhel} case and the corresponding timesteps), the Wasserstein distance measures and the accompanying physical diagnostics can be applied to isolate the dynamical mechanisms underlying individual fluctuation events. This approach permits intermittent bursts to be quantitatively linked to physical parameters such as energy dissipation, kinetic energy, and kinetic helicity, thereby establishing a direct connection between topological signatures and measurable flow properties. Such localized analysis provides a refined pathway for extending the applicability of the proposed methodology to the systematic quantification of intermittent events in turbulent flows.

\section{Conclusion}
\label{sec:conclusion}
Taken together, these independent observations; consistency across the Wasserstein-distance measures, variations in physical quantities, and the coordinated breaking and reformation of contour trees, demonstrate that the proposed framework captures intermittent behavior across space and time, resolving both strong and weak turbulent fluctuations and enabling a quantitative assessment of intermittency in a given flow configuration.  The precise mechanisms driving the emergence of these turbulent fluctuations, however, require deeper investigation. We also provide a probable way this analysis could be performed on data obtained from Particle Image Velocimetry (PIV) experiments details of which could be found in \cref{sec:support-info}. 

Looking ahead, we plan to extend this analysis to three-dimensional magnetohydrodynamic (MHD) datasets. The coupled dynamics of velocity and magnetic fields are expected to exhibit richer intermittent behavior, providing an opportunity to probe more complex multiscale interactions. This extension will also enable the investigation of magnetic reconnection and magnetic helicity, both of which possess an inherently topological character and are therefore naturally suited to the proposed framework.

\section*{Acknowledgements}

The simulation data were generated  using the TCIS–TIFR Hyderabad super-computing facility and we acknowledge both Prof. Prasad Perlekar and TCIS-TIFR Hyderabad for the same. We also acknowledge the support of BRNS under Grant No.~39/14/08/2017-BRNS.
We also thank Prof. Rahul Pandit and Prof. Prasad for useful comments that helped improve the work.

\printbibliography

\newpage
\section{Methodology}
\label{sec:methods}

\subsection{Preliminary Data Analysis}
\label{subsec:data-analysis}
As a preliminary validation step, we examine whether the numerical simulations reproduce the expected turbulence characteristics reported in the literature.  Although the energy spectra are not shown here, all rotation–helicity configurations display clear power-law behavior consistent with theoretical predictions. In particular, the ~\emph{norotnohel} configuration exhibits a slope close to the classical Kolmogorov \(k^{-5/3}\) scaling expected for homogeneous and isotropic 3D turbulence. The ~\emph{highrotminhel} case shows a steeper slope of approximately \(k^{-2.87}\), while the ~\emph{maxrotnohel} configuration follows a \(k^{-7/3}\) scaling; both behaviours are consistent with the two-dimensionalisation tendencies of rapidly rotating flows reported in earlier studies~\cite{chakraborty2007signatures}. These results confirm that our simulations remain physically realistic and correctly capture the spectral dynamics associated with variations in rotation and helicity.  

\begin{figure}[H]
    \centering
    
    \begin{subfigure}[b]{0.45\linewidth}
        \centering
        \includegraphics[width=\linewidth]{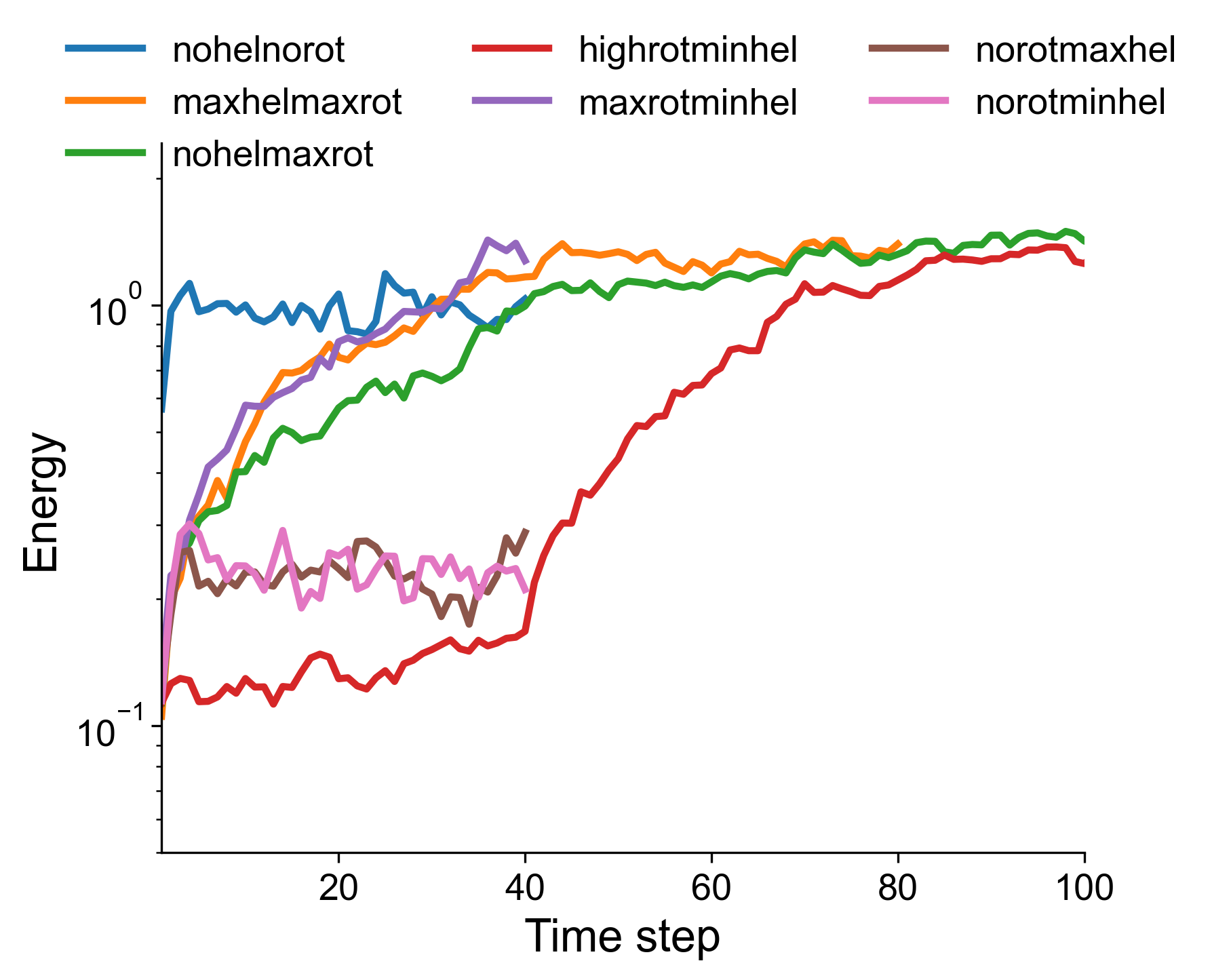}
        \caption{Energy vs.\ $t$}
        \label{fig:energy}
    \end{subfigure}
    \hfill
    \begin{subfigure}[b]{0.45\linewidth}
        \centering
        \includegraphics[width=\linewidth]{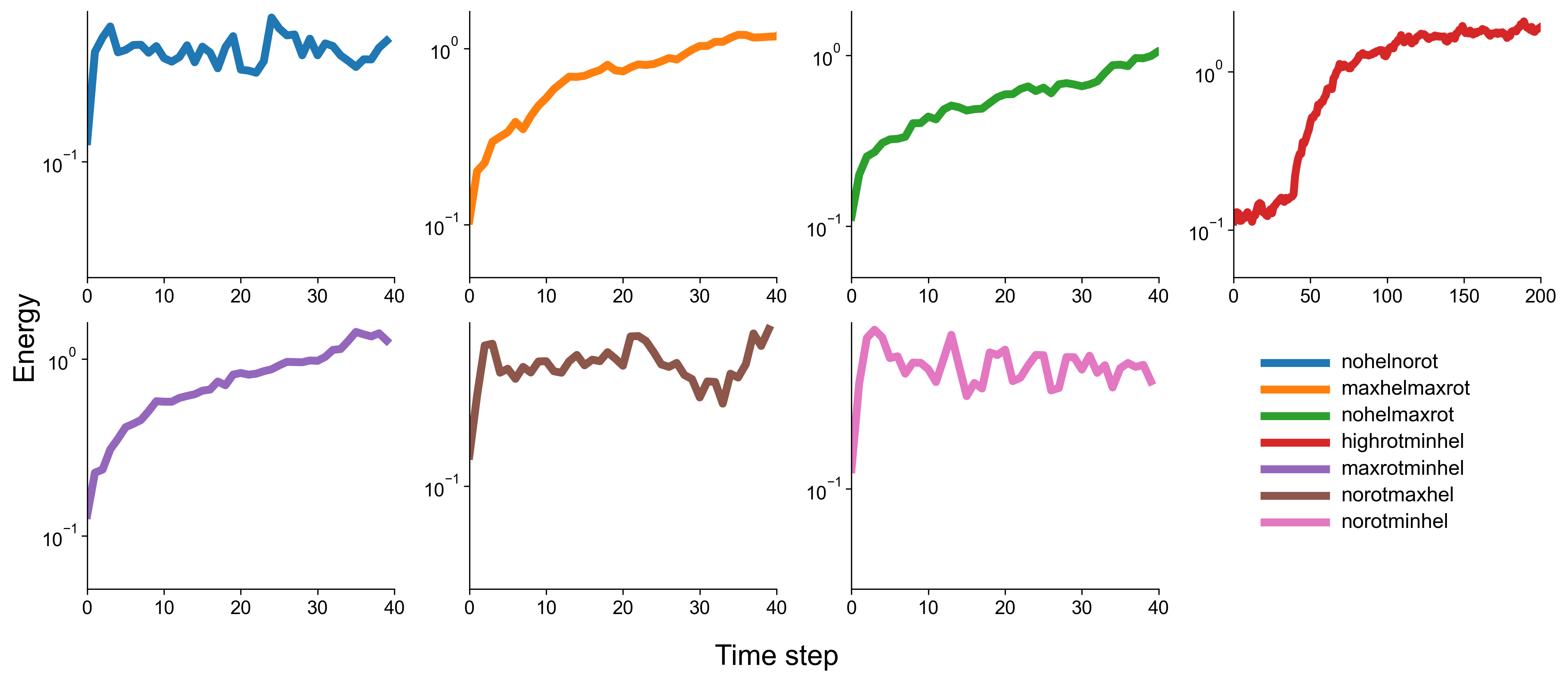}
        \caption{Energy for seven configurations plotted separately}
        \label{fig:energy-multi}
    \end{subfigure}

    \vspace{1em}

    \begin{subfigure}[b]{0.45\linewidth}
        \centering
        \includegraphics[width=\linewidth]{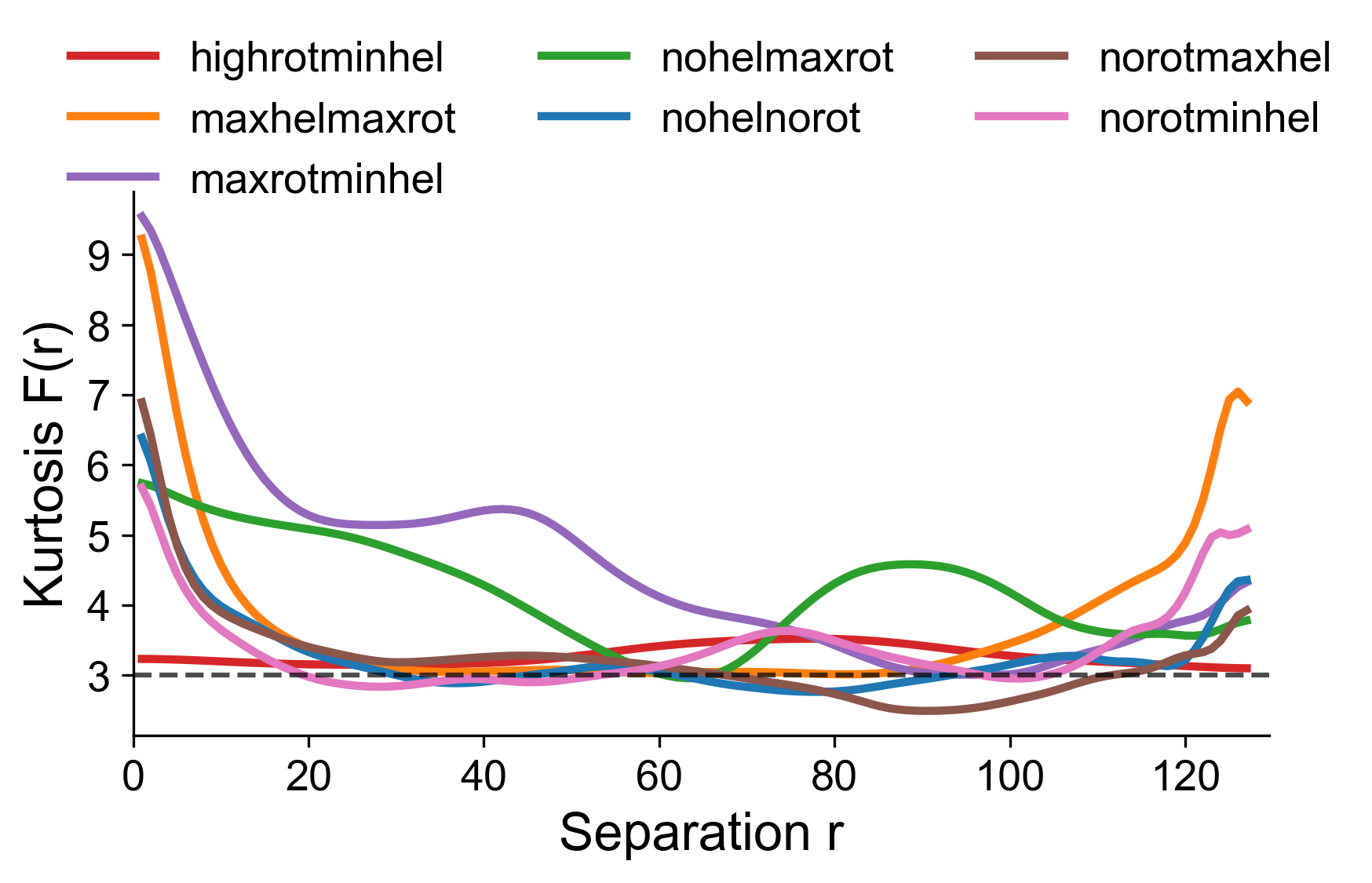}
        \caption{Kurtosis vs.\ $r$}
        \label{fig:kurtosis}
    \end{subfigure}
    \hfill
    \begin{subfigure}[b]{0.45\linewidth}
        \centering
        \includegraphics[width=\linewidth]{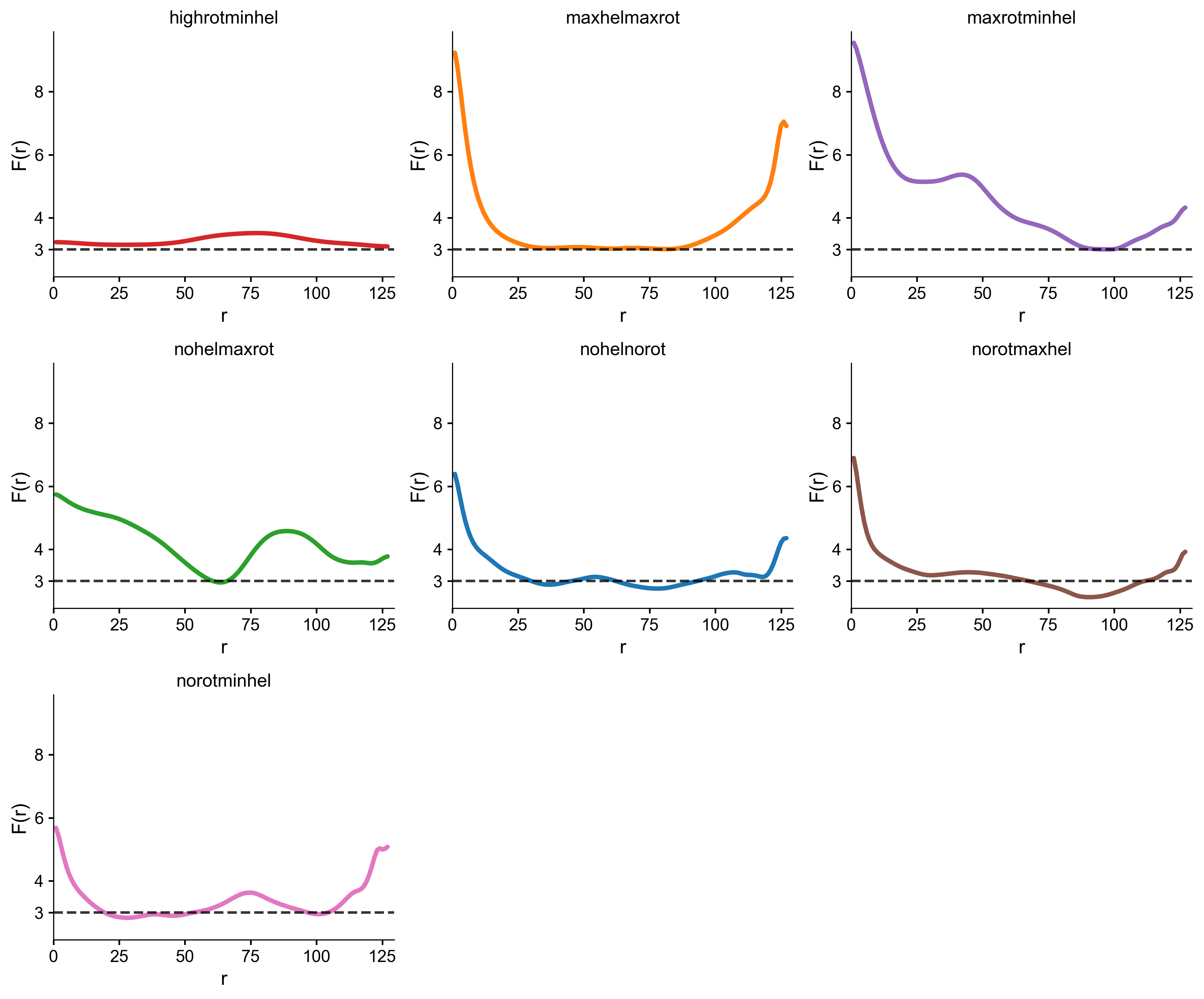}
        \caption{Kurtosis for seven configurations plotted separately}
        \label{fig:kurtosis-multi}
    \end{subfigure}

    \caption{
    (a) Energy versus time for seven different configurations.
    (b) Same energy curves shown individually for clarity.
    (c) Corresponding kurtosis curves indicating deviations from Gaussian behavior.
    (d) Same kurtosis curves shown individually for clarity.
    }
    
    \label{fig:stats-dyn-comparison}
\end{figure}

Since our primary objective is to investigate intermittency in these turbulent flows, we first verify if the systems have reached a quasi–steady state by plotting the total energy as a function of time for all configurations. The resulting energy–time plots (~\cref{fig:energy} and ~\cref{fig:energy-multi}) confirm that each configuration attains a quasi-steady turbulent state after an initial transient period. We then analyze the turbulent dynamics within this regime for each configuration by examining the kurtosis, defined as $\langle S_4 \rangle / \langle S_2 \rangle^2$, where $\langle S_4 \rangle$ and $\langle S_2 \rangle$ are the fourth- and second-order structure functions of the velocity field $u$, with respect to the spatial separation $r$. As shown in \cref{fig:kurtosis}, the kurtosis values deviate significantly from the Gaussian reference value of 3 across a wide range of scales, revealing the presence of intense fluctuations. These heavy-tailed distributions are consistent with strong intermittent behavior in the simulated turbulent flows. These kurtosis plots reveal that intermittent events are more pronounced at both small and large scales, whereas the intermediate scales exhibit comparatively weaker intermittency. This scale-dependent variation is a hallmark of turbulent intermittency and signifies clear deviations from the classical Kolmogorov (1941) framework~\cite{Kolmogorov1941}, which assumes statistical self-similarity and Gaussian fluctuations at all scales. 

We now present a brief overview of TDA and describe how it is employed in our analysis of turbulent flows.

 \subsection{Topological Data Analysis}
 \label{subsec:tda}

Topology concerns the systematic study and classification of shapes, while computational topology focuses on algorithms and data structures for extracting such information from discrete data. Building on these foundations, TDA provides tools for studying complex, high–dimensional datasets by examining their underlying topological structure. Rigorous mathematical treatments of TDA can be found in Adler et al.~\cite{adler2010persistent}, Edelsbrunner and Harer~\cite{edelsbrunner2010computational}, Adler and Taylor~\cite{adler2011topological}, and Edelsbrunner~\cite{edelsbrunner2014short}. Other notable introductions are provided by Makarenko et al.~\cite{makarenko2018}, Park et al.~\cite{park2013betti}, and Pranav et al.~\cite{pranav2017topology}.

However, in the present work, to incorporate TDA, the dataset is first represented as a scalar field, i.e., a function that assigns a scalar value to every point in the spatial domain. For example, vorticity or local length scale can be viewed as scalar fields defined over the fluid domain.

To analyze these scalar fields, we employ topological methods implemented in the Topology ToolKit (TTK)~\cite{tierny2017ttk}. In contrast to conventional statistical measures, which characterize data through a finite number of moments, TDA captures structural information by identifying and tracking topological features across multiple scales. In our three-dimensional domain, these features include connected components (0-dimensional, or 0D), loops or tunnels (1-dimensional, or 1D), and enclosed voids (2-dimensional, or 2D). Features of dimension higher than two are not physically interpretable for the present dataset.

The analysis of the scalar field is performed through a \emph{topological filtration}, in which the underlying scalar field is progressively thresholded, causing topological features to appear (be ``born'') or disappear (``die'') as the threshold varies. The birth-death points of these features encodes the multiscale geometry of the flow, enabling robust comparison of topological behaviour across different timesteps. 

 However to give a brief overview, the algebraic foundation of TDA lies in homology. Let \( f : \mathbb{X} \rightarrow \mathbb{R} \) be a scalar field defined on a three–dimensional manifold \( \mathbb{X} \). For each dimension \( n \), the homology group \( H_n(\mathbb{X}) \) is generated by equivalence classes of \( n \)-dimensional \emph{cycles}, i.e., boundary–free structures such as connected regions, closed loops, or enclosed voids. Two cycles are equivalent if one can be continuously deformed into the other. For a scalar threshold \( h \in \mathbb{R} \), the corresponding level set is
\[
f^{-1}(h) = \{\, \mathbf{x} \in \mathbb{X} \mid f(\mathbf{x}) = h \,\},
\]
which may exhibit these topological features.



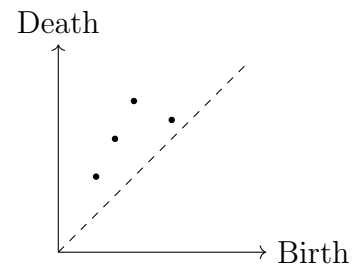
\begin{wrapfigure}{r}{0.35\linewidth}
    \centering
    \begin{tikzpicture}[scale=0.5]
        \draw[->] (0,0) -- (5.5,0) node[right] {Birth};
        \draw[->] (0,0) -- (0,5.5) node[above] {Death};
        \draw[dashed] (0,0) -- (5,5);
        \filldraw (1,2) circle (2pt);
        \filldraw (1.5,3) circle (2pt);
        \filldraw (2,4) circle (2pt);
        \filldraw (3,3.5) circle (2pt);
    \end{tikzpicture}
    \caption{Schematic persistence diagram; points away from the diagonal indicate persistent, topologically significant features.}
    \label{fig:pd}
\end{wrapfigure}

As the threshold \( h \) is varied in either the increasing or decreasing direction, the sublevel sets \( \{\,f(\mathbf{x}) \leq h\,\} \) or the superlevel sets \( \{\,f(\mathbf{x}) \geq h\,\} \) are formed. 
This process generates a nested sequence of topological spaces, known as a \emph{filtration}, within which topological features (e.g., connected components, loops, and voids) appear and subsequently may vanish or persist upto infinity. 
If a feature is created at level \( b \) and disappears at level \( d \), its \emph{persistence} is defined as \( p = d - b \). 
The collection of points \( (b,d) \) across all features of a level set \( f\) constitutes the \emph{persistence diagram} (see ~\cref{fig:pd}). The diagonal corresponds to the set \( \{(b,d) \mid b = d\} \); features far from this diagonal have longer lifetimes and are thus interpreted as more robust. Thus,
Persistence diagrams provide a quantitative representation of how topological features evolve throughout the entire filtration. For a detailed discussion on persistence diagrams, see \cite{edelsbrunner2010computational}.

In practice, filtrations are realized through \emph{complex construction}. A scalar field sampled on a grid can be represented by a cubical complex, where vertices, edges, squares, and cubes correspond to $0$-, $1$-, $2$-, and $3$-dimensional cells. As $h$ increases, cells are added in order, producing a nested sequence of complexes. By tracking changes in homology across this sequence, persistence diagrams are constructed.

Persistence diagrams are ususally compared quantitatively using two standard metrics. 
The first is the \emph{bottleneck distance}, which measures the greatest mismatch between corresponding features. 
For two persistence diagrams \( D_1 \) and \( D_2 \), it is defined as
\[
d_B(D_1, D_2) = \inf_{\gamma} \sup_{x \in D_1} \|x - \gamma(x)\|_{\infty},
\]
where \( \gamma \) ranges over all bijections between the points of \( D_1 \) and \( D_2 \). 

\newpage
The second metric is the \emph{Wasserstein distance} of order \( q \), which instead accumulates discrepancies across all matched features:
\[
d_{W,q}(D_1, D_2) = 
\left( 
\inf_{\gamma} 
\sum_{x \in D_1} 
\|x - \gamma(x)\|_{\infty}^q 
\right)^{1/q},
\]
where \( q \ge 1 \).

Another useful topological descriptor is the \emph{contour tree}, which tracks how the connected components of level sets evolve as the function value changes~\cite{carr2003topology,carr2003computing}.  
Formally, as the threshold \( h \) varies, the level sets may appear, merge, split, or disappear. The contour tree encodes these topological transitions in a simplified graph: nodes represent critical points (minima, maxima, and saddles), and edges correspond to intervals of \( h \) over which the connectivity of level sets remains unchanged.  
By reducing a scalar field to a one-dimensional tree structure, contour trees provide an interpretable summary of global topological evolution and are widely used in visualization and scientific data analysis. For a detailed discussion of contour tree computation and its topological interpretation, see \cite{carr2003computing}.

In this work, topological analysis is performed on three–dimensional scalar fields derived from the given velocity field of the turbulent flow, like vorticity and local length scale. While global features capture the overall organization of the flow, such as the number of large connected regions or dominant vortex structures, study of intermittency in turbulence requires sensitivity to localized and transient phenomena. Persistence diagrams, combined with Wasserstein distances, enable this by separating long–lived (persistent) structures from short–lived noise–like features. Together, they provide a robust framework for characterizing the topological signatures of strong non-linear fluctuations that lead to intermittency in three–dimensional hydrodynamic turbulence.

\subsection{Our approach}

\textbf{Scalar fields used.} We consider a three-dimensional turbulent velocity field defined on a spatial grid. From this field, we compute two scalar observables: the vorticity magnitude $|\omega| = |\nabla \times \mathbf{u}|$ and a local characteristic length scale $L(x,t) = |\mathbf{u}(x,t)|^3/\varepsilon(x,t)$. The energy dissipation rate is given by $\varepsilon = 2\nu \sum_{i,j} S_{ij}^2$, where $S_{ij} = \frac{1}{2}\left(\frac{\partial v_i}{\partial x_j} + \frac{\partial v_j}{\partial x_i}\right)$ is the symmetric strain-rate tensor. Spatial derivatives are evaluated using central differences.

\textbf{Persistence diagrams and Wasserstein distance.} Topological analysis is performed on these scalar fields using a cubical-complex representation of the grid. For each timestep, filtrations are constructed from the scalar fields, capturing the birth and death of topological features across thresholds. Both sublevel and superlevel filtrations are considered, and the resulting persistence diagrams are analyzed separately for features of different dimensions (0D, 1D, and 2D). This procedure is applied to all snapshots in each dataset, yielding on an average 40 persistence diagrams per case. Temporal variations in topology are quantified by computing pairwise Wasserstein distances between persistence diagrams, resulting in $40 \times 40$ distance matrices that are visualized as heatmaps to identify intervals of significant structural variation. The analysis is further performed for different values of the order of the Wasserstein distance  $q$, with corresponding results presented in Supplementary material (Appendix ~\cref{sec:fourth-order-wd} and ~\cref{sec:zero-and-two-wd}).

\textbf{Contour trees.} Contour trees are computed using the local length scale as the scalar field. Owing to its larger dynamic range compared to the vorticity magnitude, the length-scale field is used to capture structural changes in the flow. Critical points of $L$, including minima, maxima, and saddle points, are identified by tracking changes in connectivity as the scalar threshold is varied. These events are encoded as nodes and edges in the contour tree, and the evolution of branches is analyzed to assess their correspondence with the occurrence of STFs. By capturing how connected regions merge and split across scalar thresholds, contour trees summarize global structural changes in the field and complement the persistence-based analysis.

\textbf{Length-scale classification.} The local length scales $L$ are further classified into five regimes based on characteristic turbulent scales reported in the literature. This classification is used to identify the scale ranges associated with STFs and to localize intermittent structures in the flow.

\textbf{Topological feature evolution.} We compute the number of topological features of different dimensions (0D, 1D, and 2D) as a function of time to quantify their temporal variation.

\textbf{Physical observables.} In addition, the temporal evolution of physical observables, including the energy dissipation rate, kinetic energy, and kinetic helicity, is analyzed to assess their correlation with the identified STFs. Using the velocity field data, we compute the kinetic energy and kinetic helicity, while the local energy dissipation rate is obtained from prior calculations.

\newpage
\section{Supplementary Material}
\label{sec:support-info}
\begin{figure}[h!]
    \centering
    \begin{subfigure}[t]{0.32\linewidth}
        \centering
        \includegraphics[width=\linewidth]{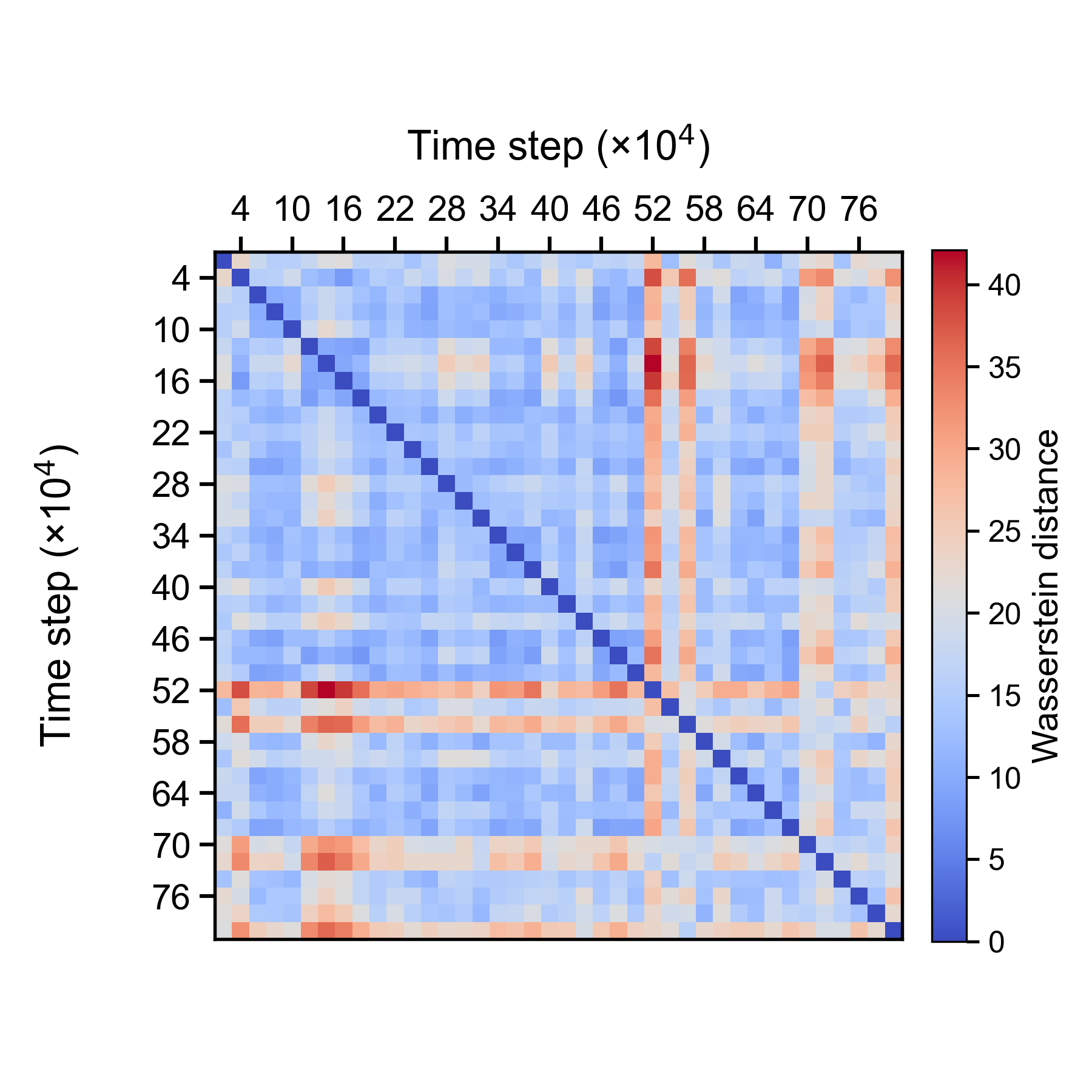}
        \caption{y-z plane}
        \label{fig:curlx}
    \end{subfigure}
    \hfill
    \begin{subfigure}[t]{0.32\linewidth}
        \centering
        \includegraphics[width=\linewidth]{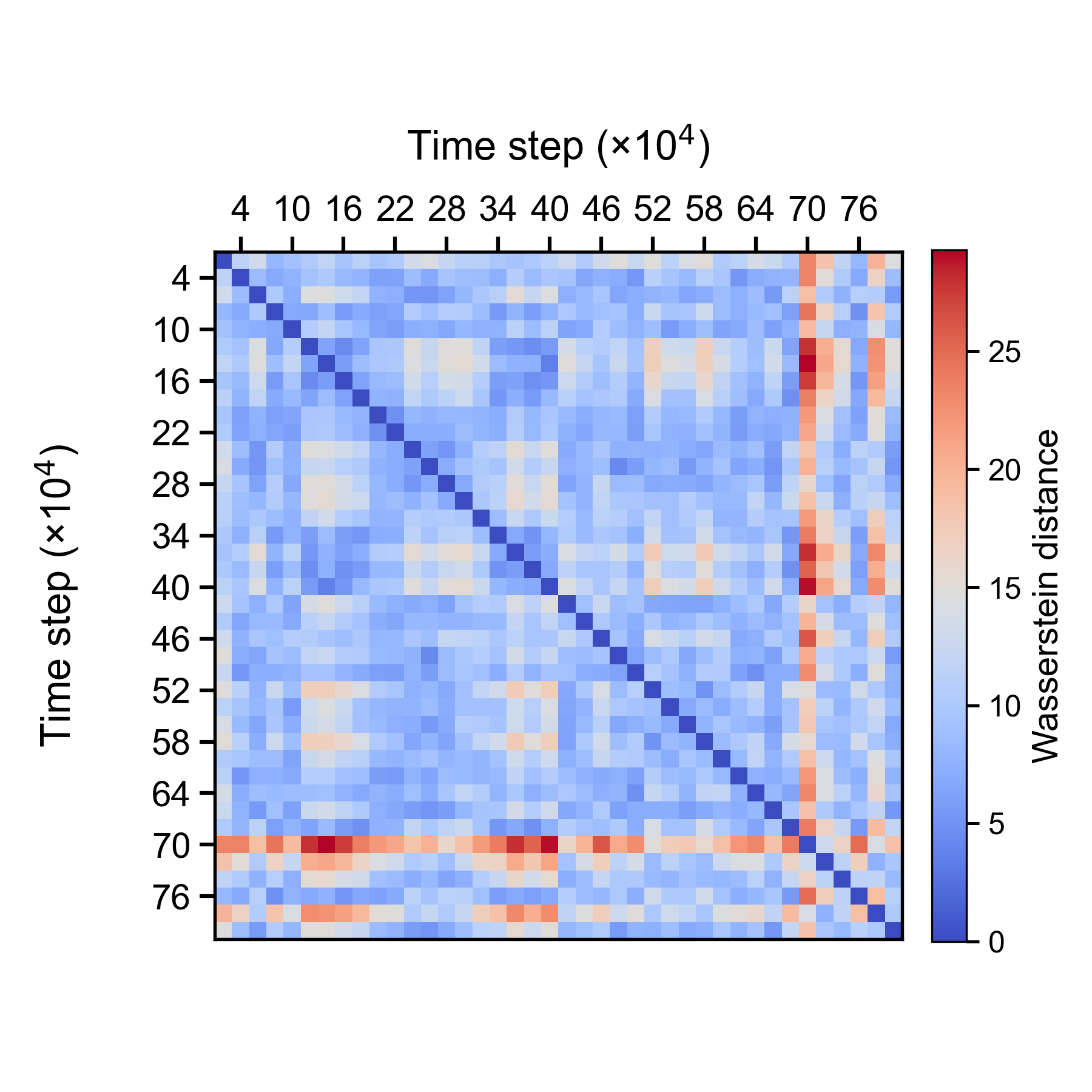}
        \caption{x-z plane}
        \label{fig:curly}
    \end{subfigure}
    \hfill
    \begin{subfigure}[t]{0.32\linewidth}
        \centering
        \includegraphics[width=\linewidth]{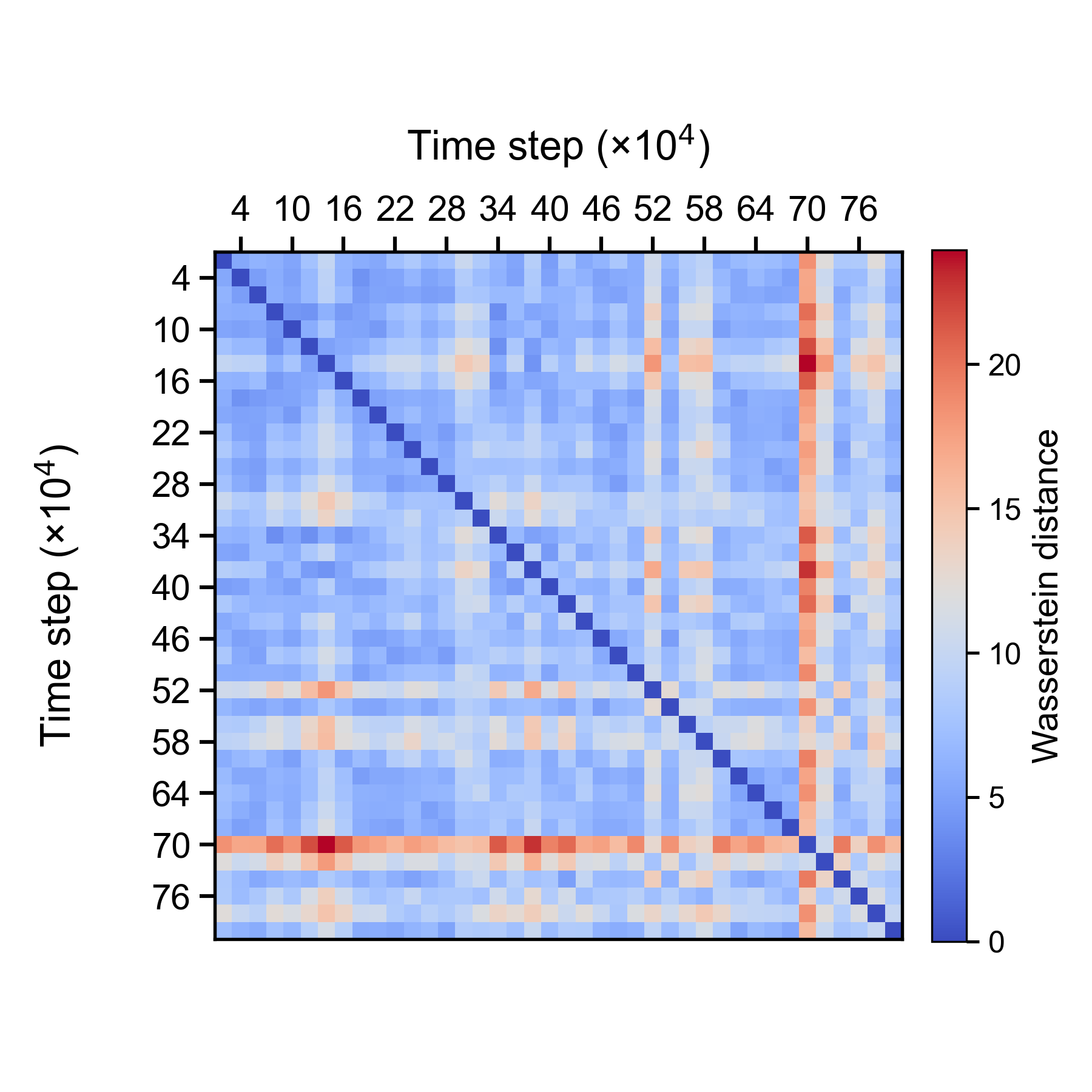}
        \caption{x-y plane}
        \label{fig:curlz}
    \end{subfigure}
    \caption{1st-order Wasserstein distance matrices for curl components in the dataset.}
    \label{fig:wasserstein_piv}
\end{figure}

\begin{figure}[h!]
    \centering
    \begin{subfigure}[t]{0.32\linewidth}
        \centering
        \includegraphics[width=\linewidth]{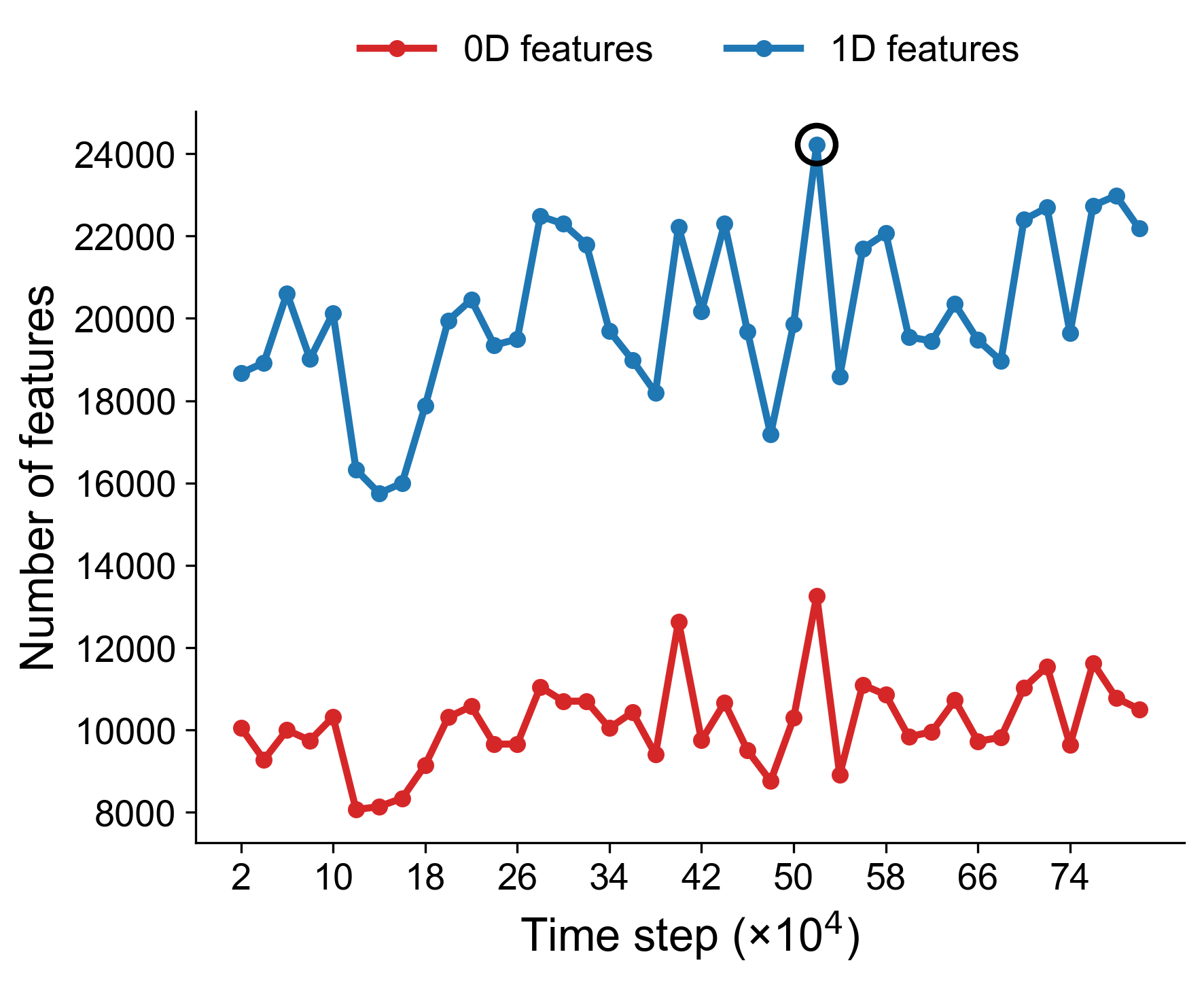}
        \caption{Sublevel persistence counts for (y--z plane).}
        \label{fig:curlx_sub}
    \end{subfigure}
    \hfill
    \begin{subfigure}[t]{0.32\linewidth}
        \centering
        \includegraphics[width=\linewidth]{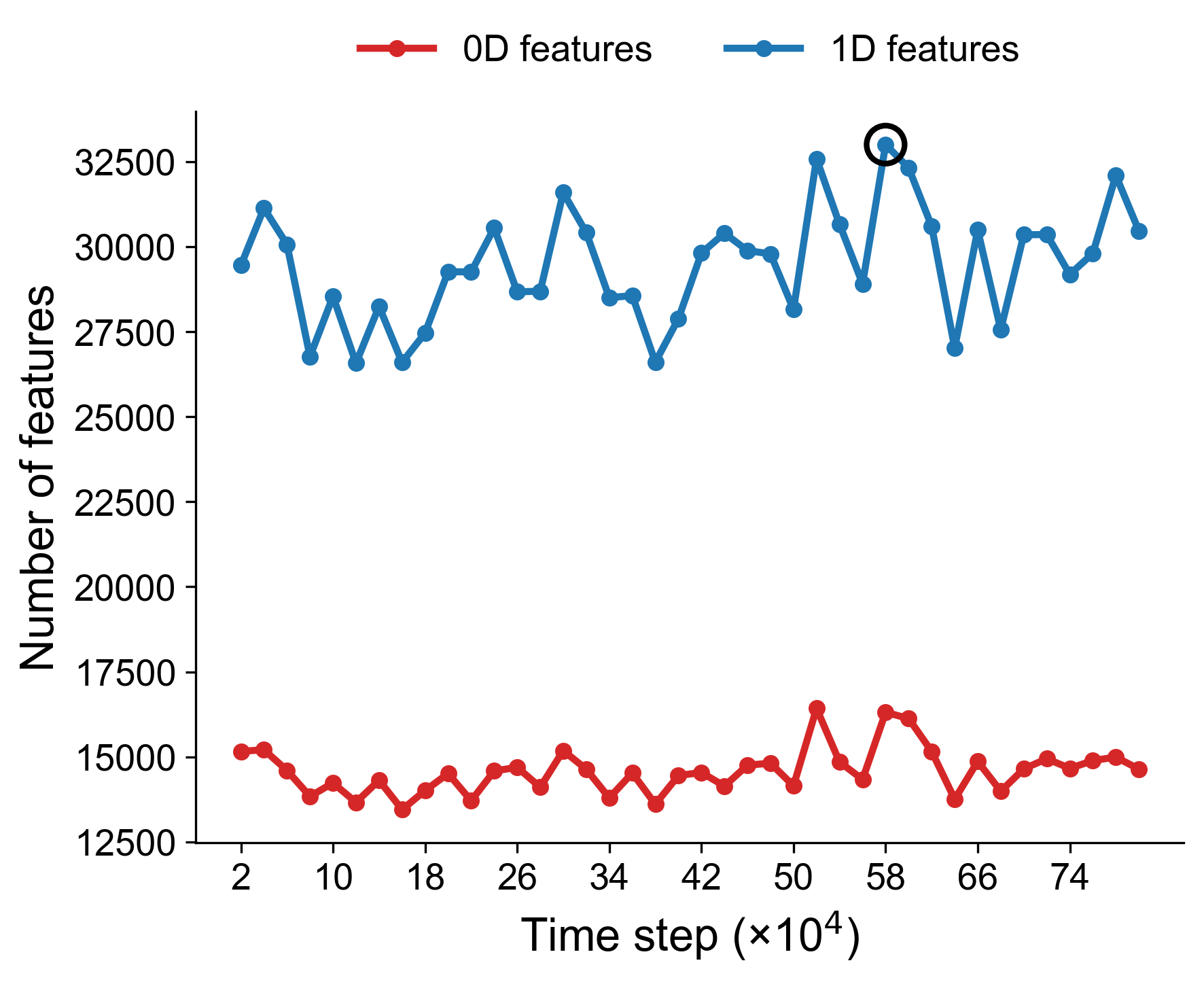}
        \caption{Sublevel persistence counts for (x--z plane).}
        \label{fig:curly_sub}
    \end{subfigure}
    \hfill
    \begin{subfigure}[t]{0.32\linewidth}
        \centering
        \includegraphics[width=\linewidth]{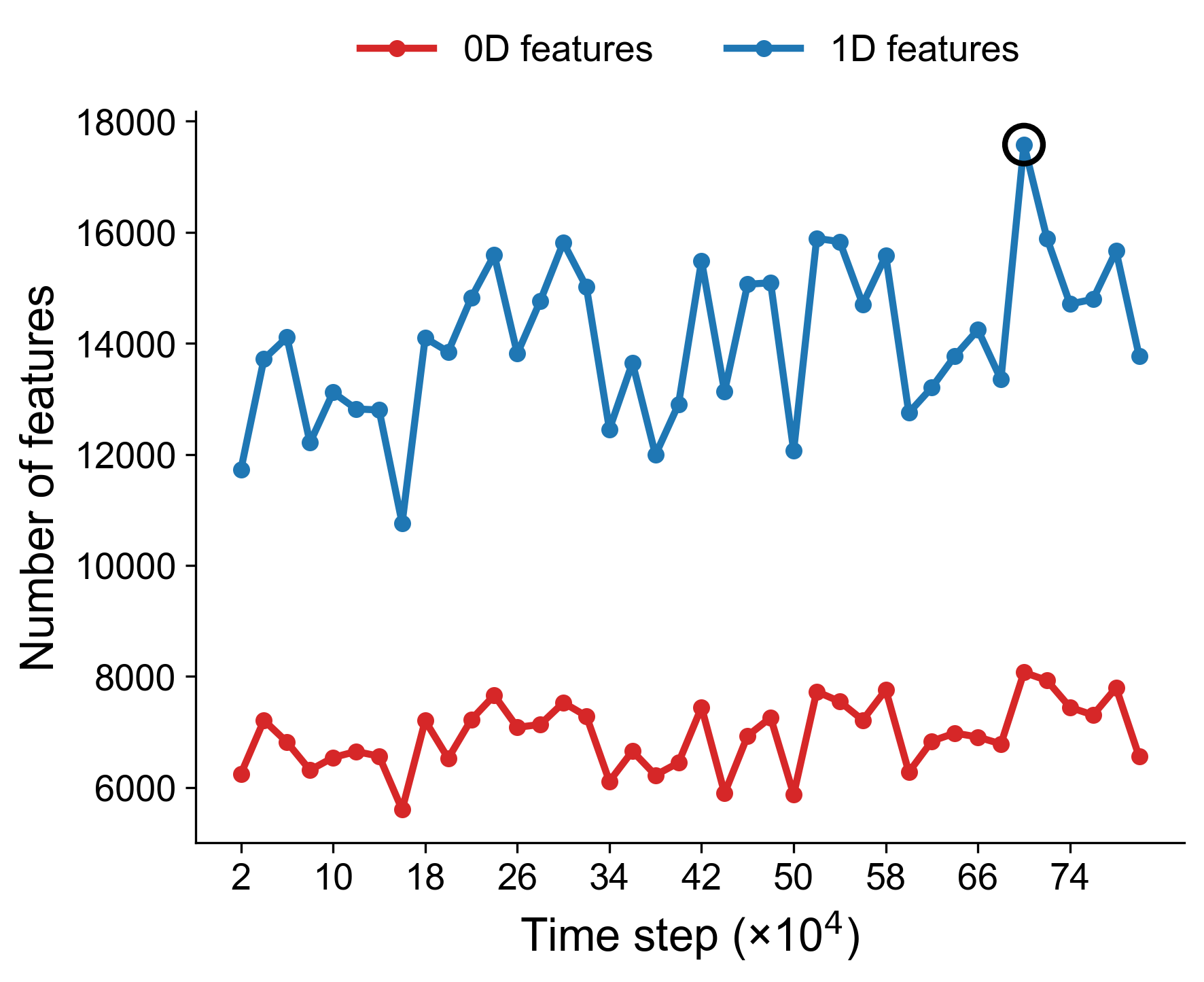}
        \caption{Sublevel persistence counts for (x--y plane).}
        \label{fig:curlz_sub}
    \end{subfigure}
    
    \vskip\baselineskip
    \begin{subfigure}[t]{0.32\linewidth}
        \centering
        \includegraphics[width=\linewidth]{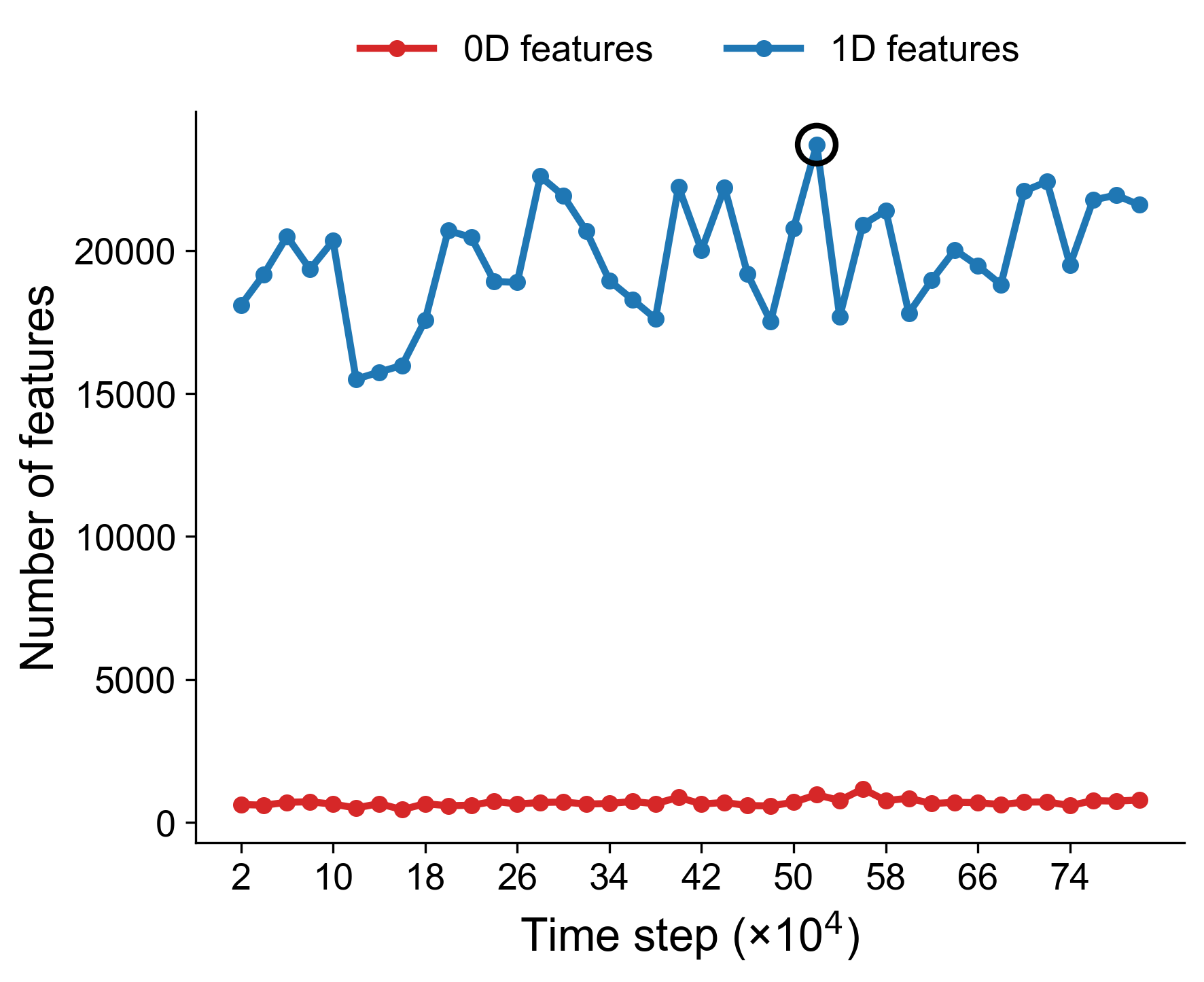}
        \caption{Superlevel persistence counts for (y--z plane).}
        \label{fig:curlx_sup}
    \end{subfigure}
    \hfill
    \begin{subfigure}[t]{0.32\linewidth}
        \centering
        \includegraphics[width=\linewidth]{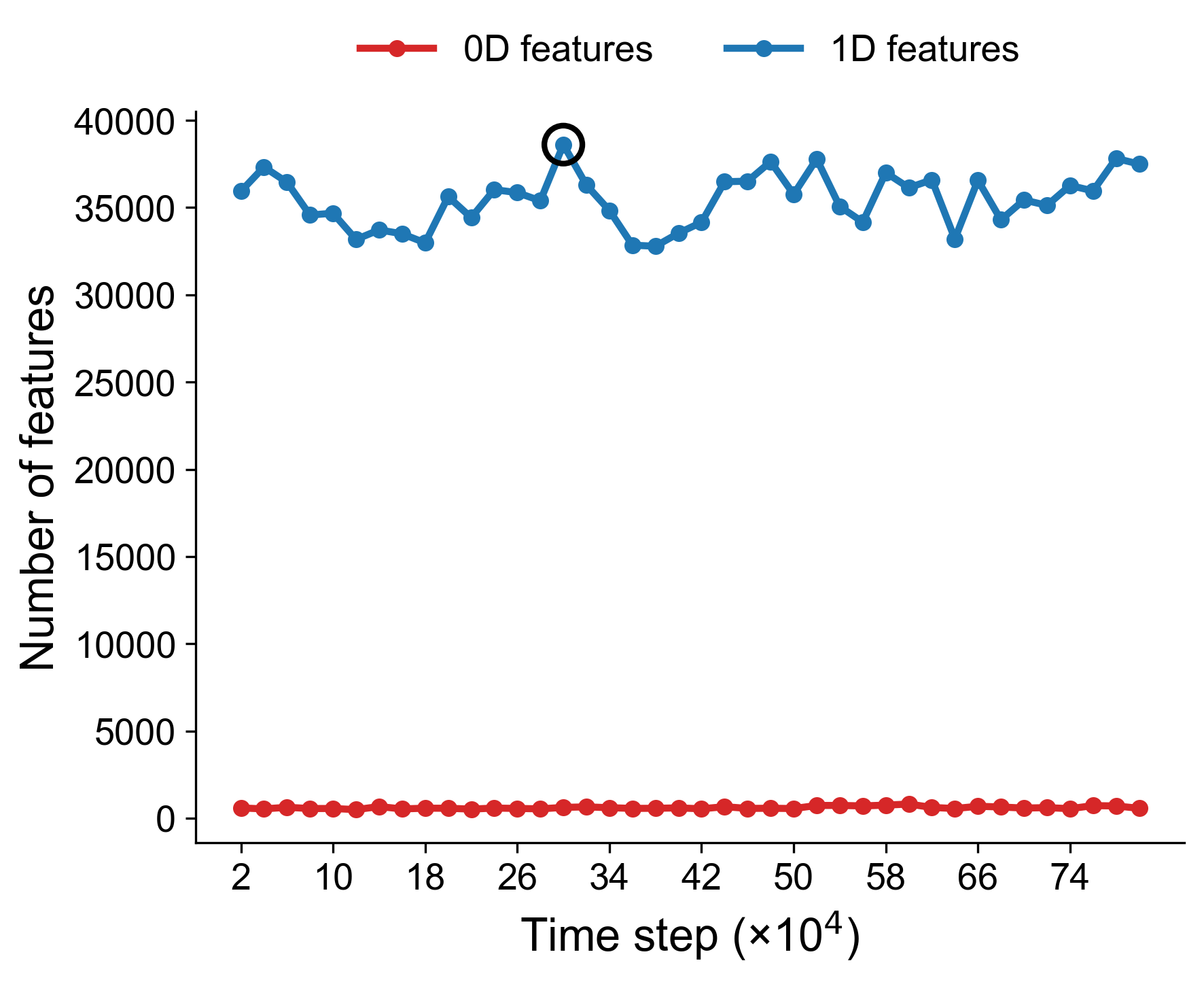}
        \caption{Superlevel persistence counts for (x--z plane).}
        \label{fig:curly_sup}
    \end{subfigure}
    \hfill
    \begin{subfigure}[t]{0.32\linewidth}
        \centering
        \includegraphics[width=\linewidth]{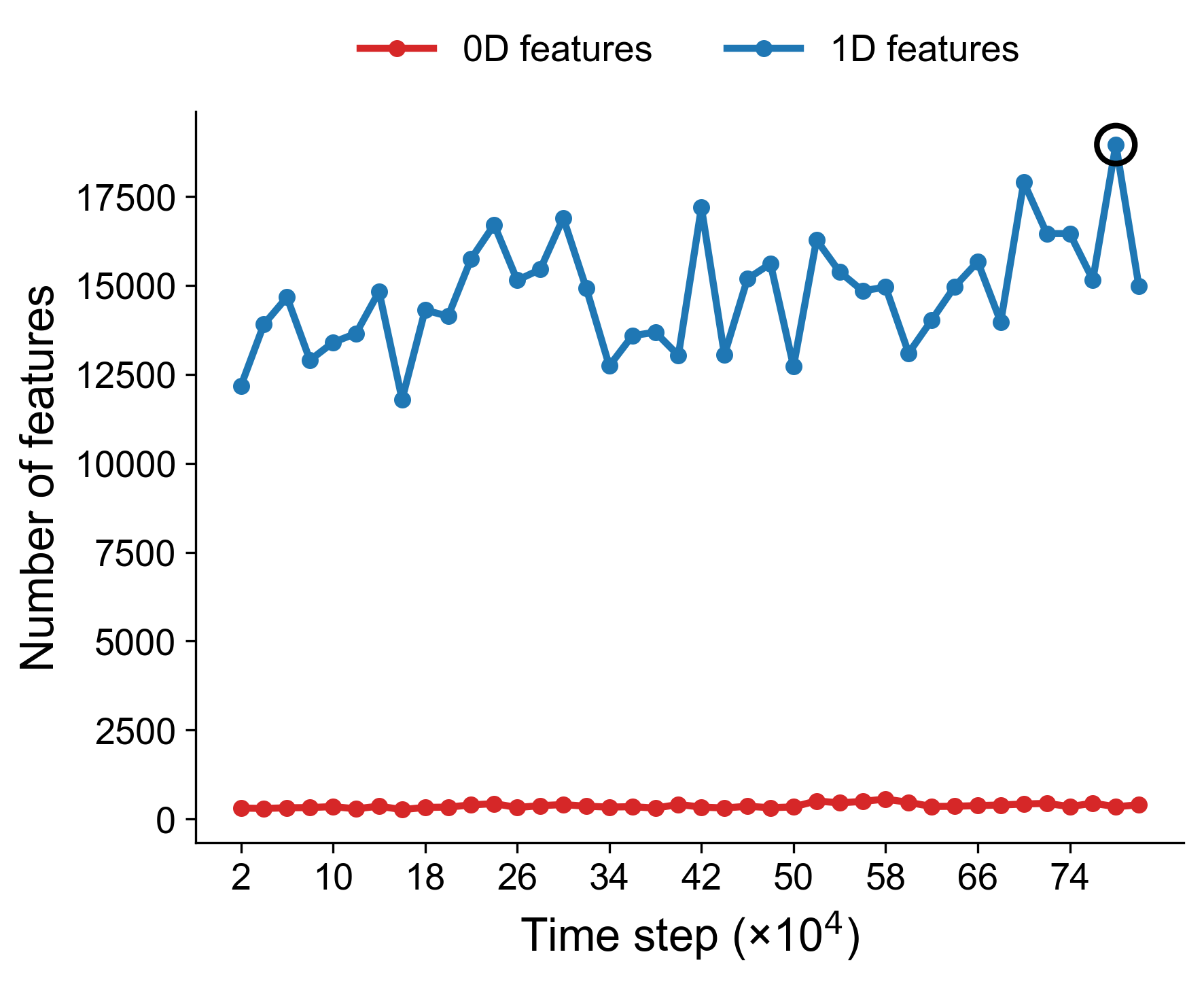}
        \caption{Superlevel persistence counts for (x--y plane).}
        \label{fig:curlz_sup}
    \end{subfigure}

    \caption{Number of topological features as a function of timestep. The top row shows sublevel set filtrations, while the bottom row shows superlevel set filtrations, for curl components across different planes.}
    \label{fig:persistence_counts}
\end{figure}

\begin{figure}[!htbp]
    \centering
    \includegraphics[width=\textwidth,height=\textheight,keepaspectratio]{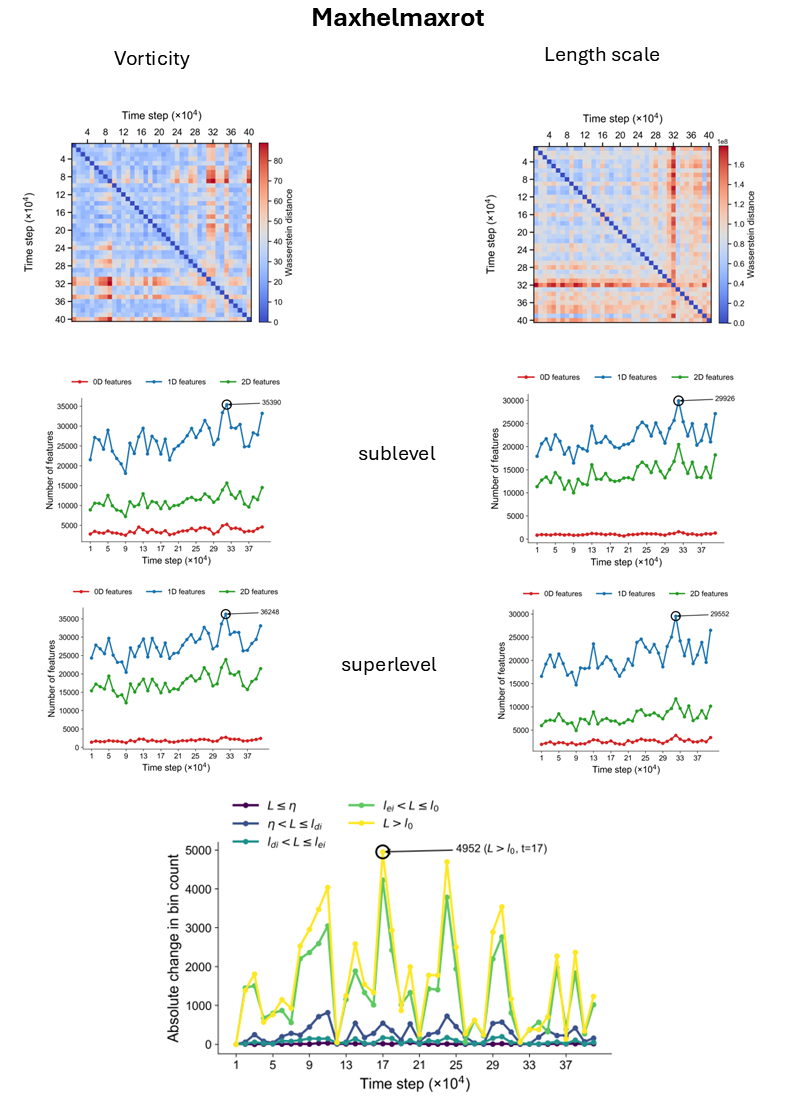}
    \caption{Maxhelmaxrot case}
    \label{fig:maxhelmaxrot1}
\end{figure}

From an experimental perspective, this framework is readily adaptable to two-dimensional Particle Image Velocimetry (PIV) datasets~\cite{ChandraDutta2025,Uchytil2025MHD,Steiner2025VortexRing}. Here, we demonstrate how two-dimensional slices of the available three-dimensional dataset can be analyzed using the same pipeline. This approach reveals directional influences of turbulent fluctuations, confirming the applicability of our analysis in two-dimensional scenarios as well (~\cref{fig:wasserstein_piv} and ~\cref{fig:persistence_counts}).

\newpage
Intermittency signatures in two dimensional scenario is explored by performing \\two–dimensional slices of the vorticity field in all three coordinate planes: $\nabla\times\mathbf{u}$ in the $y$–$z$ plane (corresponding to $\omega_x$), the $x$–$z$ plane (corresponding to $\omega_y$), and the $x$–$y$ plane (corresponding to $\omega_z$) of the ~\emph{highrotminhel} case. For each case, persistence diagrams were computed and the temporal evolution was quantified using Wasserstein distance heatmaps.
As shown in ~\cref{fig:wasserstein_piv}, the $y$--$z$ and, to a lesser extent, the $x$--$y$ planes show a marked increase in Wasserstein distances over the interval $t = 52$--$58$, whereas the $x$--$z$ plane exhibits a comparable variation at later times, consistent with the three-dimensional results. This confirms that the two–dimensional slices preserve the intermittency signatures identified in the full 3D fields.

Among these, the $x$--$y$ plane (associated with $\omega_z$) displays the clearest and most localized change, indicating that the strongest reorganization of vorticity occurs along the rotation axis. This is consistent with the physics of rotating turbulence, where vortex tubes preferentially align parallel to the axis of rotation.\\ 
Furthermore,~\cref{fig:persistence_counts} shows that the most significant variations occur in the 1D topological features, corresponding to vortex tubes, across all three planes. Their enhanced temporal fluctuations reinforce the conclusion that intermittent events in rotating turbulence are mediated primarily by the creation, stretching, and decay of filamentary vortex structures aligned with the rotation axis.\\
Additionally, we explored the role of the ``Wasserstein distance order ($q$)'' , which reflects how large-scale discrepancies in the system influence the underlying topology. Higher values of $q$ emphasize larger-scale variations. By extending our analysis to $q = 4$, we found broad agreement with our earlier results, thereby adding confidence to the reliability of the method.\\

In this study, we primarily focused on the Wasserstein distances of one-dimensional topological features, as they are physically meaningful: loops correspond to vortex tubes, whose gradual evolution reveals intrinsic properties of the flow. While we also computed zero-dimensional features (connected components), these largely reflect clusters of vortices with similar values of physical parameters such as vorticity, which are less informative from a topological perspective. Similarly, two-dimensional features, corresponding to vortex sheets, appeared far less frequently in our results. A more detailed breakdown of these higher- and lower-dimensional features is provided in Appendix H (Details of the Appendix can be found in \cref{tab:appendix_overview}).

\begin{table}[H]
\centering
\begin{tabular}{|c|p{9cm}|p{3cm}|}
\hline
\textbf{Appendix Section} & \textbf{Details of the Content} & \textbf{Figure(s)} \\ 
\hline
A  & maxhelmaxrot & Fig.~15--21 \\ 
\hline
B  & maxrotminhel & Fig.~22--28 \\ 
\hline
C  & nohelmaxrot & Fig.~29--35 \\ 
\hline
D & nohelnorot & Fig.~36--42 \\ 
\hline
E & norotmaxhel & Fig.~43--49 \\ 
\hline
F & norotminhel & Fig.~50--56 \\ 
\hline
G & 4th order Wasserstein distance heatmaps (all cases) & Fig.~57--63 \\ 
\hline
H & 0D and 2D Wasserstein distance heatmaps (highrotminhel) & Fig.~64--71 \\ 
\hline
\end{tabular}
\caption{Overview of appendix sections, associated content, and corresponding figures.}
\label{tab:appendix_overview}
\end{table}

\clearpage
\section{Case Study: A}
\label{sec:maxhelmaxrot}

\clearpage

\begin{figure}[!htbp]
    \centering
    \includegraphics[width=\textwidth,height=\textheight,keepaspectratio]{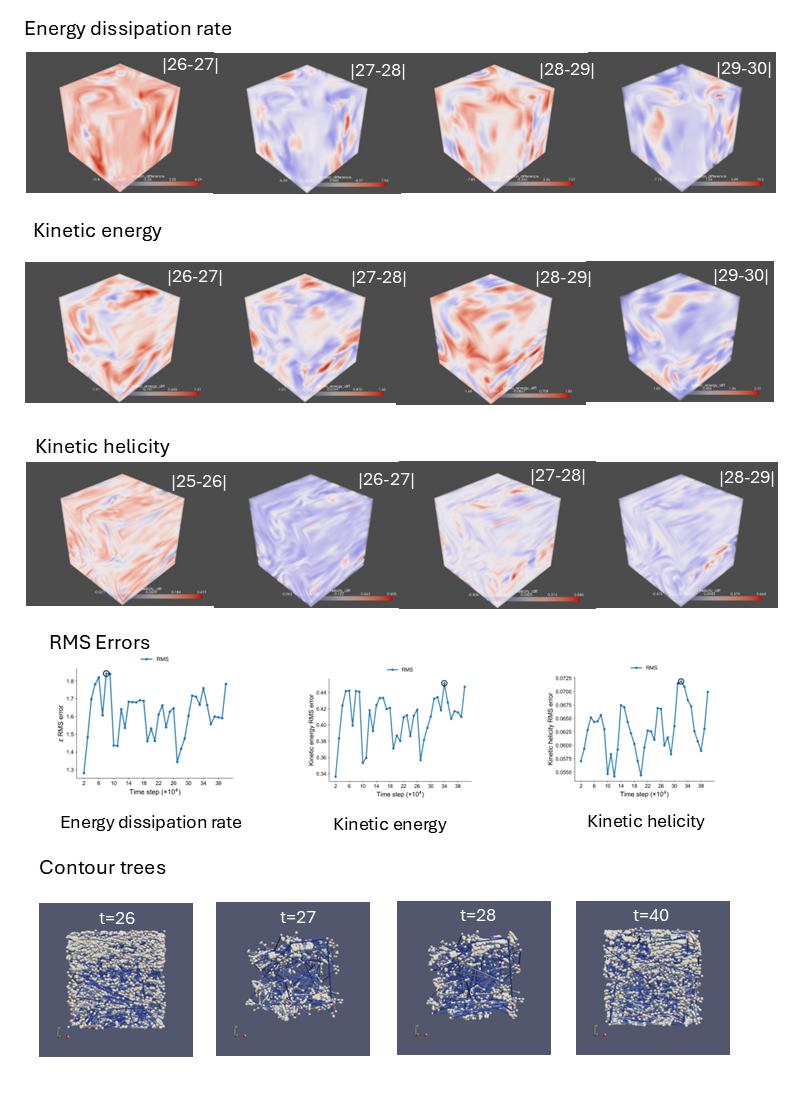}
    \caption{Maxhelmaxrot case}
    \label{fig:maxhelmaxrot2}
\end{figure}

\clearpage
\section{Case Study: B}
\label{sec:maxrotminhel}

\begin{figure}[!htbp]
    \centering
    \includegraphics[width=\textwidth,height=\textheight,keepaspectratio]{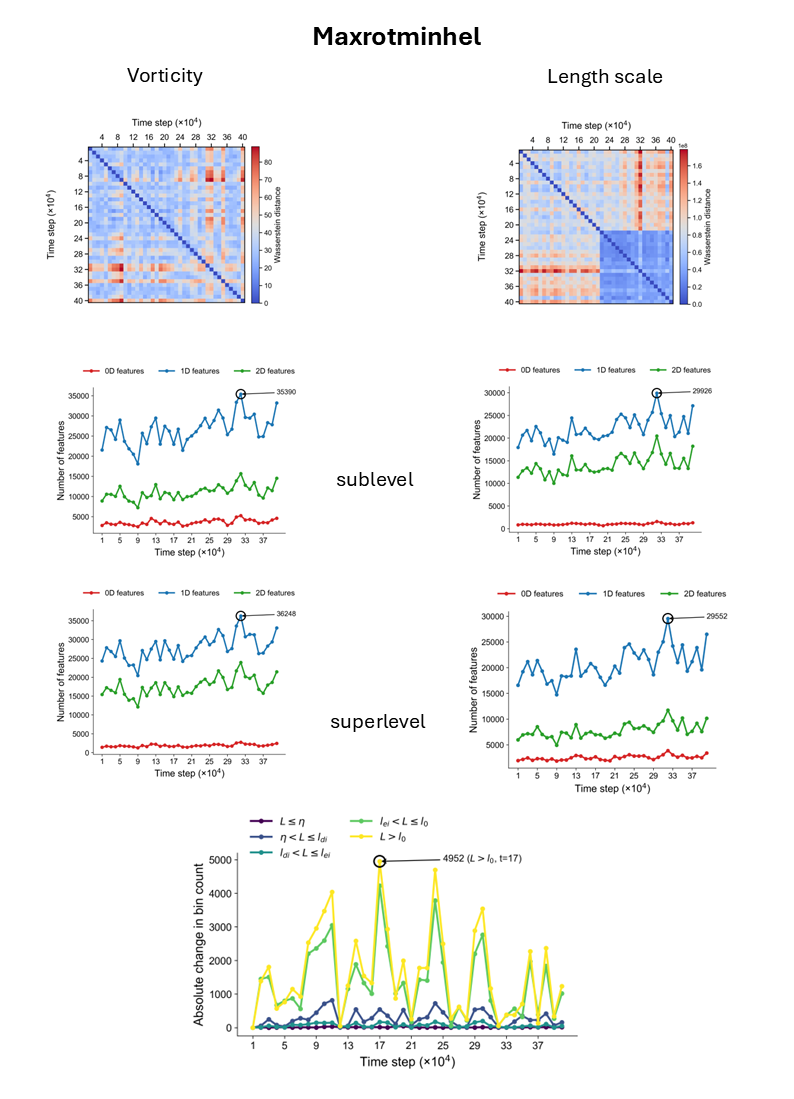}
    \caption{Maxrotminhel case}
    \label{fig:maxrotminhel1}
\end{figure}

\clearpage

\begin{figure}[!htbp]
    \centering
    \includegraphics[width=\textwidth,height=\textheight,keepaspectratio]{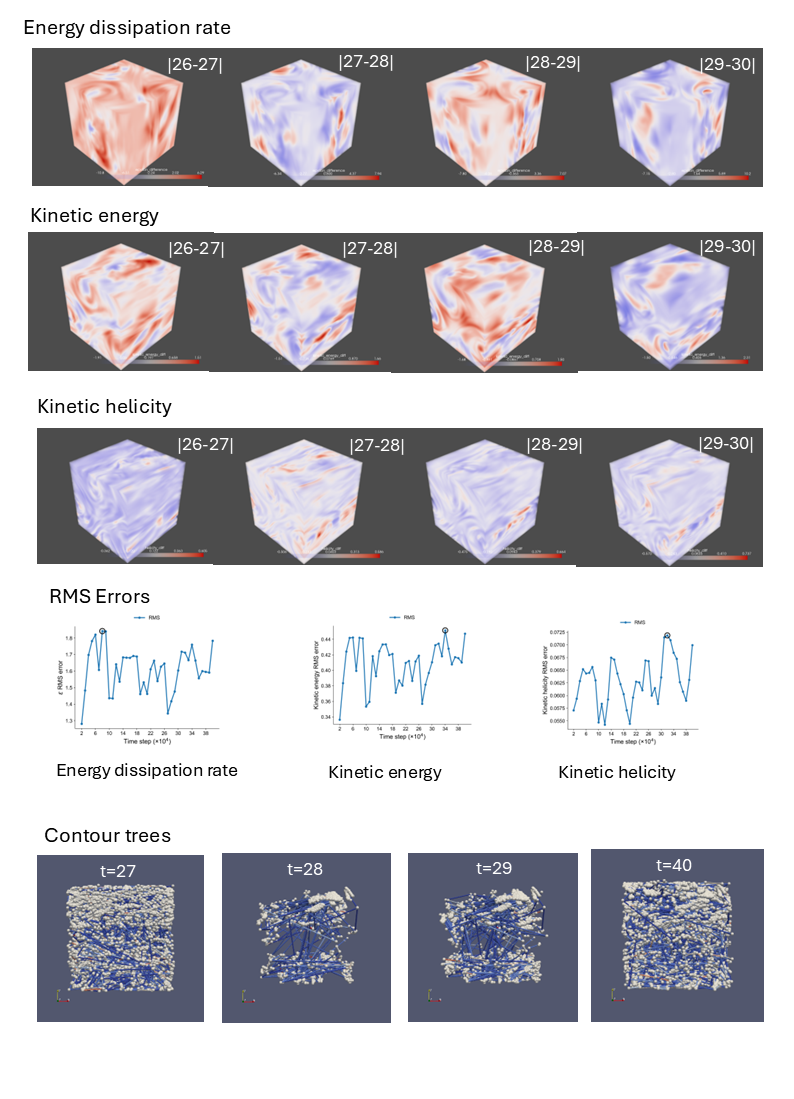}
    \caption{Maxrotminhel case}
    \label{fig:maxrotminhel2}
\end{figure}

\clearpage
\section{Case Study: C}
\label{sec:nohelmaxrot}

\begin{figure}[!htbp]
    \centering
    \includegraphics[width=\textwidth,height=\textheight,keepaspectratio]{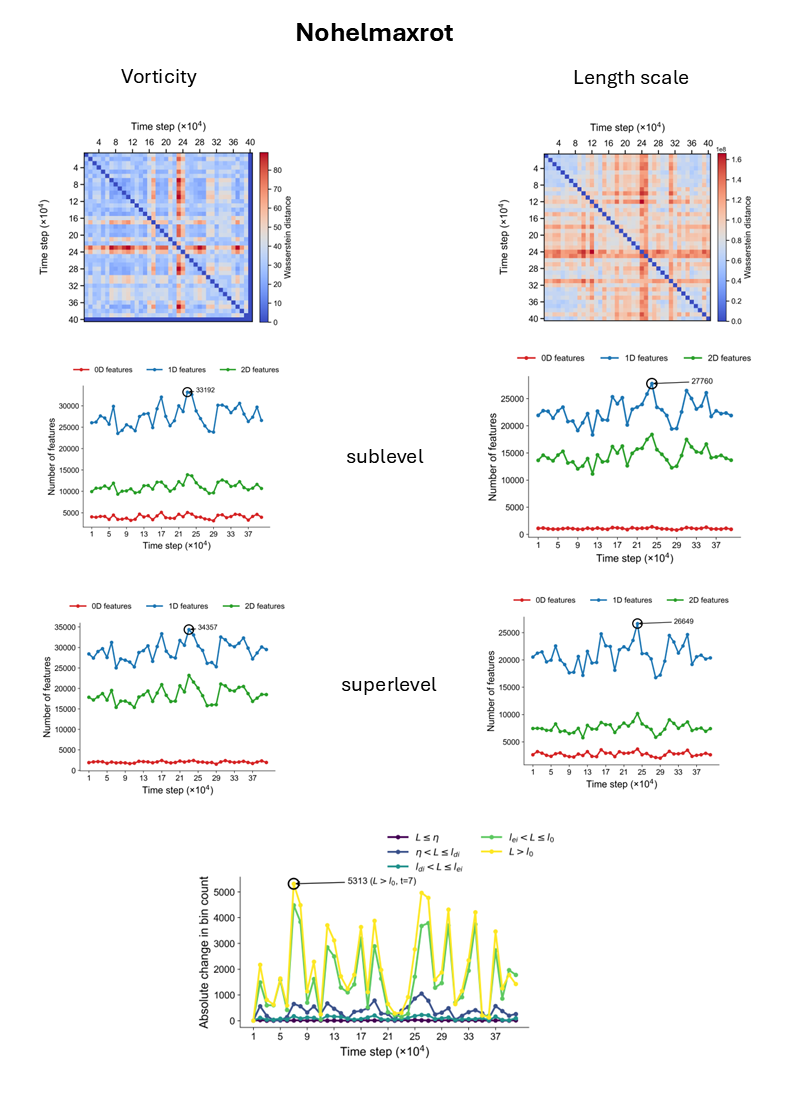}
    \caption{Nohelmaxrot case}
    \label{fig:nohelmaxrot1}
\end{figure}

\clearpage

\begin{figure}[!htbp]
    \centering
    \includegraphics[width=\textwidth,height=\textheight,keepaspectratio]{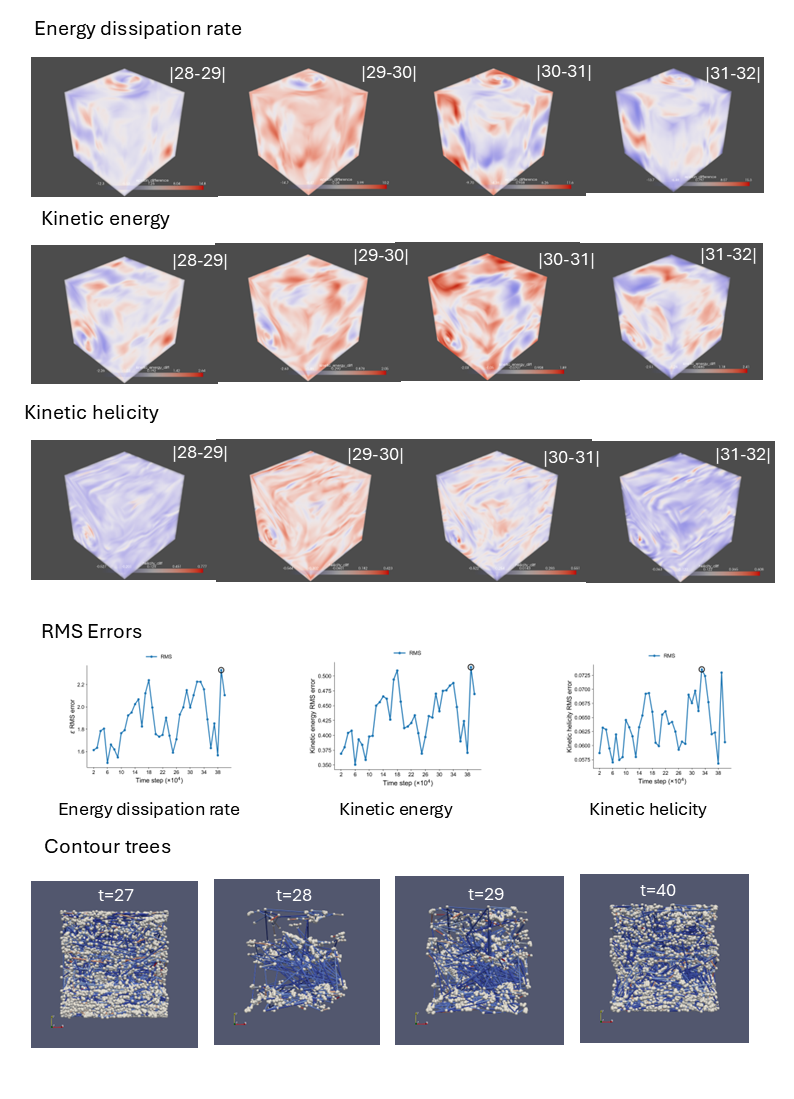}
    \caption{Nohelmaxrot case}
    \label{fig:nohelmaxrot2}
\end{figure}

\clearpage
\section{Case Study: D}
\label{sec:nohelnorot}

\begin{figure}[!htbp]
    \centering
    \includegraphics[width=\textwidth,height=\textheight,keepaspectratio]{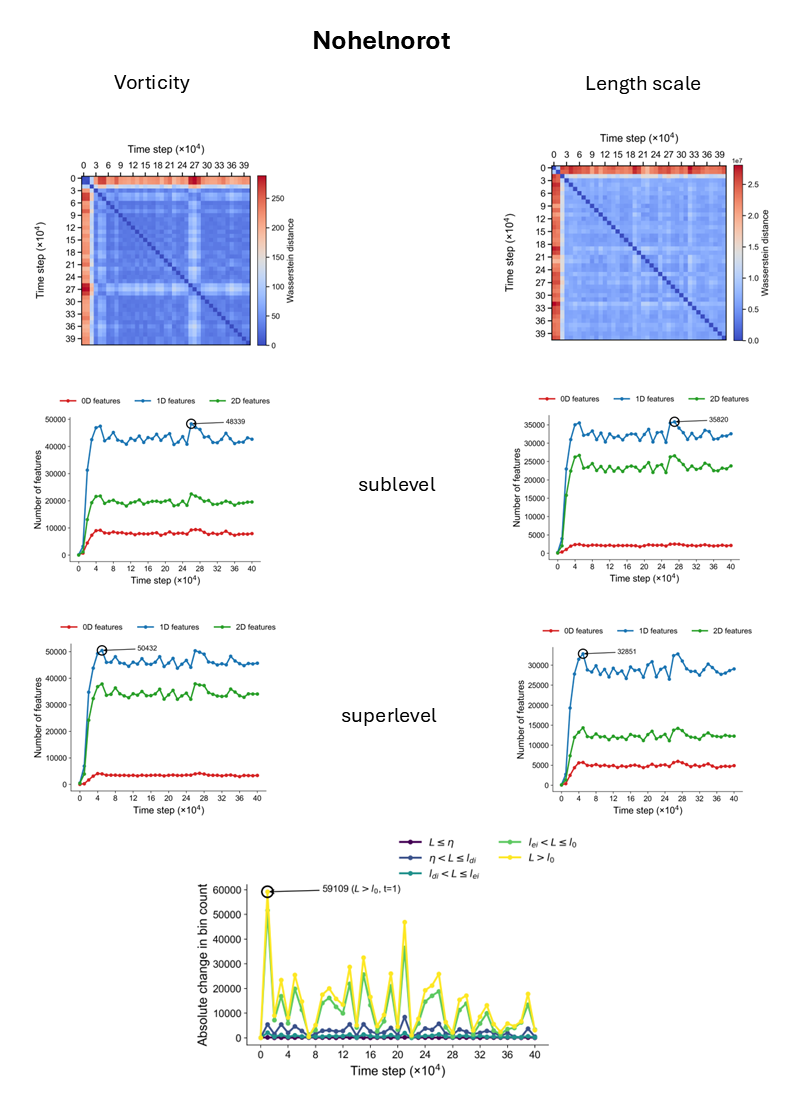}
    \caption{Nohelnorot case}
    \label{fig:nohelnorot1}
\end{figure}

\clearpage

\begin{figure}[!htbp]
    \centering
    \includegraphics[width=\textwidth,height=\textheight,keepaspectratio]{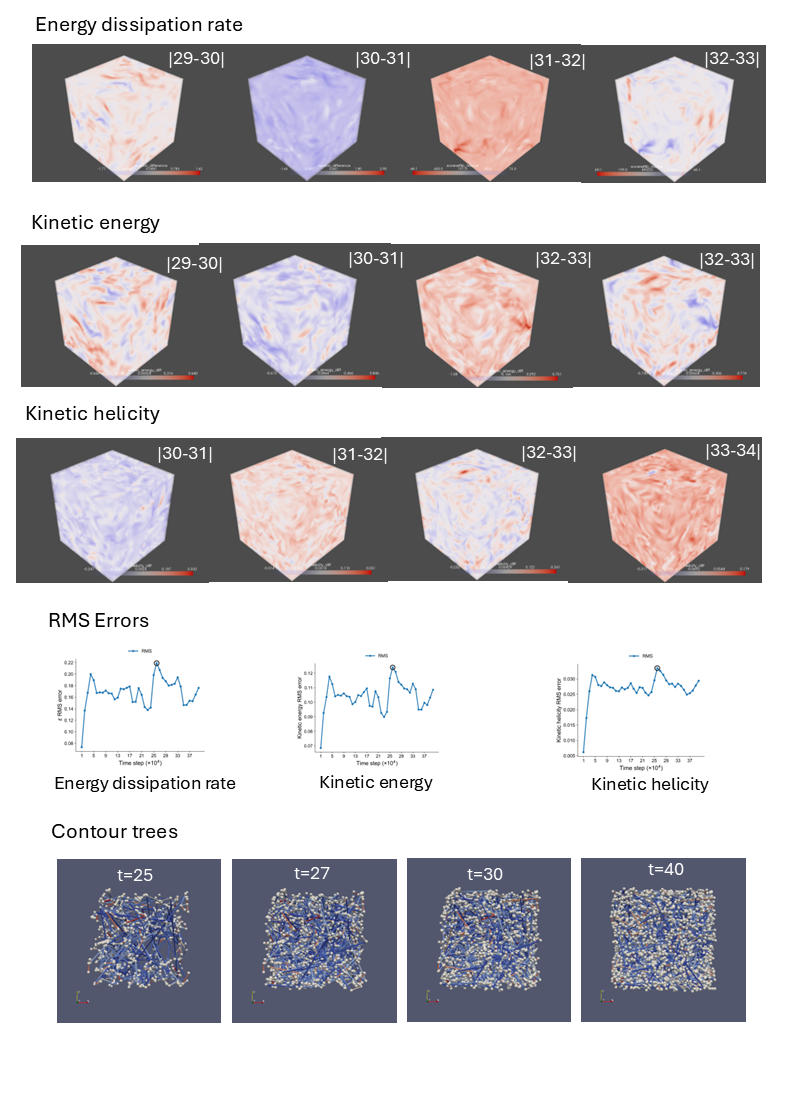}
    \caption{Nohelnorot case}
    \label{fig:nohelnorot2}
\end{figure}

\clearpage
\section{Case Study: E}
\label{sec:norotmaxhel}

\begin{figure}[!htbp]
    \centering
    \includegraphics[width=\textwidth,height=\textheight,keepaspectratio]{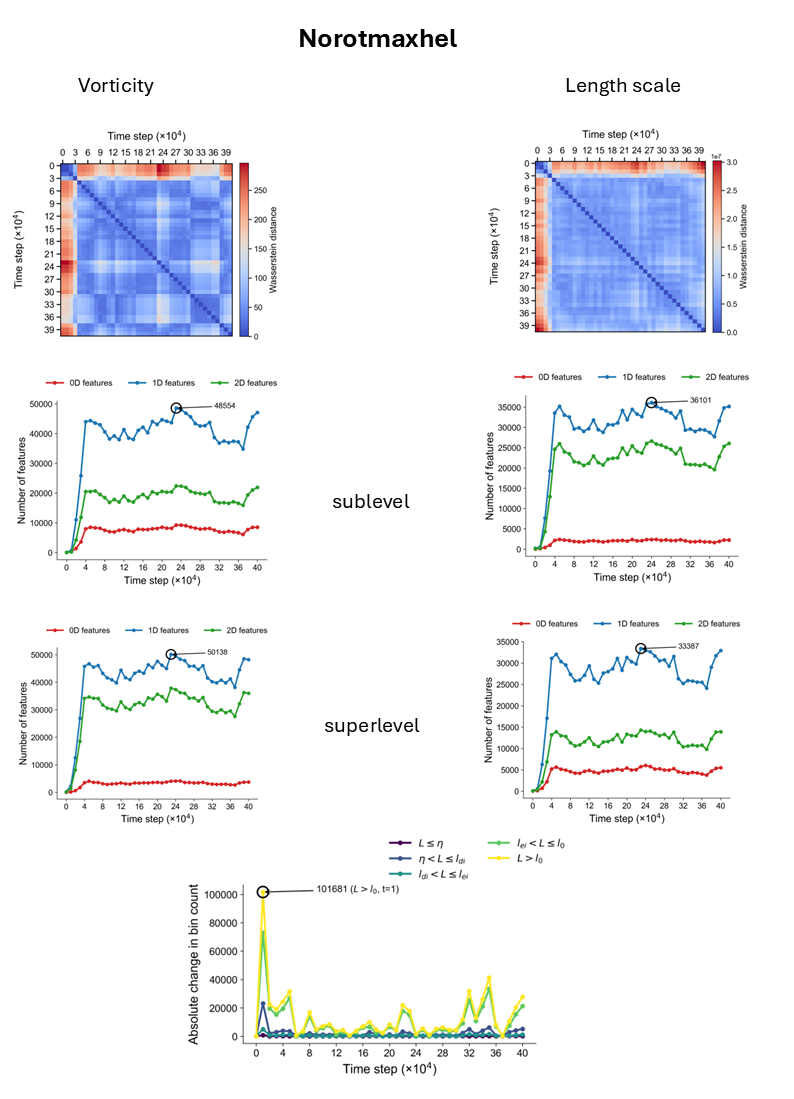}
    \caption{Norotmaxhel case}
    \label{fig:norotmaxhel1}
\end{figure}

\clearpage

\begin{figure}[!htbp]
    \centering
    \includegraphics[width=\textwidth,height=\textheight,keepaspectratio]{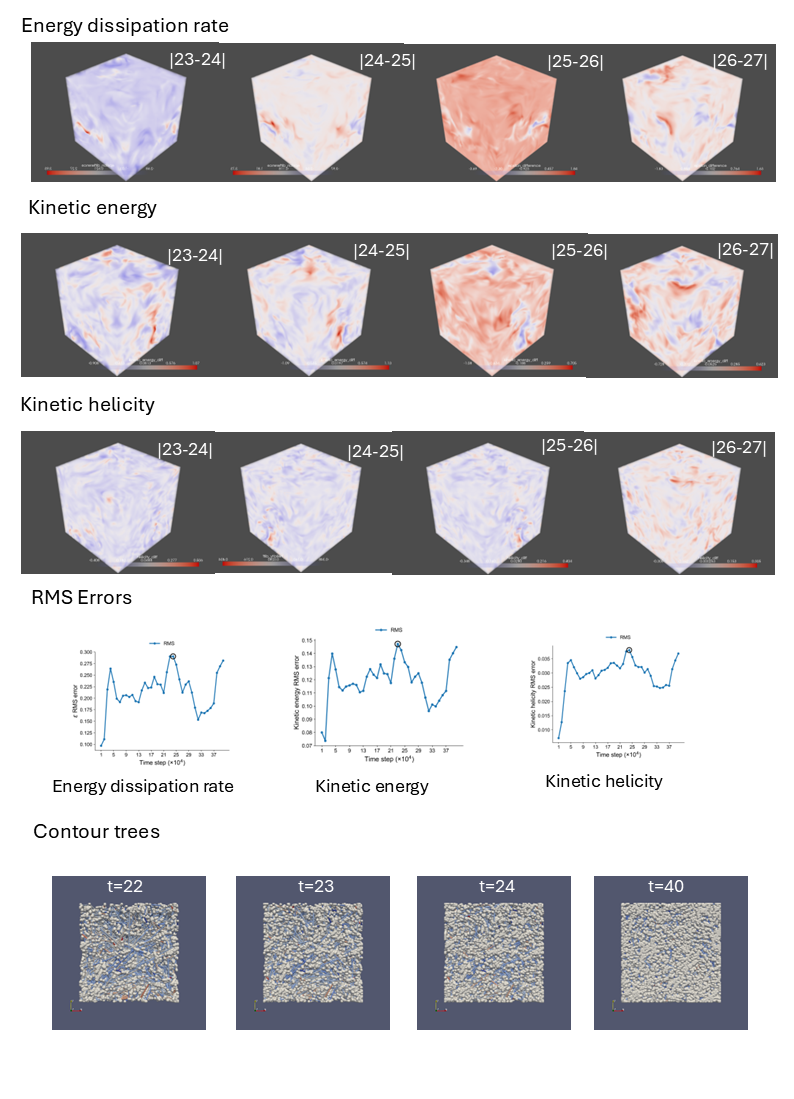}
    \caption{Norotmaxhel case}
    \label{fig:norotmaxhel2}
\end{figure}

\clearpage
\section{Case Study: F}
\label{sec:norotminhel}

\begin{figure}[!htbp]
    \centering
    \includegraphics[width=\textwidth,height=\textheight,keepaspectratio]{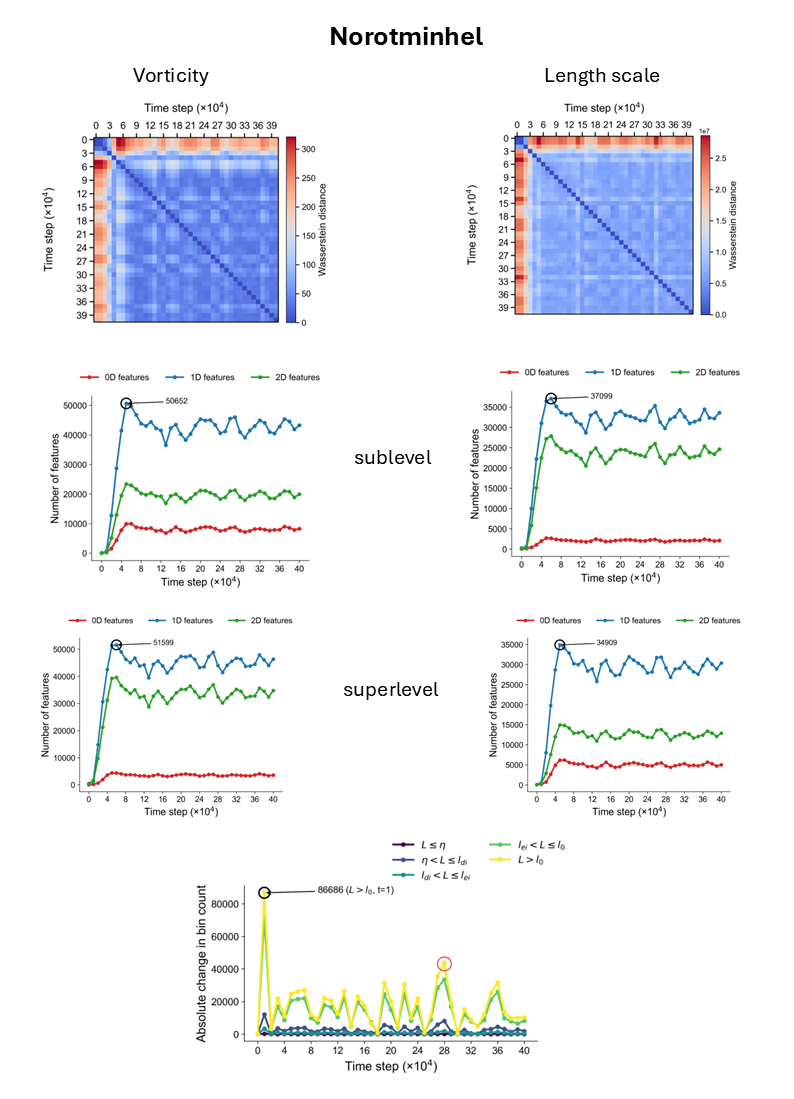}
    \caption{Norotminhel case}
    \label{fig:norotminhel1}
\end{figure}

\clearpage

\begin{figure}[!htbp]
    \centering
    \includegraphics[width=\textwidth,height=\textheight,keepaspectratio]{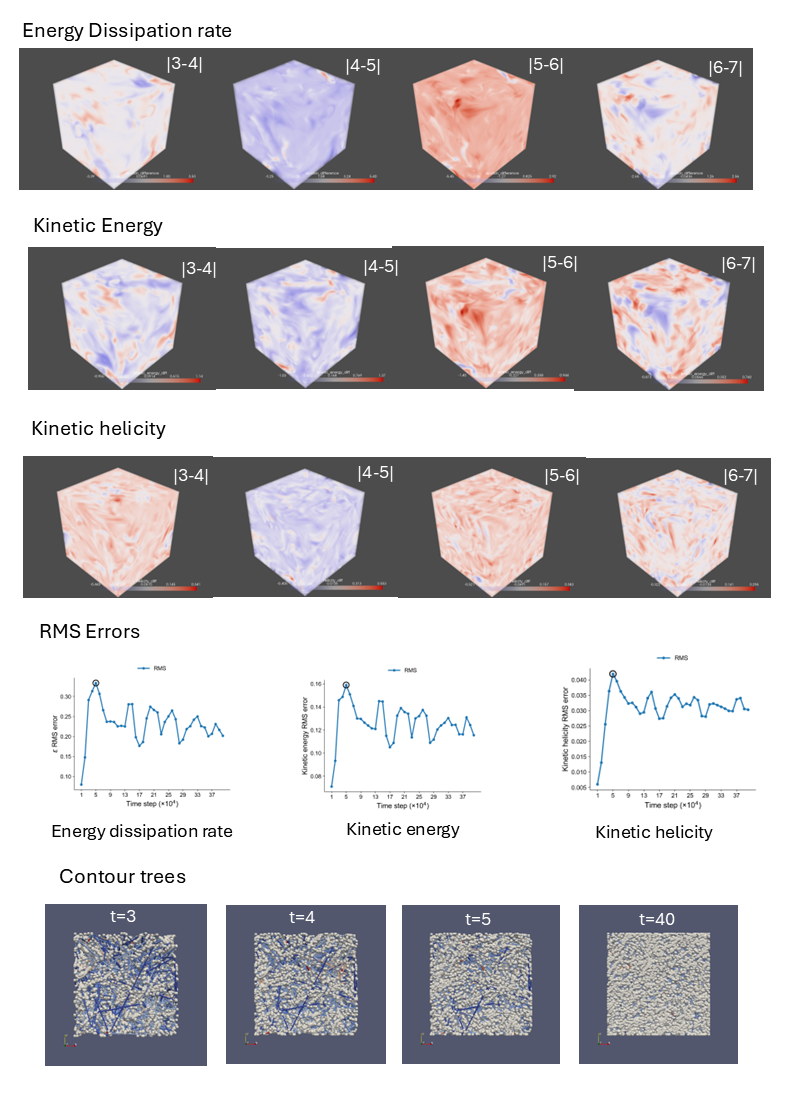}
    \caption{Norotminhel case}
    \label{fig:norotminhel2}
\end{figure}

\clearpage
\section{Case Study: }
\label{sec:fourth-order-wd}

\begin{figure}[!htbp]
    \centering
    \includegraphics[width=\textwidth,height=\textheight,keepaspectratio]{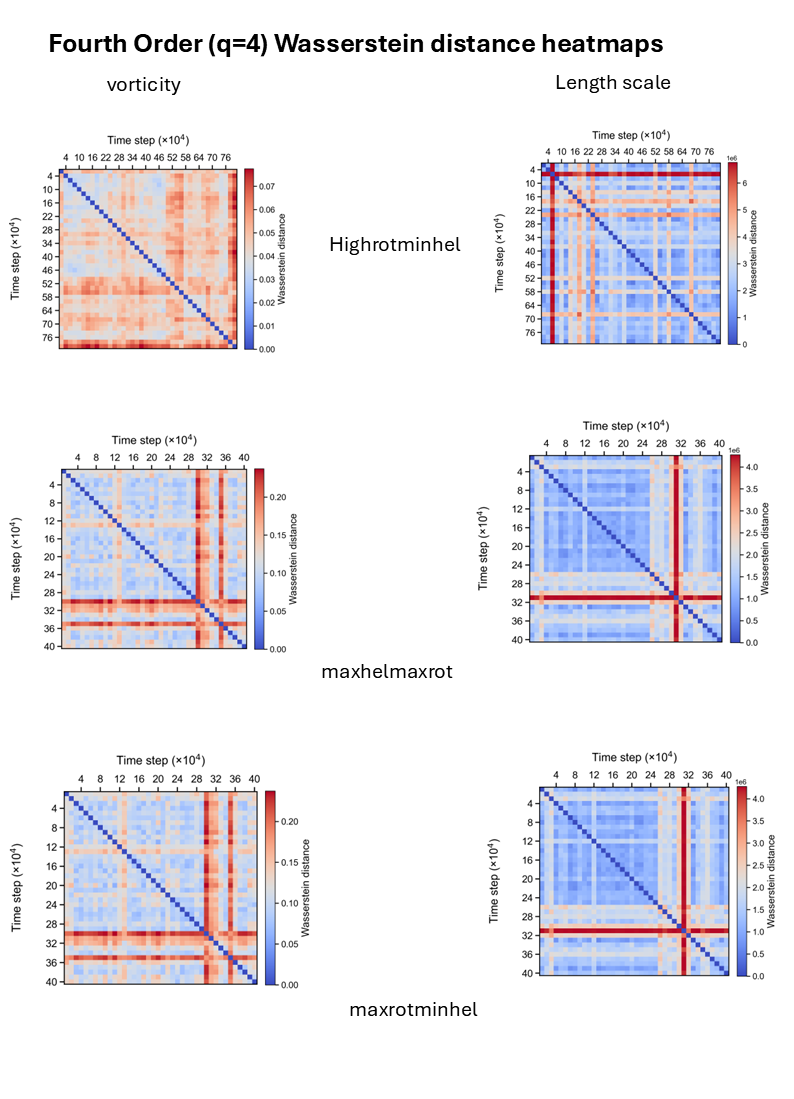}
    \caption{4th order WD}
    \label{fig:4d_1}
\end{figure}

\clearpage

\begin{figure}[!htbp]
    \centering
    \includegraphics[width=\textwidth,height=\textheight,keepaspectratio]{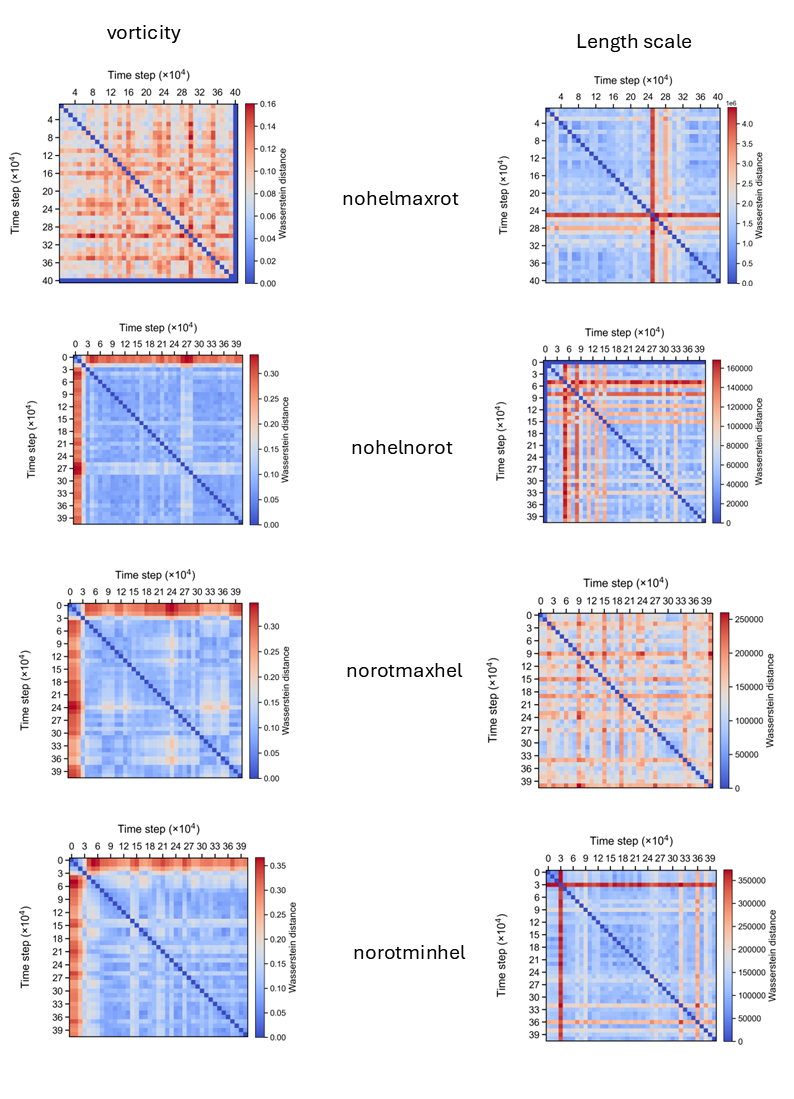}
    \caption{4th order WD}
    \label{fig:4d_2}
\end{figure}

\section{Case Study:} 
\label{sec:zero-and-two-wd} 
\begin{figure}[!htbp] 
\centering \includegraphics[width=\textwidth,height=\textheight,keepaspectratio]{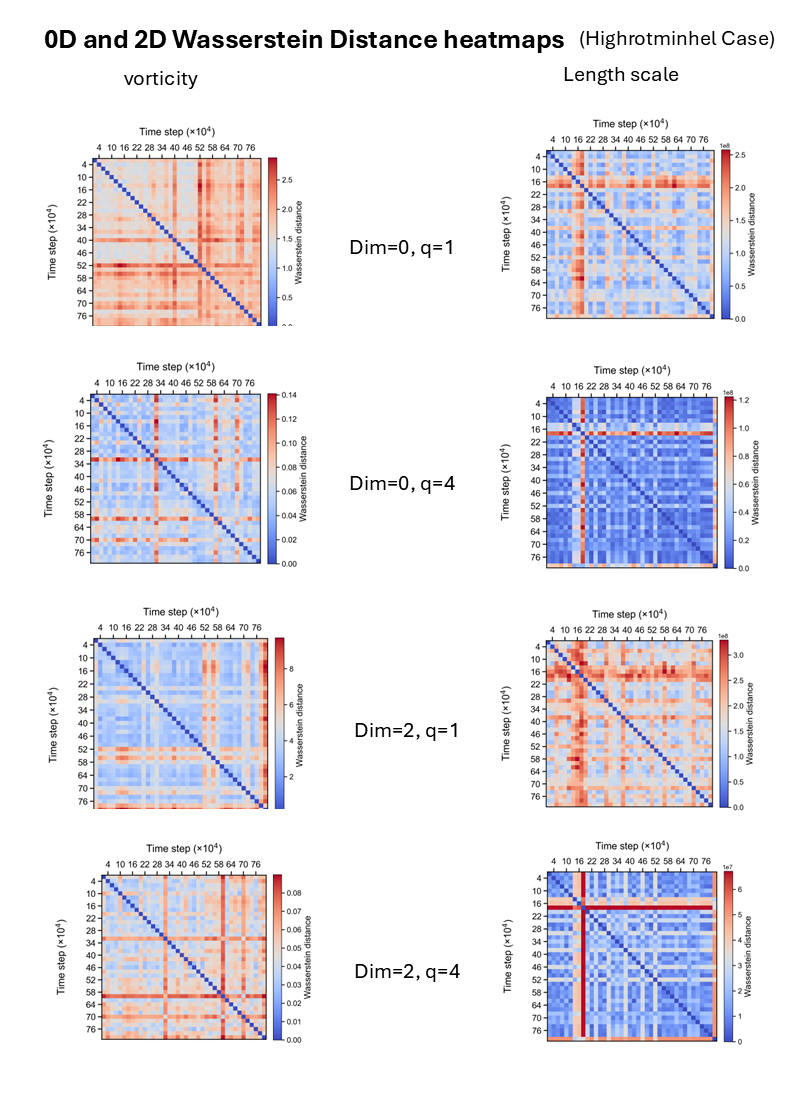} 
\caption{0D and 2D WD heatmaps} 
\label{fig:02d_heatmap} 
\end{figure}
\end{document}